\documentclass[a4paper,12pt]{article}
\usepackage[utf8]{inputenc}
\usepackage{cancel}
\usepackage{ulem}
\usepackage{amsfonts}
\usepackage{amssymb}
\usepackage{adjustbox}
\usepackage{graphicx}
\usepackage{amsmath,bm}
\usepackage{enumerate}
\usepackage{comment}
\usepackage{mathtools}
\usepackage{tikz} \usetikzlibrary{calc}
\usepackage[countmax]{subfloat}
\usepackage{xcolor}
\usepackage{float}
\usepackage{color}
\usepackage{siunitx}
\usepackage{array}
\usepackage{booktabs}
\usepackage{tabularx}
\usepackage{here}
\usepackage{cite}
\usepackage{subcaption}
\usepackage{mwe}
\usepackage{multirow}

\usepackage{mathrsfs}
\usepackage{float,epsfig}
\usepackage{dcolumn}
\usepackage{pgfplots}
\usepackage{graphicx}
\usepackage{bm}
\usepackage{amsmath,amssymb,amsthm}
\usepackage[colorlinks=true,linkcolor=blue,citecolor=red]{hyperref}
\usepackage[a4paper,top=2cm,bottom=2cm,left=2cm,right=2cm,marginparwidth=2cm]{geometry}

\definecolor{lime}{HTML}{A6CE39}
\newcommand{\orcidicon}{%
	\begin{tikzpicture}
	\draw[lime, fill=lime] (0,0)
	circle [radius=0.16]
	node[white] {{\fontfamily{qag}\selectfont \tiny ID}};
	\draw[white, fill=white] (-0.0625,0.095)
	circle [radius=0.007];
	\end{tikzpicture}   \hspace{-2mm}
}

\newcommand\orcidHasan{{\href{https://orcid.org/0000-0001-7408-0910}{\orcidicon}}}
\newcommand\orcidKarima{{\href{https://orcid.org/0000-0001-5419-8516}{\orcidicon}}}
\newcommand\orcidZakaria{{\href{https://orcid.org/0009-0006-7872-4713}{\orcidicon}}}

\title{\bf  Modeling HF-QPOs in Microquasars and AGNs: Charged Particles around Black Holes with CDM Halos
}

\author{
	Z. Ahal\orcidZakaria\!\!$^{1}$\thanks{zakaria.ahal@edu.uiz.ac.ma},  
	H.  El Moumni\orcidHasan\!\!$^1$\thanks{h.elmoumni@uiz.ac.ma }, K. Masmar\orcidKarima\!\!$^{1}$\thanks{k.masmar@uiz.ac.ma (Corresponding author)}\\
{\small $^{1}$ LPTHE, Physics Department, Faculty of Sciences, Ibnou Zohr University, Agadir, Morocco. }
}
\date{\today}
\begin{document} 
	\maketitle
 \begin{abstract}

HF QPOs are among the most intriguing phenomena observed in LMXBs containing BHs or neutron stars. In this work, we investigate charged particles' dynamics in the nearby of a Schwarzschild-like BH embedded in a uniform magnetic field and surrounded on all sides by CDM. Thereby gaining deeper insight into the influence of magnetic and DM distributions on observable phenomena near compact objects. We first present a modified metric, which incorporates the effects of a CDM, and we explore how both DM and magnetic fields influence the effective potential, stable circular orbits, and escape conditions for ionized particles. Employing a Hamiltonian formalism, we analyze the energy boundaries and ISCO, demonstrating that CDM causes an outward shift of the ISCO, while magnetic fields tend to pull it closer to the event horizon.

We compute the fundamental oscillation frequencies—radial, latitudinal, Keplerian, and Larmor—and demonstrate how their variation depends on the combined influence of CDM and magnetic field strength. The resulting frequency structure allows us to identify resonance radii associated with HF QPOs, particularly those in 3:2 ratios observed in microquasars. We assess several theoretical models for QPO generation, including the Epicyclic Resonance (ER0, ER1, ER2, ER3, ER4, ER5), Relativistic Precession (RP0, RP1, RP2), Tidal Disruption (TD), and Warped Disk (WD) models. A comparative fit of observational data from GRS 1915+105, H1743-322, XTE 1550-564, and GRO 1655-40 demonstrates that the ER4 model provides the best match for HF QPOs in the presence of moderate magnetic flux $B' = {-0.01,\,0,\,0.01}$. In contrast, for supermassive black holes in some active galactic nuclei (AGNs), the model that best fits the majority of the data depends on the magnetic field strength: ER0 for $B' = 0$, and ER5 for $B' = \pm 0.01$.

Our results highlight the importance of including both magnetic and dark matter (DM) effects in strong-field astrophysics and support the use of HF QPOs as sensitive probes of BH environments. This study opens new perspectives for exploring particle dynamics, accretion disk structure, and observational signatures of DM near compact objects.
\end{abstract}
	
    \tableofcontents

\section{Introduction}
High-frequency quasi-periodic oscillations (HF QPOs) observed in the X-ray flux of low-mass X-ray binaries (LMXBs) consisting of BHs (BHs) or neutron stars offer a valuable insight into the physics of strong gravitational fields \cite{Remillard:2006fc,Belloni:2012sv,Yuan:2014gma}. These oscillations, typically appear as twin peaks in the power spectral density of X-ray observations, exhibit characteristic frequencies in hundreds of hertz. Remarkably, in many cases, the captured frequencies of HF QPOs appear in a commensurate 3:2 ratio, suggesting the presence of resonant processes in the inner accretion disk region \cite{Abramowicz:2001bi, Abramowicz:2004rr}.

The study of HF QPOs is of significant astrophysical importance, as their frequencies are believed to be directly linked to the fundamental properties of the compact object, including its mass, spin, and the spacetime geometry near the event horizon \cite{Motta:2013wga}. Various theoretical frameworks have been suggested to explain the origin of HF QPOs, ranging from relativistic precession and disk oscillation models to non-linear resonances between different oscillation modes in the accretion flow \cite{Stella:1999sj, boshkayev2023quasiperiodic,Abramowicz:2004je,boshkayev2023numerical}. More recently, the role of electromagnetic interactions, particularly in systems with strong magnetic fields, has been explored as a potential contributor to HF QPOs formation \cite{Stuchlik:2014foa,Churilova:2023swu}.

In addition to classical general relativity frameworks, modifications to black hole (BH) spacetimes arising from quantum gravity corrections and alternative theories of gravity have been investigated through HF QPOs \cite{Bambi:2015kza,Mustafa:2024kjy}. Quantum-corrected BHs, regular BHs with modified cores, and deviations from Kerr or Schwarzschild solutions can introduce perturbations in the epicyclic frequencies of accreting particles, thereby influencing the HF QPOs spectrum \cite{McClintock:2006xd,Johannsen:2010ru}. Observational constraints on these models indirectly test fundamental physics beyond Einstein's theory of general relativity.

The Cold DM (CDM) paradigm, which posits that DM is composed of non-relativistic, collisionless particles, has proven remarkably effective at explaining large-scale cosmological phenomena. It unveils an excellent fit to the observed temperature anisotropies in the Cosmic Microwave Background (CMB), the matter power spectrum, and the distribution and abundance of galaxy clusters \cite{Planck:2015fie,Bullock:2017xww}. According to this model, DM must be electrically neutral or, if charged, have a charge significantly less than that of the electron—unless in the exceptionally massive particles \cite{Davidson:1993sj,Gupta:2021rod,Feng:2023ubl,Fung:2023euv}. This leaves open the possibility of DM being made of millicharged particles characterized by a fractional electric charge, where such particles naturally arise in various extensions of the Standard Model \cite{Berlin:2023gvx,Bertone:2004pz}. These hypothetical particles may be either fermionic or bosonic.

Although the CDM model reliably predicts the hierarchical expansion of the cosmic structure on large scales, it faces challenges at galactic and subgalactic scales. Discrepancies have emerged from comparisons between simulations and observational data, particularly in the context of the Local Group's dwarf satellite galaxies \cite{Bullock:2017xww}. These satellites, being dark-matter-dominated systems, serve as critical laboratories for testing the microphysics of DM. The advent of ultra-faint dwarf galaxy discoveries has addressed some earlier concerns, such as the “missing satellites” problem \cite{Kim:2017iwr}, but newer issues have arisen. Notably, simulations predict an overabundance of dense subhaloes that are too massive to correspond to the brightest observed of the Milky Way satellites, a problem known as “too big to fail” \cite{Boylan-Kolchin:2011qkt}. Additionally, the observation that satellite galaxies around both the Milky Way and Andromeda tend to lie in vast, rotating planar structures remains an unexplained anomaly within the standard CDM framework \cite{Qiao:2022cge,Zhang:2024hle}.

\textcolor{black}{The influence of CDM on compact astrophysical objects is often introduced through large-scale galactic density profiles, such as the Fukugita–Nakamura–Waga (F-N-W) profile or the widely adopted Navarro–Frenk–White (NFW) one. While these models are primarily designed for galaxy-scale structures, several studies have explored the implications of CDM and $\Lambda$CDM cosmologies on black holes, ranging from stellar-mass black holes in microquasars to supermassive black holes at galactic centers. Besides, one key process involves the accretion of dark matter onto seed black holes at the centers of galaxies, potentially contributing to their hierarchical growth via mass absorption mechanisms, as discussed in \cite{ilyin2004dark}. Despite the absence of direct observational evidence confirming this scenario, Man Ho Chan and Chak Man Lee recently reported indications of dark matter density spikes surrounding both the stellar-mass black hole XTE J1118+40 and supermassive black holes, offering indirect support for CDM accumulation in compact objects \cite{chan2023indirect}. Furthermore, theoretical predictions have suggested a correlation between black hole mass and the mass of the surrounding dark matter halo \cite{booth2010dark}. According to the NFW profile, CDM forms a cuspy density distribution at small galactocentric radii, implying a natural accumulation of mass in the central regions. The rising behavior of the circular velocity profile near the galactic core, also predicted by the NFW model, supports the idea of significant mass concentration, which may enhance the gravitational influence of central black holes \cite{navarro1997universal}.}

The Event Horizon Telescope (EHT) has marked a milestone in BH research by capturing the first images of BHs. The 2019 image of $\textrm{M}87^\star$ and the 2022 image of $\textrm{Sgr A}^\star $ revealed bright, ring-like structures surrounding dark centers, offering the most direct visual evidence of BH existence to date \cite{EventHorizonTelescope:2019dse,EventHorizonTelescope:2022wkp}. BHs are not anticipated to exist in complete isolation. In realistic astrophysical settings, they are typically embedded in dynamic and complex environments. In particular, there is compelling evidence that supermassive BHs play a fundamental role in powering active galactic nuclei \cite{Rees:1984si, Richstone:1998ky}, where the complex surroundings consist of ionized plasma, electromagnetic fields, and DM \cite{Sadeghian:2013laa}. Consequently, the presence of a dark-matter halo near these compact objects is anticipated to influence the motion of surrounding matter and may leave observable signatures in the form of altered gravitational wave emissions from such galactic environments \cite{Barausse:2014tra}.

Particle dynamics near BHs is of considerable interest due to the valuable insights they provide into the geometry and nature of the surrounding spacetime\cite{nozari2025circular}. A particularly notable feature is the existence of a limiting radius, known as the innermost stable circular orbit (ISCO), beyond of which stable circular motion cannot exist. When a particle reaches this boundary, it cannot maintain a stable orbit and is eventually drawn into the BH\cite{Pugliese:2010ps,Liu:2017fjx,Zhang:2018eau}. In this sense, a substantial body of work has examined the charged particles behavior near BHs situated within external magnetic force \cite{kolovs2015quasi,kolovs2023charged,stuchlik2016acceleration,sun2021dynamics}. Observations suggest that strong magnetic fields exist near BHs, varying in mass. For instance, fields can reach magnitudes of $10^8$ G around stellar-mass BHs and about $10^4$ G for supermassive ones, whereas $\textrm{Sgr A}^\star$ is estimated to have a weaker field of approximately $10$--$100$ G \cite{piotrovich2010magneticfieldsblackholes,kolovs2015quasi}. The influence of magnetic fields on particle motion near BHs is often modeled using the Wald solution, which reveals chaotic trajectories for charged particles due to deviations from the equatorial plane \cite{wald1974black}. 

Beyond the Schwarzschild BH model, which reveals an ISCO at $ 6 M$ \cite{kolovs2015quasi}, more complex models involving Kerr BHs surrounded by various magnetic field configurations have also been explored \cite{rana2019astrophysically,stuchlik2019magnetized,shahzadi2021epicyclic}. These studies highlight the need to account for surrounding matter fields, as BHs are nonexistent in a vacuum. One such key component of the astrophysical environment is DM (DM), whose existence was first inferred from the rotational curves of spiral galaxies, where BHs are commonly found at their centers \cite{shaymatov2021effect}.

The combination of gravitational, electromagnetic, and DM effects offers a more realistic setting to analyze BH environments \cite{narzilloev2020dynamics,ashraf2025observational,shaymatov2021effect}.
Furthermore, both magnetic fields and DM significantly influence the dynamics of the accretion disk. Strong magnetic fields are known to be responsible for relativistic jets, while DM affects the disk's luminosity \cite{boshkayev2020accretion,kurmanov2022accretion,boshkayev2022accretion}. Among the observable features of accretion disks are high-frequency quasiperiodic oscillations (HF QPOs), which reflect the properties of the gravitational field near BHs. HF QPOs have been detected in microquasars such as GRS $\textrm{1915+105}$, $\textrm{H1743-322}$, $\textrm{XTE 1550-564}$ and $\textrm{GRO 1655-40}$ \cite{torok2005orbital,kolovs2015quasi}, and their origin is closely related to the particles trajectories around compact objects and continues to be a topic of extensive study \cite{vrba2023charged,xamidov2025probing,ashraf2025observational}. \textcolor{black}{Besides, HF QPOs have also been detected in supermassive black holes within active galactic nuclei \cite{smith2021confrontation}. Several studies have been conducted to explain these HF QPOs in the presence of dark matter associated with supermassive black holes in AGNs \cite{stuchlik2021supermassive}. In addition, dark matter models, such as the shell DM scenario, have been applied to test AGNs hosting supermassive black holes \cite{stuchlik2022geodesic}.}  

It is particularly compelling to investigate the interplay between gravity, magnetic fields (MF), and Dark Matter (DM) in the surroundings of BHs, as these components significantly alter the spacetime geometry. Various models of DM halos have been suggested, including perfect fluid DM (PFDM), scalar field DM (SFDM), warm DM (WDM), and cold DM (CDM) \cite{dubinski1991structure,navarro1996structure}. In this work, we focus on the Schwarzschild BH Exposed to a uniform magnetic field and surrounded by a CDM halo, as a representative scenario for exploring the charged particles dynamics under the coupled influence of gravitational, electromagnetic, and DM fields.

This study primarily seeks to contribute to this interesting area of gravitational physics. Indeed, we attempt to understand how charged-particle behavior is affected by the simultaneous presence of strong magnetic fields and high DM density. We first present a modified Schwarzschild metric that incorporates the effects of a CDM halo. Then, using the Hamiltonian formalism, we obtain the effective potential governing the trajectory of a charged particle, allowing us to analyze its energy, angular momentum, and orbital stability.

Furthermore, we extend our investigation to examine the quasi-periodic oscillations (QPOs) of charged particles in this environment. Specifically, we aim to characterize the epicyclic frequencies of particles orbiting a Schwarzschild BH embedded in a uniform magnetic field and surrounded by CDM, thereby gaining deeper insight into the influence of these fields on observable phenomena near compact objects.

This paper is structured as follows. In Sec.\ref{sec2}, we revisit the Schwarzschild metric modified by the presence of CDM and discuss its key geometrical implications. Sec.\ref{sec3} is committed to the study of charged particle dynamics in the vicinity of the modified Schwarzschild BH, embedded in a uniform magnetic field and surrounded by CDM. We analyze the effective potential, angular momentum, energy, and the ISCO under various conditions, including the presence or absence of CDM, \textcolor{black}{and investigate how varying CDM density influences the ionized Keplerian disk under different polarities of the Lorentz force.”}. 
In Sec.\ref{sec4}, we examine the harmonic oscillations of charged particles around the magnetized BH in a CDM environment. Special emphasis is put on the behavior of ionized particles, jet-like trajectories, and the resonant radii associated with twin-peak HF QPOs, particularly focusing on the 3:2 ratio between upper and lower frequencies. 
Sect.\ref{sec5} presents astrophysical estimations of HF QPOs, where we perform a comparative analysis by fitting observational data from microquasars \textcolor{black}{as well as from supermassive black holes in AGNs} within different theoretical models. The final section concludes the paper by summarizing the main findings and discussing their broader astrophysical implications.

\section{Schwarzschild BH solution surrounded by the CDM: some geometrical implications}\label{sec2}

\paragraph{}In order to obtain spherically symmetric black hole (BH) solution modified by the presence of a Cold Dark Matter (CDM) halo, where The CDM distribution is modeled using the Navarro–Frenk–White (NFW) density profile, derived from large-scale cosmological simulations within the $\Lambda$CDM framework \cite{Navarro:1995iw, Navarro:1996gj}. First, we set the spacetime geometry of pure DM, given in the following form
\begin{equation}
    ds^2 = -f^\text{DM}(r)\,dt^2 + \frac{1}{g^\text{DM}(r)}\,dr^2 + r^2 (d\theta^2 + \sin^2\theta\, d\phi^2),
\end{equation}
The rotational velocity of a particle in the equatorial plane of a spherically symmetric spacetime is obtained by the coefficient function $f^\text{DM}(r)$ as follows \cite{Xu_2018},
\begin{equation}\label{rotational velocity}
    V_\text{DM}(r)^2 = r \frac{\,d \ln(\sqrt{f^\text{DM}(r)})}{dr}
\end{equation}
In which the metric coefficient is denoted by $f^\text{DM}(r)=g^\text{DM}(r)$. This assumption is due to the consideration of the weak impact of pure dark matter gravity compared to the black hole,\cite{Xu_2018}. 
The radial density distribution is indicated by
\begin{equation}
    \rho_\text{NFW}(r)= \frac{\rho_s}{\frac{r}{R_s}\left( 1+ \frac{r}{R_s}\right)^2},
\end{equation}
where $\rho_s$ is the characteristic (core) density and $R_s$ is the scale (characteristic) radius of the halo. The mass of dark matter enclosed within a radius $r$ \cite{matos2005general}, 
\begin{equation}
    M_\text{DM}(r)= 4 \pi \int_0 ^r \rho_{NFW}(r) r^2 \,d r=4 \pi \rho_s R_s^3 \left(\ln\left(1 + \frac{r}{R_s} \right)-\frac{\frac{r}{R_s}}{\left(1 + \frac{r}{R_s} \right)} \right).
\end{equation}
Introducing such a dark matter mass profile into Kepler's third law, the rotational velocity becomes
\begin{equation}\label{velocity}
    V_\text{DM}(r) = \sqrt{\left(4 \pi \rho_s G R_s^3 \right)  \frac{1}{r} \left[  \ln \left(1 + \frac{r}{R_s} \right)-\frac{\frac{r}{R_s}}{\left(1 + \frac{r}{R_s} \right)}  \right]}.
\end{equation}
Solving Eqs.\eqref{rotational velocity},\eqref{velocity}, we obtain the metric function $f^\text{DM}(r)$ in the following form 
\begin{equation}
    f^\text{DM}(r) = \left[ 1 + \frac{r}{R_s}\right]^{-\frac{8 \pi G \rho_{s} R_s^3}{r}}.
\end{equation}
NFW predicts that cold dark matter (CDM) tends to accumulate in the central regions of galaxies~\cite{navarro1997universal}. Moreover, CDM enhances the density in the vicinity of black holes~\cite{Xu_2018}. Since black holes are typically located at galactic centers, it is therefore natural to consider them embedded within dark matter halos.
In the vacuum Schwarzschild spacetime, characterized by the blackening function $f^\text{Sch}(r)=g^\text{Sch}(r)$, the Ricci tensor vanishes, $R_{\mu\nu}^\text{Sch}=0$. By contrast, when incorporating galactic CDM models, one obtains $R_{\mu\nu}^\text{DM} \neq 0$, with the non-vanishing components of $R_{\mu\nu}^\text{DM}$ already defined in~\cite{Xu_2018}.

The Einstein field equations can be modified to incorporate the influence of a DM halo surrounding the BH by introducing an effective energy-momentum tensor, leading to the form
\begin{equation}\label{the energy-momentum tensor + DM}
    G_{\mu}^{\nu} = k^2 \left( T_{\mu}^{\nu {\text{Sch}}} + T_{\mu}^{\nu \text{DM}} \right).
\end{equation}
We are now in a position to propose an extended form of the Schwarzschild metric that incorporates the gravitational influence of a surrounding CDM halo, given by:
\begin{equation}
    \,d s^2 = -\left( f^{\text{Sch}}(r) + f^\text{DM}(r)\right)\,d t^2 + \left(f^{\text{Sch}}(r) + f^\text{DM}(r) \right)^{-1} \,d r^2 + r^2 \,d r^2 + r^2 \sin^2{\theta} 
\end{equation}
Including both energy-momentum tensors ensures that the solution simplifies to the standard Schwarzschild metric without a DM halo.
According to \cite{Xu_2018}, the resulting modified Schwarzschild metric, accounting for the gravitational effects of a CDM halo, takes the form of the blackening function:
\begin{equation}\label{modified metric}
f(r)=f^{\text{DM}}(r) + f^{\text{Sch}}(r)= \left( 1 + \frac{r}{R_s} \right)^{- \frac{8\pi G \rho_s R_s^3}{c^2 r}} - \frac{2GM}{rc^2} .
\end{equation}
Where finally, we can write the Schwarzschild spacetime metric including CDM as
\begin{equation}\label{CDM metric}
    \,ds^2 = -f(r)\,dt^2 + \frac{1}{f(r)} \,d r^2 + r^2\,d\theta+r^2 \sin^2{\theta} \,d\phi
\end{equation}
Let us introduce the parameter ${\bm \kappa} = \rho_s R_s^3$, which, together with $R_s$, characterizes the CDM profile. These parameters play a crucial role in determining the behavior and physical influence of the dark matter distribution. 
For the combined system of a Schwarzschild black hole embedded in a CDM halo, the Ricci tensor corresponding to the metric defined in Eq.~\eqref{CDM metric}, under the conventions $c=G=1$, takes the following form:
\begin{eqnarray}\nonumber
\label{}
R_{tt}&=&\frac{4 \pi  \bm{\kappa} \left(\mathcal{R}^*\right)^{-\frac{16 \pi  \bm{\kappa}}{r}} \left(r-2 M \left(\mathcal{R}^*\right)^{\frac{8 \pi  \bm{\kappa}}{r}}\right)}{r^5 (r+R_s)^2} \nonumber \\
&\quad& \times \left(r^2 (8 \pi  \bm{\kappa}+r)+8 \pi  \bm{\kappa}(r+R_s) \log \left(\mathcal{R}^*\right) \left((r+R_s) \log \left(\mathcal{R}^*\right)-2 r\right)\right),
\\
\label{}
R_{rr}&=&\frac{4 \pi  \bm{\kappa} \left(8 \pi  \bm{\kappa} (r+R_s) \log \left(\mathcal{R}^*\right) \left(2 r-(r+R_s) \log \left(\mathcal{R}^*\right)\right)-r^2 (8 \pi \bm{\kappa}+r)\right)}{r^3 (r+R_s)^2 \left(r-2 M \left(\mathcal{R}^*\right)^{\frac{8 \pi  \bm{\kappa}}{r}}\right)},
\\
\label{}
R_{\theta\theta}&=&\frac{\left(\mathcal{R}^*\right)^{-\frac{8 \pi   \bm{\kappa}}{r}} \left(-8 \pi   \bm{\kappa} (r+R_s) \log \left(\mathcal{R}^*\right)-r (-8 \pi   \bm{\kappa}+r+R_s)\right)}{r(r+R_s)}+1,
\\
\label{}
R_{\phi\phi}&=&\frac{\sin ^2(\theta ) \left(\mathcal{R}^*\right)^{-\frac{8 \pi   \bm{\kappa}}{r}} \left(r (r+R_s) \left(\left(\mathcal{R}^*\right)^{\frac{8 \pi   \bm{\kappa}}{r}}-1\right)-8 \pi   \bm{\kappa}(r+R_s) \log \left(\mathcal{R}^*\right)+8 \pi   \bm{\kappa} r\right)}{r (r+R_s)},
\end{eqnarray}
where $\mathcal{R}^* = (\frac{r+R_s}{R_s})$, and the other components of the $R_{\mu\nu}$ are equal to zero. Thus, preserving the Ricci tensor spherical symmetry, the background geometry of the black hole remains unaffected.

\begin{figure}[H]        
    \includegraphics[width=\textwidth]{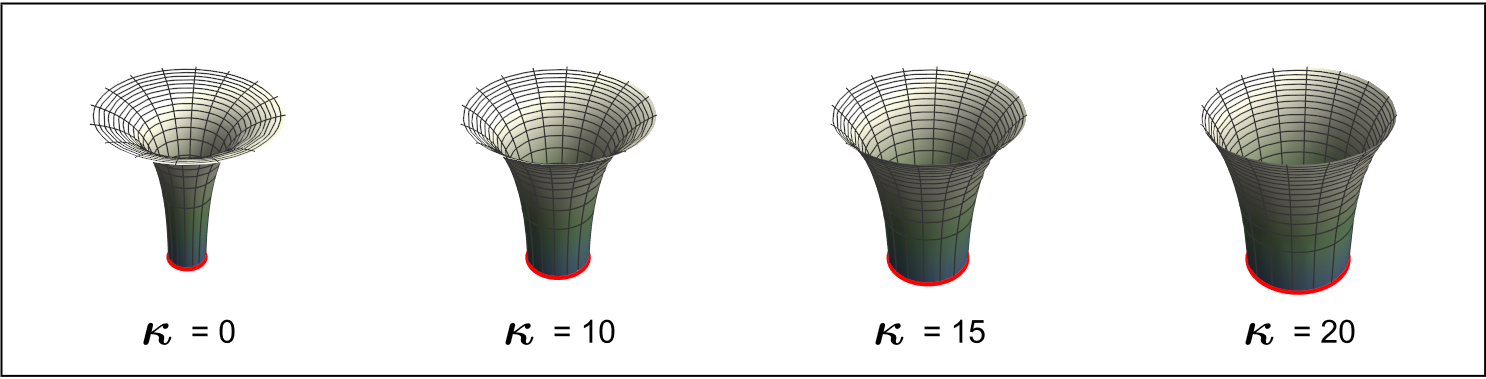} 
    \caption{\footnotesize{\it Embedding diagram within different values  dark-matter parameter ${\bm \kappa}$. The red circle represents the BH event horizon. }}\label{fig:1}
    \label{embedding diagrams K}
\end{figure}

To provide the geometrical consequences of dark matter on the black hole geometry, Fig.\ref{fig:1} presents the embedding diagram of a black hole for several values of the parameter $\bm{\kappa}$. Throughout the remainder of this paper, we fix the characteristic scale $R_s$, so that the impact of dark matter is entirely encoded in variations of $\bm{\kappa}$.

Initially, we examine the limiting case where DM is absent to explore the physical implications of the modified metric incorporating DM. In this scenario, the metric diminishes to the standard Schwarzschild solution, which the singularity located at $r=0$ and possesses an event horizon located at $r_h=2M$. When DM is present, the spacetime geometry is influenced by the parameter ${\bm \kappa}$, which governs the modifications introduced by the surrounding halo.
In particular, as the parameter ${\bm \kappa}$ increases, both the size of the BH core and the radius of the event horizon grow.
 The location of the event horizon is determined by solving the following equation:
\begin{equation}
   \left( 1 + \frac{r}{R_s} \right)^{- \frac{8\pi G \bm{\kappa}}{c^2 r}} - \frac{2GM}{rc^2} = 0.
\end{equation}
Having revisited the BH solution embedded in a CDM and unveiled the effect of the ${\bm \kappa}$ parameter on the spacetime geometry, our attention is drawn to disclose the charged particle dynamics in such a magnetized BH background.

\section{Particle dynamics in magnetized BH surrounded by CDM halos}\label{sec3}

\subsection{Magnetized BH solution with CDM}

\paragraph{}This study focuses on the orbital dynamics of particles with $q\neq 0$ and mass $m$ in the vicinity of a BH embedded within a uniform magnetic field, taking into account the presence of a surrounding CDM halo. The underlying spacetime is represented by the blackening function in Eq.\eqref {CDM metric}. 

Such a stationary and axial symmetry unveils timelike and spacelike killing vectors obtained by the equation \cite{wald1974black,kolovs2015quasi}
\begin{equation}
    \partial \xi_{\alpha,\beta} + \partial \xi_{\beta,\alpha} = 0.
\end{equation}
The expression for the four-vector potential $A^\mu$, which describes the test electromagnetic field, is expressed as \cite{wald1974black,kolovs2015quasi}
\begin{equation}
    A^\mu = C_1 \xi^\mu_{(t)} + C_2 \xi^\mu_{(\phi)},
\end{equation}
with $C_{1,2}$ are integration constants.
Giving that the BH is situated in a uniform magnetic field $B$, the associated electromagnetic field tensor $F$ must satisfy the conditions established by Wald \cite{wald1974black}. Specifically, $F$ must be stationary and axially symmetric, regular (i.e., non-singular) across the event horizon, and asymptotically approach a uniform magnetic field of strength $B$ at spatial infinity \cite{wald1974black}. Under these conditions, the electromagnetic field tensor be formulated as
\begin{equation}
    F = \frac{1}{2} B \left( d\xi_{(\phi)} + \frac{2J}{M} d\xi_{(t)} \right),
\end{equation}
in which $J$ and $M$ represent the mass and angular momentum of the BH, respectively.

Accordingly, the four-vector potential is given by \cite{wald1974black}
\begin{equation}
    A^\mu = \frac{1}{2} B \left( \xi^\mu_{(\phi)} + \frac{2J}{m} \xi^\mu_{(t)} \right)
\end{equation}
With regard to a non-rotating (Schwarzschild) BH, where $J=0$, the four-vector potential simplifies to
\begin{equation} A^\mu = \frac{1}{2} B, \xi^\mu_{(\phi)}, 
\end{equation}
With $\xi^\mu_{(\phi)} = \partial / \partial \phi$.
 The covariant component $A_\phi$ of the potential, then takes the form
\begin{equation}
    A_\phi = \frac{1}{2} B g_{\phi\phi},
\end{equation}
where $g_{\phi\phi}$ is the corresponding component of the spacetime metric Eq.\eqref{CDM metric}. 
The motion of a particle with mass $m$ and charge $q$ in this background is governed by the Lorentz force equation
\begin{equation}
    m \Dot{u}^\mu = q F^{\mu}_\nu u^{\nu}.
\end{equation}
$\Dot{u}^\mu $ refers to the particle four-velocity with the normalization condition $ u^\mu u_\mu = -1$ , the dot refers to the derivative over a proper time $\tau$. The electromagnetic tensor given by $F_{\mu\nu} = A_{\nu,\mu}-A_{\mu,\nu} $ and highlighting its antisymmetric nature.


\textcolor{black}{It's commonly known that magnetic fields in the vicinity of black holes may arise from several astrophysical sources. They can be generated by electric currents within the surrounding accretion disks or originate from external environments such as galactic-scale magnetic fields or the magnetic influence of a binary companion, particularly if the companion is a highly magnetized object like a magnetar \cite{Kovar:2016kqh,Tursunov:2018wgx}. Observations of the supermassive black hole $\textrm{Sgr A}^\star$, located at the center of the Milky Way, indicate the presence of an ordered magnetic field with an intensity ranging between $10$ and $100$ Gauss in regions just a few gravitational radii from the event horizon \cite{Eatough:2013nva}. Similarly, around the central black hole of the galaxy $\textrm{M}87$, magnetic field strengths have been estimated to be on the order of $\sim 100$ Gauss \cite{Doeleman:2012zc}. More broadly, studies across various active galactic nuclei report average magnetic field intensities reaching up to $\sim 10^4$ Gauss \cite{Daly:2019srb}.}

\textcolor{black}{It is important to emphasize that, within the framework of general relativity, these magnetic field strengths are still considered weak. This classification is based on the relative magnitude of the electromagnetic stress-energy tensor, which remains sufficiently small such that the influence of the magnetic field on the spacetime curvature can be neglected. A useful estimate for this general-relativistic threshold is given by the expression \cite{stuchlik2020influence}
\begin{equation}
B \ll B_{\text{GR}} = 10^{18} \left( \frac{10 M_{\odot}}{M} \right) \text{G},
\end{equation}
where $M$ is the mass of the black hole. For all astrophysical cases considered in this study, this condition is well satisfied, justifying the treatment of the magnetic field as a test field on a fixed gravitational background. 
When accounting for the combined effects of a magnetic field (MF) and a CDM halo, it is important to distinguish the relative contributions of each component to the overall spacetime geometry. In our framework, the energy-momentum tensor associated with the CDM halo is significantly more dominant than that of the magnetic field. Consequently, we consider the magnetic field as a test field: it is treated as a perturbative element that does not alter the CDM distribution or the underlying spacetime geometry induced by the dark matter.}

\subsection{Charged particle dynamics: Hamiltonian and effective potential}
The dynamics of a charged particle around a BH with CDM located within a uniform magnetic field can be conveniently explained using the Hamiltonian formalism \cite{kolovs2015quasi}. The general form of the Hamiltonian is given by
\begin{equation}\label{Hamiltonian formalis}
    H = \frac{1}{2} g^{\alpha\beta} \left(  \pi_\alpha - q A_\alpha \right)\left(  \pi_\beta - q A_\beta \right) + \frac{1}{2} m^2,
\end{equation}
where $g^{\alpha\beta}$ are the components of the inverse metric tensor, corresponding to the modified Schwarzschild metric introduced in Eq.\eqref{CDM metric}. The quantity $\pi_\alpha$ represents the generalized (canonical) four-momentum, which is linked to the kinetic four-momentum $p^\alpha = m u^\alpha$ by
\begin{equation}\label{equation de pi}
    \pi^\mu = p^\mu + q A^\mu,
\end{equation}
with $u^\alpha$ denoting the particle's four-velocity.
Constructing the effective potential from the Hamiltonian formalism is crucial to understanding the trajectory of charged particles in the vicinity of magnetized BHs. This topic has been extensively studied in the literature; see, for example, Refs.\cite{frolov2010motion,kolovs2015quasi,qi2023charged}. In the framework of this study, we specifically concentrate on the behavior of charged particles around a non-rotating, uncharged BH surrounded by a CDM halo and operating within a uniform magnetic field.

The Hamiltonian equations of motion are given by
\begin{equation}
    m\frac{dx^\mu}{d\tau} = p^\mu = \frac{\partial H}{\partial \pi_\mu},\quad \frac{d\pi_\mu}{d\tau} = -\frac{\partial H}{\partial x^\mu}.
\end{equation}
Applying Eq.\eqref{equation de pi}, conserved quantities corresponding to the spacetime symmetries are obtained. The energy $E$ and axial angular momentum $L$ of a particle are given by
\begin{equation}
    E = - \pi_t = g_{tt}( m u^t + q A^t) = m g_{tt} \Dot{t},
\end{equation}
\begin{equation}\label{angular momentum*}
    L = \pi_\phi = m g_{\phi\phi} \left( \Dot{\phi} + \frac{qB}{2m} \right).
\end{equation}
The remaining components of the motion are governed by
\begin{align}
    \Dot{r} = g^{rr} p_r \quad \text{ and }\quad
    \Dot{\theta} = g^{\theta\theta} p_\theta.
\end{align}

The dynamical equation governing charged particles' motion in Cartesian coordinates $(x,y,z)$ can be derived through coordinate transformation.
\begin{align}\label{coordinates 3D}
    &x = r \cos{\phi}\sin{\theta} ,\quad  y = r \cos{\theta}\sin{\phi} , \quad z = r \cos{\theta}.
\end{align}
Finally, it is convenient to define the specific (per unit mass) quantities \cite{Frolov:2010mi}
\begin{align}
    E' = \frac{E}{m},\quad
    L' = \frac{L}{m}, \quad
    B' = \frac{qB}{2m}.
\end{align}
The Hamiltonian given in Eq.\eqref{Hamiltonian formalis} can be rewritten as represented below
\begin{equation}
    H = \frac{1}{2}g^{rr} p_r^2 + \frac{1}{2}g^{\theta\theta}  p_\theta^2+\frac{1}{2} m^2 g^{tt}\left[ V_\text{eff}(r,\theta) - E'^2 \right]
\end{equation}
where $ V_\text{eff}(r,\theta;L',B',\bm{\kappa})$  indicates the effective potential, defined as
\begin{equation}\label{effective potential}
     V_\text{eff}(r,\theta;L',B',\bm{\kappa}) \equiv g_{tt}\left[ 1 + \left( \frac{L'}{r \sin{\theta}} - B' r \sin{\theta} \right)^2 \right].
\end{equation}
The term enclosed in parentheses represents the central force enhancement of the effective potential, arising from the combined effects of $B'$ and  $L'$.

Moreover, the system exhibits a double parity symmetry under the following transformation:
\begin{align}\label{symtery}
    (L',B') \leftrightarrow (-L',-B'), 
\end{align}

\begin{figure}[!ht]
    \centering
    \centering
    \subfloat[]{\includegraphics[width=0.33\textwidth]{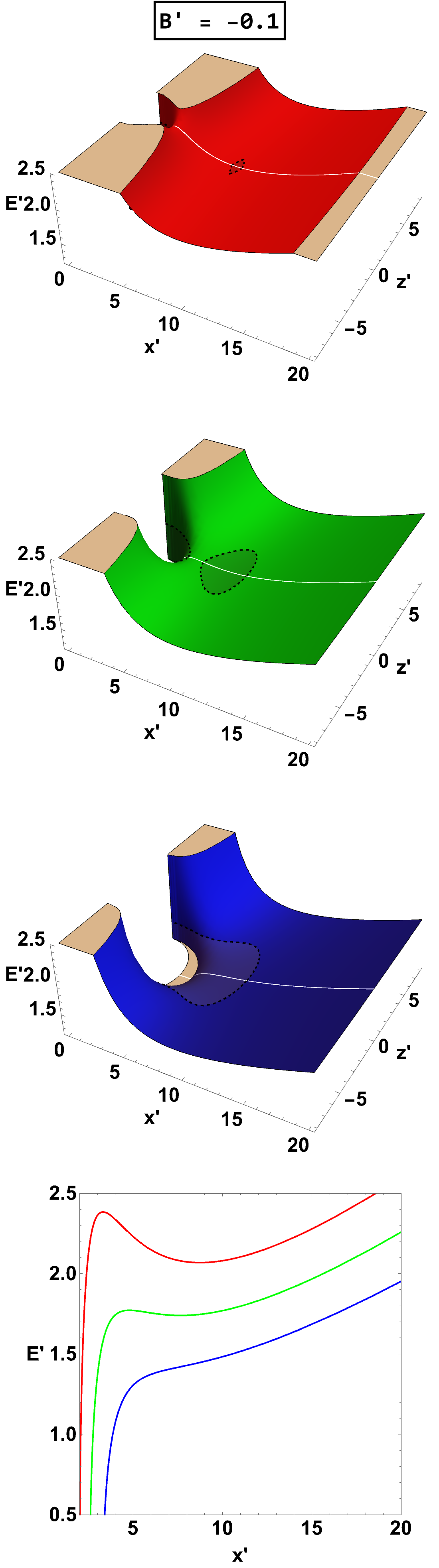}}
    \subfloat[]{\includegraphics[width=0.33\textwidth]{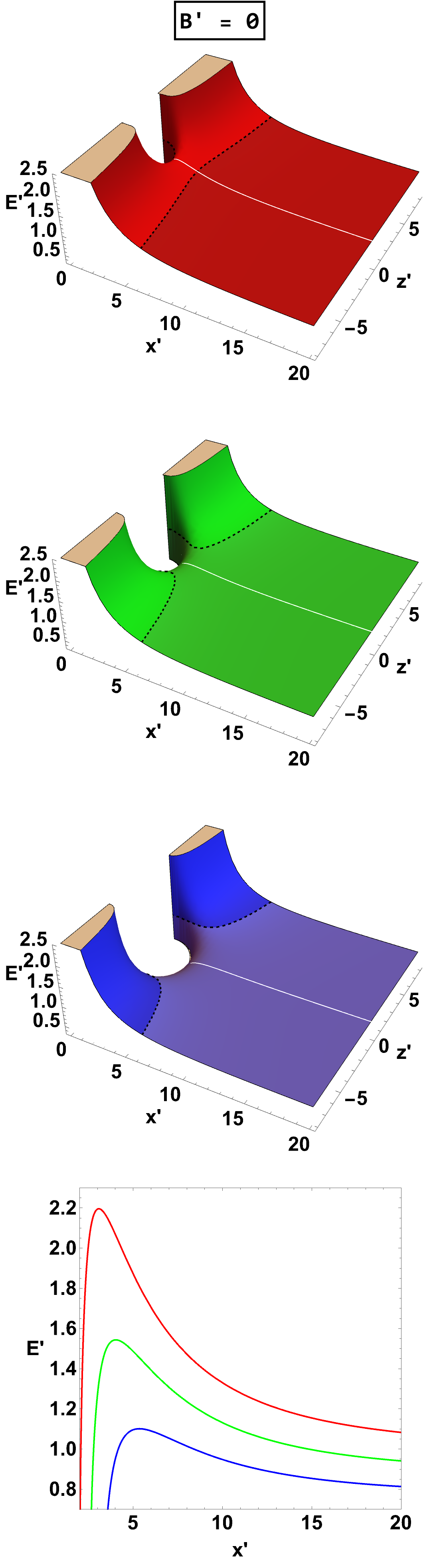}}
    \subfloat[]{\includegraphics[width=0.33\textwidth]{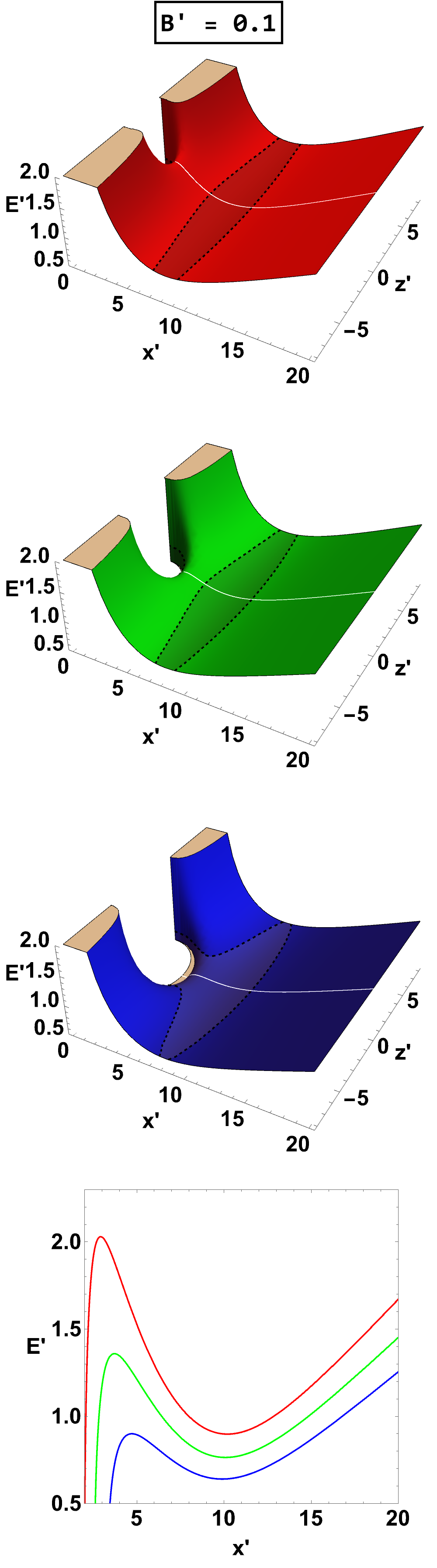}}
    \caption{\footnotesize{\it The effective potential in the equatorial plane is shown for various values of $\bm{\kappa}$ and magnetic field strength $B'=0,\pm0.1$. The first row corresponds to $\bm{\kappa}=0$, the second row to $\bm{\kappa}=10$, and the third row to $\bm{\kappa}=20$. The fourth row presents a two-dimensional projection at $z'=0$. The first three rows display three-dimensional representations of the effective potential, while the fourth row provides a comparative $2D$ slice for fixed angular momentum $L'=11$.}}
    \label{energy K differs Rs=1000}
\end{figure}

The symmetry equation indicates that both $L'$ and $B'$ may have either positive or negative sign. The angular momentum sign $L'$ determines the direction of the charged particle trajectory around the BH: A positive (negative) $L'$ corresponds to a counterclockwise (clockwise) motion viewed from above the BH. Similarly, the sign of $B'$ indicates the orientation of the magnetic field vector $\vec{B}'$
, with a positive (negative) $B'$ pointing upward (downward) along the $z$-axis. The analysis of $V_\text{eff}$ provides essential information that complements the solution of the equations of motion for charged particles \cite{kolovs2015quasi,frolov2010motion}. When $L'$ and $B'$ share the same sign, the resulting Lorentz force acts repulsively on the charged particle. In contrast, if $L'$ and $B'$ have opposite signs, the Lorentz force becomes attractive. 
Moreover, the movement of charged particles is constrained by energy conservation, where turning points occur under the condition $p_r = p_\theta = H = 0$. In this sense, $V_\text{eff}$ satisfies 
\begin{equation}
    E'^2 =  V_\text{eff}(r,\theta;L',B',\bm{\kappa}).
\end{equation}
The variation of the effective potential in the presence of a uniform magnetic field, which incorporates the effects of CDM, is illustrated in Fig.\ref{energy K differs Rs=1000}. The results are presented for angular momentum $L'=11$ with $B'>0$, $B'<0$ in a vanishing $B'=0$. 

It is observed that the magnetic field significantly affects the $V_\text{eff}$ at large distances from the BH. However, near the event horizon, gravitational effects dominate over the magnetic influence in the equatorial plane $\theta=\pi/2$. More specifically, as the parameter ${\bm \kappa}$ increases at a fixed characteristic radius $R_s$, the gravitational pull becomes more stronger, leading to a suppression of $B'$ influence close to the event horizon. The dark region in Fig.\ref{energy K differs Rs=1000}, bounded by the dashed black line, represents the stability region for charged particle orbits. The impact of CDM is visible as this stable region shifts toward the BH, indicating that charged particles are increasingly trapped closer to the event horizon. Furthermore, the event horizon radius itself increases with higher values of ${\bm \kappa}$, as previously discussed in Sec.\ref{sec2} and shown in Fig.\ref{embedding diagrams K}. 
In the cases of $B'=\pm 0.1$, the stability region acquires a parabolic form, reflecting the balance between gravitational attraction and magnetic repulsion in this scheme.  In addition, positive magnetic field values produce a larger stability region compared to negative values.

A fundamental property of the BH background under consideration is its axial symmetry. In the presence of both a gravitational field and an external uniform magnetic field, the spacetime remains invariant under rotations about the symmetry axis. As a result, $V_\text{eff}$ governing the path of charged particles is independent of the azimuthal coordinate $\phi$. This symmetry significantly simplifies the analysis, allowing $V_\text{eff}$ to be treated as a two-dimensional function $V_\text{eff}(r,\theta)$. Alternatively, it can be represented using spherical coordinates $(x, z)$, defined in Eq.\eqref{coordinates 3D}, which are particularly convenient for visual representation and numerical evaluation. The structure and behavior of the effective potential in this reduced configuration space are clearly illustrated in the lower panels of Fig.\ref{energy K differs Rs=1000}, highlighting the role of symmetry in constraining the dynamics of charged particles in such a background. 
Our investigation focuses on understanding how the presence of a CDM halo, in conjunction with a uniform magnetic field, influences the effective potential. Parameters $\bm{\kappa}$, which characterize the density profile of the CDM distribution, have a significant impact on shaping the spacetime geometry and consequently the effective potential experienced by charged particles. Notably, the effective potential exhibits distinct regions: a negative branch located within the BH interior and a positive branch outside the event horizon. The crossover between these two regimes occurs precisely at the event horizon, which acts as a natural boundary separating the interior and exterior dynamics. As illustrated in Fig.\ref{fig:1}, the the event horizon radius is strongly influenced by the parameter $\bm{\kappa}$. For example, where $\bm{\kappa}=0$, $V_\text{eff}$ is zero at $r=2$, positive for $r>2$, and negative for $r<2$, consistent trough the classical Schwarzschild solution. However, when CDM is included—for example, with $\bm{\kappa}=10$ and $R_s=1000$ the event horizon shifts outward. In this case, the effective potential is negative for $r<2.57063$, positive for $r>2.57063$, and vanishes at $r\approx2.57063$. This shift indicates that the gravitational influence of the CDM halo modifies the BH's causal structure. Therefore, our analysis of the $V_\text{eff}$ is restricted to the event horizon exterior region, where stable orbits and observationally relevant dynamics occur.

The stationary points of the $V_\text{eff}$ that govern the orbital motion of charged particles around a BH subjected to a uniform magnetic field and surrounded by CDM are determined. 
by solving the following equations:
\begin{align}
    &\partial_{\{r,\theta\}} V_\text{eff}(r,\theta,L',B',\bm{\kappa}) = 0.
\end{align}
All extrema of the effective potential for charged particle dynamics in the equatorial plane of a Schwarzschild BH wrapped in a uniform magnetic field—also surrounded by a CDM halo—are located in that same plane associated with $\theta=\pi/2$, as shown in \cite{kolovs2015quasi}. The other extremum corresponds to the radial coordinate $r$, which satisfies the condition for stationary orbits.
\begin{equation}\label{first extrema}
    L'^2\left( (r+R_s) \mathcal{G} +  \mathcal{N} \right) - 2 L'B'r^2  \mathcal{N} - B'^2r^4(r+R_s) \mathcal{G} + (r^2 + B^2r^4)  \mathcal{N} = 0.
\end{equation}
where the function $G$ denotes
\begin{equation}
     \mathcal{G}(r,\bm{\kappa},R_s) = 2M  - r  \mathcal{T}^{-1},
 \end{equation}
 and $ \mathcal{N}$ stands for
\begin{equation}
    \mathcal{N}(r,\bm{\kappa},R_s) = M (r + R_s) - 4 \bm{\kappa} \pi  \mathcal{T}^{-1} 
\left( r - (r + R_s) \ln\left[\frac{r + R_s}{R_s}\right] \right).
\end{equation}
Where the $\mathcal{T}$ encoded the DM term as 
\begin{equation}\label{T abriviation}
    \mathcal{T}(r)=\left( \frac{1}{R_s}(r + R_s) \right)^{\frac{8 \bm{\kappa} \pi}{r}}.
\end{equation}
Since the above equation is of the fifth order in $r$, it may admit more than five roots, corresponding to multiple extremal points. By treating Eq.\eqref{first extrema} as a second-degree equation expressed with respect to the $L'$, one can straightforwardly derive the solution. Specifically, we obtain:
\begin{equation}\label{extrema function}
    L'_{\pm}(r;B',\bm{\kappa},R_s) = \frac{B'  \mathcal{N} r^2-\frac{1}{2} \sqrt{4 B'^2  \mathcal{N}^2 r^4-4 (\mathcal{G} (r+R_S)+ \mathcal{N}) \left((B'+1)  \mathcal{N}
   r^2-B'^2 \mathcal{G} r^4 (r+R_S)\right)}}{\mathcal{G} (r+R_S)+ \mathcal{N}}.
\end{equation}

The positive root of Eq.\eqref{extrema function} corresponds to configurations with either stable or unstable circular orbits, whereas the negative root is associated via the maxima of the $V_\text{eff}$, indicating regions of instability. Furthermore, the $V_\text{eff}$ maxima and minima can be identified by analyzing its second derivative with respect to the $r$. These extremal points provide key insights into the radial positions, $E'$, and $L'$ associated with particle trajectories. In particular, the condition for the ISCO is evaluated by the inflection point, defined by $\partial_r^2 V_\text{eff} = 0$. Solving this condition yields

\begin{equation}{\label{local extrema}}
    L'_{ex}(r;B',\bm{\kappa},R_s) = \frac{2B' \mathcal{N} r \pm \sqrt{-2\mathcal{G} \mathcal{N} r + 4B'^2 \mathcal{N}^2 r^2 - 4B'^2 \mathcal{G} \mathcal{N} r^3 + 
5B'^2 \mathcal{G}^2 r^4 + 4B'^2 \mathcal{G}^2 r^3 R_s}}{\mathcal{G}}.
\end{equation}

\begin{figure}[!ht]
    \centering
    \includegraphics[width=1\linewidth]{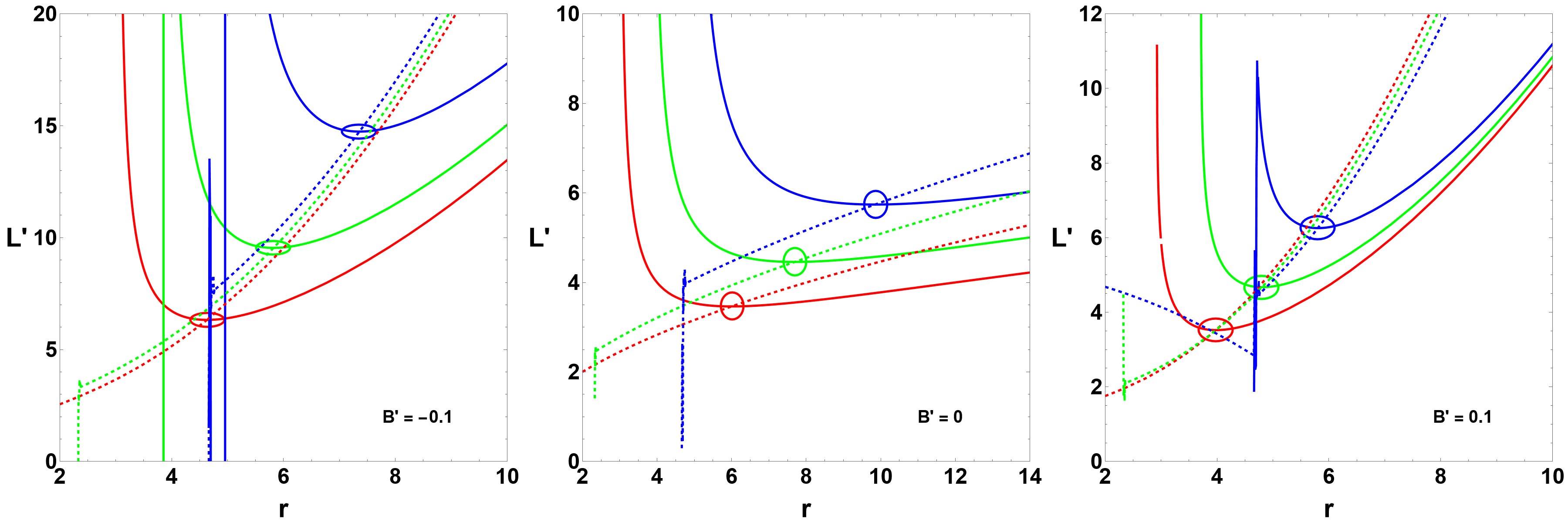}
    \caption{\footnotesize{\it Angular momentum in function of radii $L'_\pm$ (thick), $L'_{ex}$ (dashed) , $\bm{\kappa}=0$ (red), $\bm{\kappa}=10$ (blue), $\bm{\kappa}=20$ (green), within several values of $B'$. Circles representing the point of intersection of $L'_\pm$ and  $L'_{ex}$ indication the ISCO position.}}
    \label{extrema 1}
\end{figure}

Fig.\ref{extrema 1} illustrates the circular orbits of charged particles in space-time influenced by CDM and a magnetic field. 

The extrema of the functions $L_\pm(r)$ correspond to the local extrema of the effective potential in the equatorial plane. The dashed curves in the figure represent these extremal values, denoted by $ L'_{ex}(r) $, which help identify the stability characteristics of the orbits. The point where the solid and dashed curves intersect in Fig.\ref{extrema 1} marks the ISCO positions. The location of the ISCO is not universal but instead varies trough the merged influence of $B'$ and the CDM halo. Circular orbits remain stable for radial distances $r>r_\text{ISCO}$, whereas they become unstable for $r<r_\text{ISCO}$. The extent and position of these regions are directly affected through the strength and distribution of the surrounding DM.

\begin{figure}[!ht] 
    \centering
    \includegraphics[width=0.4\textwidth]{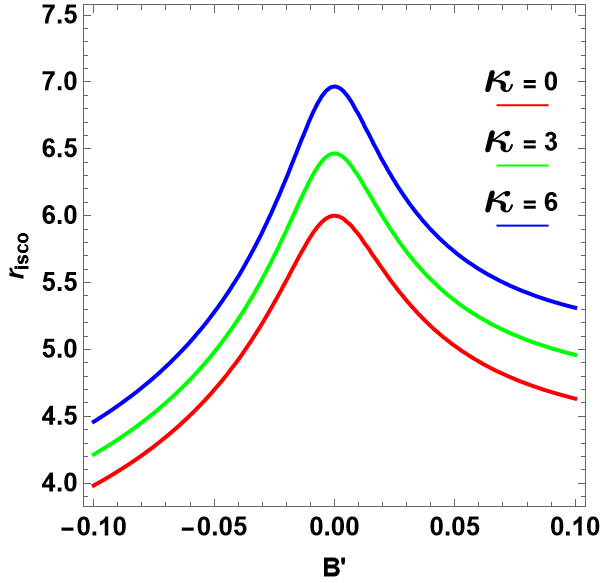} 
    \label{}
    \caption{\footnotesize{\it ISCO radius variation on function of magnetic field $B'$ in different values of $\bm{\kappa}$.}}
    \label{ISCO}
\end{figure}

Specifically, the ISCO radius is established by the intersection of the functions $ L'_{ex}(r) $ and $L'_\pm(r)$. To analyze how the magnetic field and CDM affect the ISCO location, a detailed representation is provided in Fig.\ref{ISCO}. 
The plot demonstrates that the parameter $\bm{\kappa}$, which characterizes the density of the CDM halo, alongside with $B'$, has a significant impact of the ISCO radius. Interestingly, the magnetic field and CDM exert opposing effects: an increase in the magnitude of $B'$ regardless of its sign—tends to shift the ISCO inward, closer to the event horizon, whereas an increase in $\bm{\kappa}$ pushes the ISCO outward, farther from the BH event horizon.

At large distances $r \rightarrow \infty$ and near the equatorial plane $\theta\approx\pi/2$, the path of a charged particle in a uniform magnetic field becomes radially confined. This confinement arises from the $B'r^2$ term in the effective potential, as seen in Eq.\eqref{effective potential}, which acts as a dominant repulsive barrier in the radial direction.

Under the circumstances of a repulsive Lorentz force, the effective energy boundaries open along the polar direction, permitting particles with ($q\ne0$) move off to infinity following the $z-$axis. In contrast, when the Lorentz force is attractive, it drives the particle into a spiraling trajectory, curving inward toward the BH.

Therefore, the requirement for a charged particle to move toward infinity, in the case where $B'\geq0$, can be expressed as follows:
\begin{align}\label{condition for positive B}
    E'_\text{min} \geq \sqrt{\left(1-\frac{2}{r}(1+\bm{\kappa}\pi \ln{\frac{r}{R_s}})\right)}.
\end{align}
Reflecting the lowest of the energy of the particle at the infinity.
In the case of $B' < 0$, the energetic condition of the particle asymptotically approach infinity to the following form:
\begin{equation}\label{close energy}
  E'_\text{min} (\bm{\kappa},R_s;B',L') \geq  \sqrt{\left(1-\frac{2}{r}(1+\bm{\kappa}\pi \ln{\frac{r}{R_s}})\right)\left(1 - 4B' L'\right)}. 
\end{equation}

From Eqs.\eqref{condition for positive B} and \eqref{close energy}, one can deduce the conditions under which charged particles become trapped around a magnetized BH surrounded by a CDM halo. In particular, the minimum energy $E'_\text{min}$ required for trapping is influenced by the distribution of CDM. This effect becomes increasingly significant as the CDM density intensifies, highlighting the role of the halo in shaping the energy landscape of particle motion. These bound states arise when the particle's specific energy satisfies the equality  $E'=E'_\text{min}$. Oscillatory trapped motion of charged particles is therefore permitted as long as the following condition is fulfilled
\begin{equation}
    L'_1<L'<L'_2.
\end{equation}

In the case where $B'\geq0$, the behavior of oscillating charged particles in trapped states is governed by the expression stated below for the specific angular momentum $L'_{1,2}$:
\begin{equation}
   L'_{1,2} =  \frac{r \left( B'r(-r + 2  \mathcal{T}) \pm \mathcal{T} 
\sqrt{ \left(2 - r \mathcal{T} \right)
\left(-2 + r \left( -1 +\mathcal{T} \right) \right)  }\right)}{ 2 \mathcal{T}-r}.
\end{equation}

whereas for $B'<0$, the corresponding condition for trapped states is outlined by:
\begin{equation}
    L'_{1,2}=\frac{r\left(B'r\left(-2  + r(-2 +\mathcal{T})\right) \pm \sqrt{ \left( 2 +r( 1 +  \mathcal{T})\right)
        \left(-2 + r(4 B'^2 r^2 +  \mathcal{T})\right)}
  \right)}{ r \mathcal{T}-2}.
\end{equation}
In both expressions, the function $\mathcal{T}$ encodes the influence of the CDM halo and is defined as:
$\mathcal{T}(r)=\left( \frac{1}{R_s}(r + R_s) \right)^{\frac{8 \bm{\kappa} \pi}{r}}$. These quantities characterize the angular momentum bounds required for the existence of stable, bounded oscillatory trajectory in the surroundings of a magnetized BH embedded in a CDM environment.

\begin{figure}[!ht]
    \centering
    \includegraphics[width=1\linewidth]{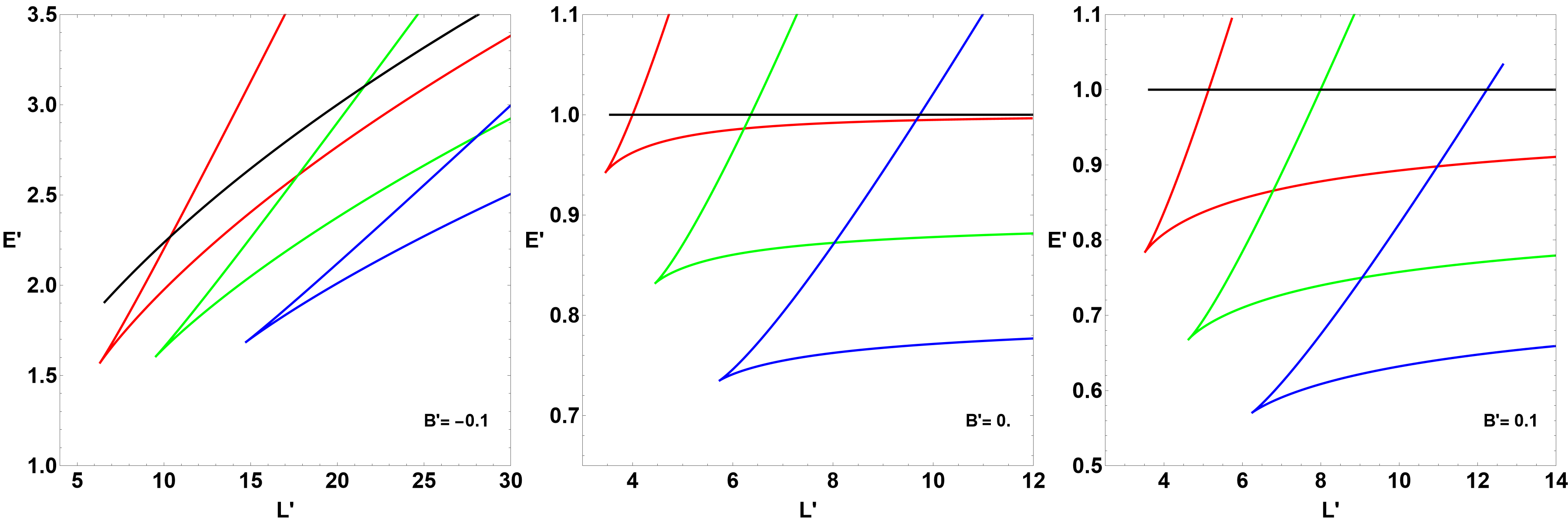}
    \caption{\footnotesize{\it Energy plotted against of angular momentum for distinct values of $\bm{\kappa}$: $\bm{\kappa}=0$ (red), $\bm{\kappa}=10$ (green), and $\bm{\kappa}=20$ (blue). The black curve represents the minimum energy required for a particle tend to infinity, shown for multiple values of $B'$.}}
    \label{energetics K}
\end{figure}

Having discussed the extrema of Eq.\eqref{extrema function} with respect to the coordinate $r$, we now turn to the behavior of the effective potential described by $L'$. Fig.\ref{energetics K} illustrates the local extrema of $V_\text{eff}$ for various values of the intensity of the $B'$ , highlighting how these features vary with changes in the DM parameter $\bm{\kappa}$ relative to $R_s$.

The point at which the potential curve deviates sharply corresponds to the specific energy at the ISCO, $E'_\text{ISCO}$. The plot demonstrates that increasing $\bm{\kappa}$ contributes to a decrease in $V_\text{eff}$ associated via the ISCO, reflecting the influence of the CDM halo.

The minimal energy configuration required for a particle to diverge to infinity is indicated by the thick black line in the Fig.\ref{energetics K}. This line separates the curves into two distinct regions that correspond to the local minima and maxima of $V_\text{eff}$. The region below the black line represents bound states, indicating where trapped particle motion is possible. In particular, the extent of this region increases with larger values of $\bm{\kappa}$, demonstrating the growing consequence of the CDM halo on the confining trajectories of particles.

\subsection{Ionized Keplerian disk}

\paragraph{}\textcolor{black}{To explore the combined effects of magnetic fields and cold dark matter on particle motion near black holes, we consider a simplified model of an Keplerian accretion disk located in the equatorial plane of a Schwarzschild black hole\footnote{ Unlike rotating black holes, which confine the accretion disk strictly to the equatorial plane, non-rotating black holes can host Keplerian disks at various central planes. In this study, we focus on a fixed-inclination configuration, fixing the disk’s inclination angle at $\theta_0 = 1.51$ radians relative to the equatorial plane.}}.
\textcolor{black}{We start by assuming that the Keplerian disk is initially composed of neutral particles ($q = 0$) on stable circular orbits around a Schwarzschild black hole surrounded by a CDM halo and immersed in a weak, asymptotically uniform magnetic field. A realistic ionization scenario is provided by the magnetic Penrose process (MPP) \cite{panis2019determination}, where a neutral particle of mass $m_1$ undergoes a decay or fragmentation event, producing two charged particles ($q_2 \ne 0$, $q_3 \ne 0$) with conserved total charge, i.e. $q_1= q_2 +q_3=0 $. The canonical momentum is conserved during such an ionization process is 
\begin{equation}
    \pi_\alpha(1) = \pi_\alpha(2) + \pi_\alpha(3).
\end{equation}
While other scenarios, e.g., Keplerian disks generated by a plasma beam, are discussed in \cite{panis2019determination}. In our simplified framework, we assume that one of the daughter particles is significantly more massive than the other, such that $m_2 \gg m_3$, as is the case in ion-electron or proton-electron decays. Therefore, the majority of the initial momentum is transferred to the heavier fragment
\begin{equation}
    p_\alpha(1) \approx p_\alpha(2) >> p_\alpha(3).
\end{equation}
This leads to an approximate conservation relation for the kinematic momentum
\begin{equation}
     p_\alpha(1) = p_\alpha(2).
\end{equation}
Using the definition of canonical momentum, the conservation law can be expressed as
\begin{equation}\label{conservation equation}
    p_\alpha(1) = p_\alpha(2) + qA_\alpha,
\end{equation}
where $A_\alpha$ is the electromagnetic four-potential due to the magnetic field.
Let the initial position of the neutral particle in the ionized Keplerian disk (IKD) be denoted by
\begin{equation}
    x^\alpha = (t,r,\theta,\phi) = (0,r,\theta,0),
\end{equation}
and the four-velocity components as
\begin{equation}
    u_\alpha=(u_t,u_r,u_\theta,u_\phi ) = (E_1,0,0,L_1)
\end{equation}
where $E_1$ and $L_1$ are the specific energy and angular momentum of the particle prior to ionization.
\textcolor{black}{
\begin{equation}
    L_1 =\frac{r_0\sqrt{(r_0+R_s) \left(4 \pi  \bm{\kappa} \log \left(\frac{r_0+R_s}{R_s}\right)+\mathcal{T}_0\right)-4
   \pi   \bm{\kappa} r_0}}{\sqrt{(r_0+R_s) \left(-4 \pi   \bm{\kappa} \log
   \left(\frac{r_0+R_s}{R_s}\right)+r_0-3 \mathcal{T}_0\right)+4 \pi   \bm{\kappa} r_0}},
\end{equation}
\begin{equation}
    E_1 = \sqrt{\frac{(r_0+R_s) (r_0-2 \mathcal{T}_0)^2}{r_0 \mathcal{T}_0 \left(-r_0 (4 \pi  \bm{\kappa}+r_0+R_s)+4
   \pi  K (r_0+R_s) \log \left(\frac{r_0+R_s}{R_s}\right)+3 \mathcal{T}_0
   (r_0+R_s)\right)}}.
\end{equation}}
}
\textcolor{black}{Where $\mathcal{T}_0$ is  nothing than Eq.\eqref{T abriviation}  with $r_0$ as argument.}
\textcolor{black}{The explicit expressions for $E_1$ and $L_1$ depend on the spacetime geometry modified by the CDM halo. These quantities are computed  in terms of the density parameter $\bm{\kappa} = \rho_s R_s^3$, the scale radius $R_s$, and the position $r_0$ of the particle on the Keplerian disk. For the ionized particle, the specific energy remains unchanged $E_2=E_1$, while the angular momentum is modified due to the presence of the magnetic field $L_2 = L_1 +qA_\phi$.}

\textcolor{black}{Since $A_t = 0$ for a purely magnetic field, only the angular momentum is affected.
It is worth emphasizing that, in a Schwarzschild background, particles on stable circular orbits cannot escape to infinity purely due to ionization, in contrast to scenarios involving rotational  spacetimes or stronger external fields \cite{panis2019determination}.}

\textcolor{black}{In Fig.\ref{keplerian disk, different values of kappa}, we illustrate the structure of the Keplerian disk in the absence of magnetic field ($B'=0$) for different values of the CDM density parameter $\bm{\kappa}$.}
\begin{figure}[!ht]
    \centering
    \subfloat[]{\includegraphics[width=0.3\textwidth]{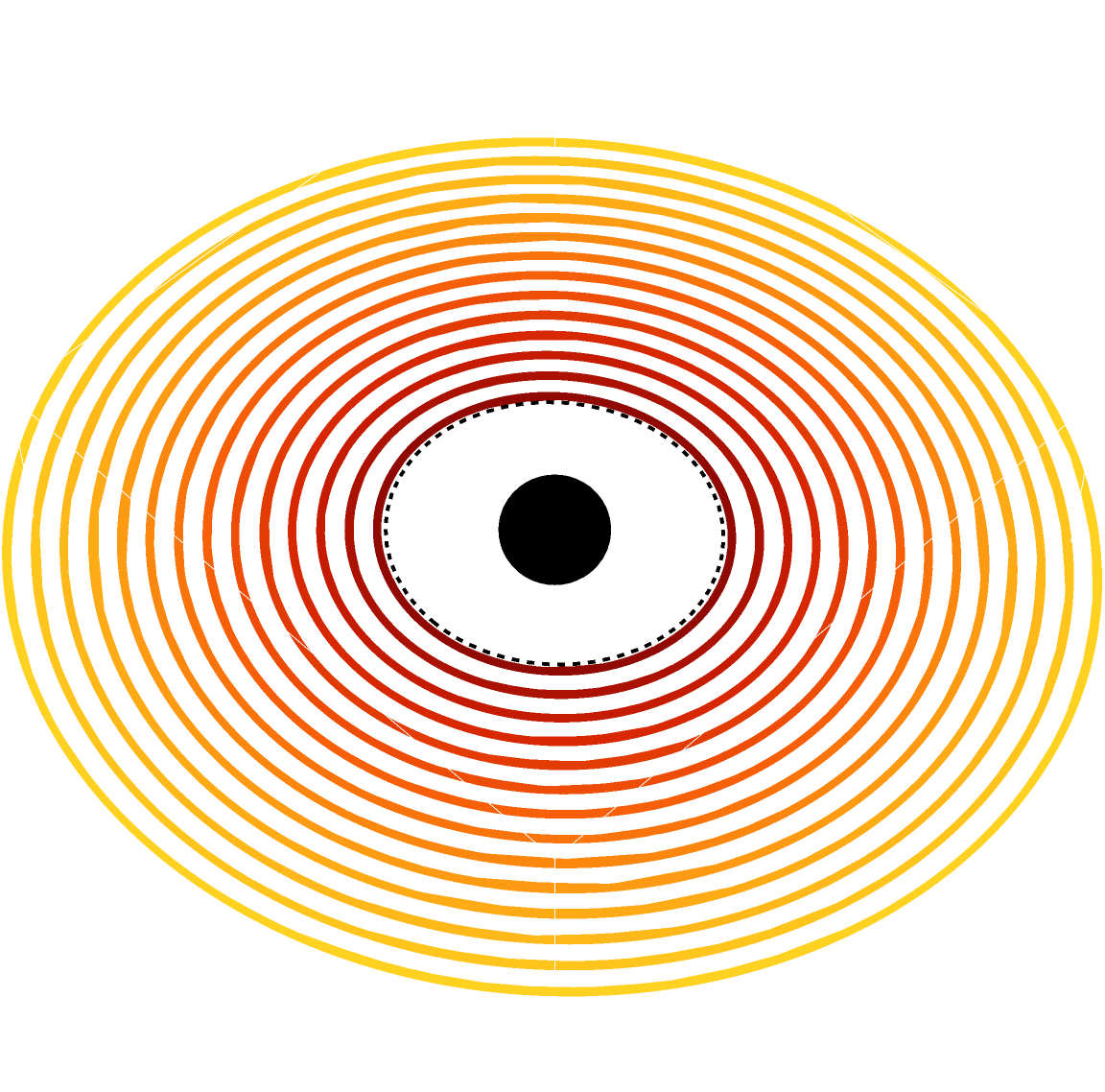}\label{}}
    \subfloat[]{\includegraphics[width=0.3\textwidth]{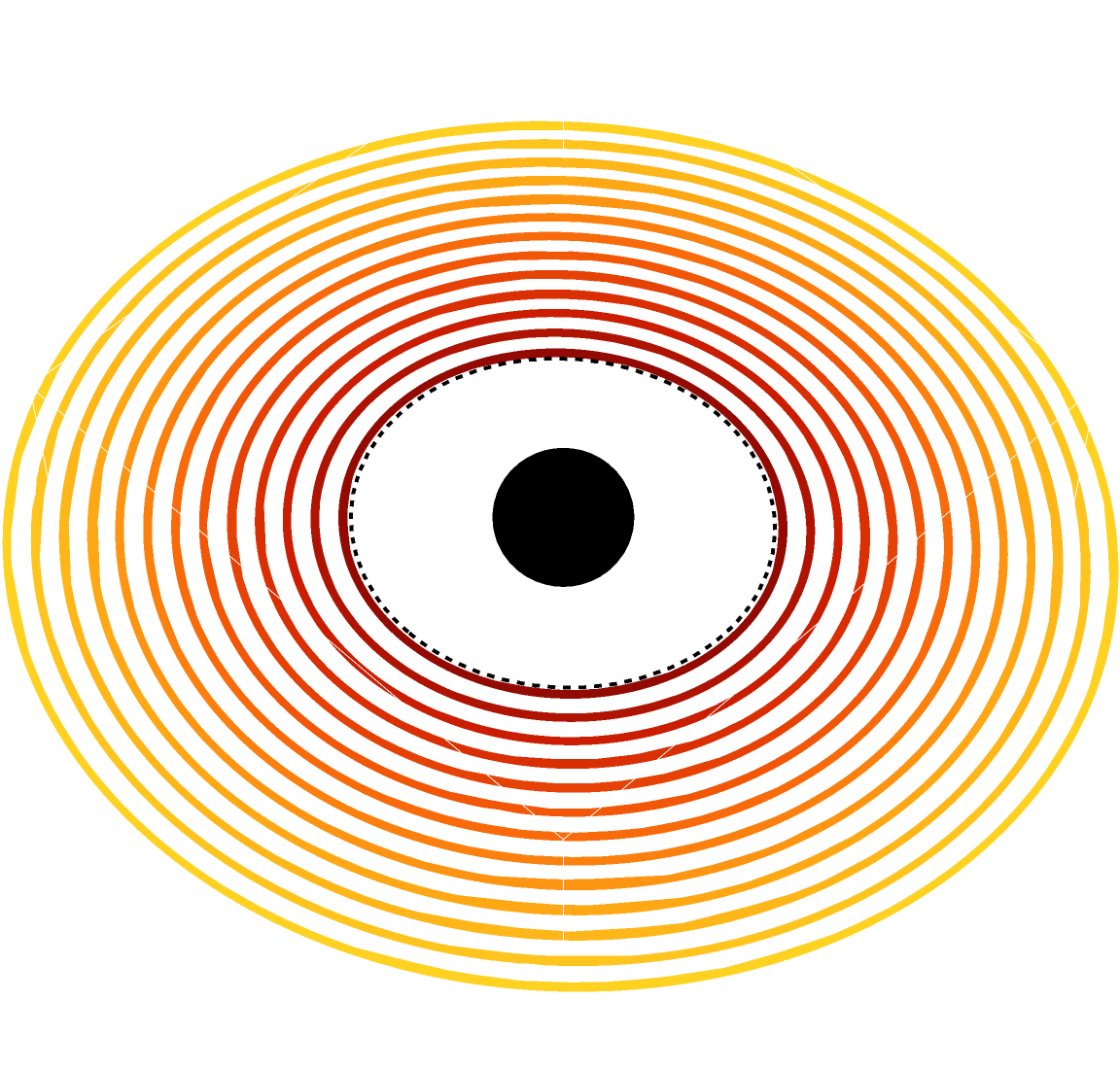}\label{}}
    \subfloat[]{\includegraphics[width=0.3\textwidth]{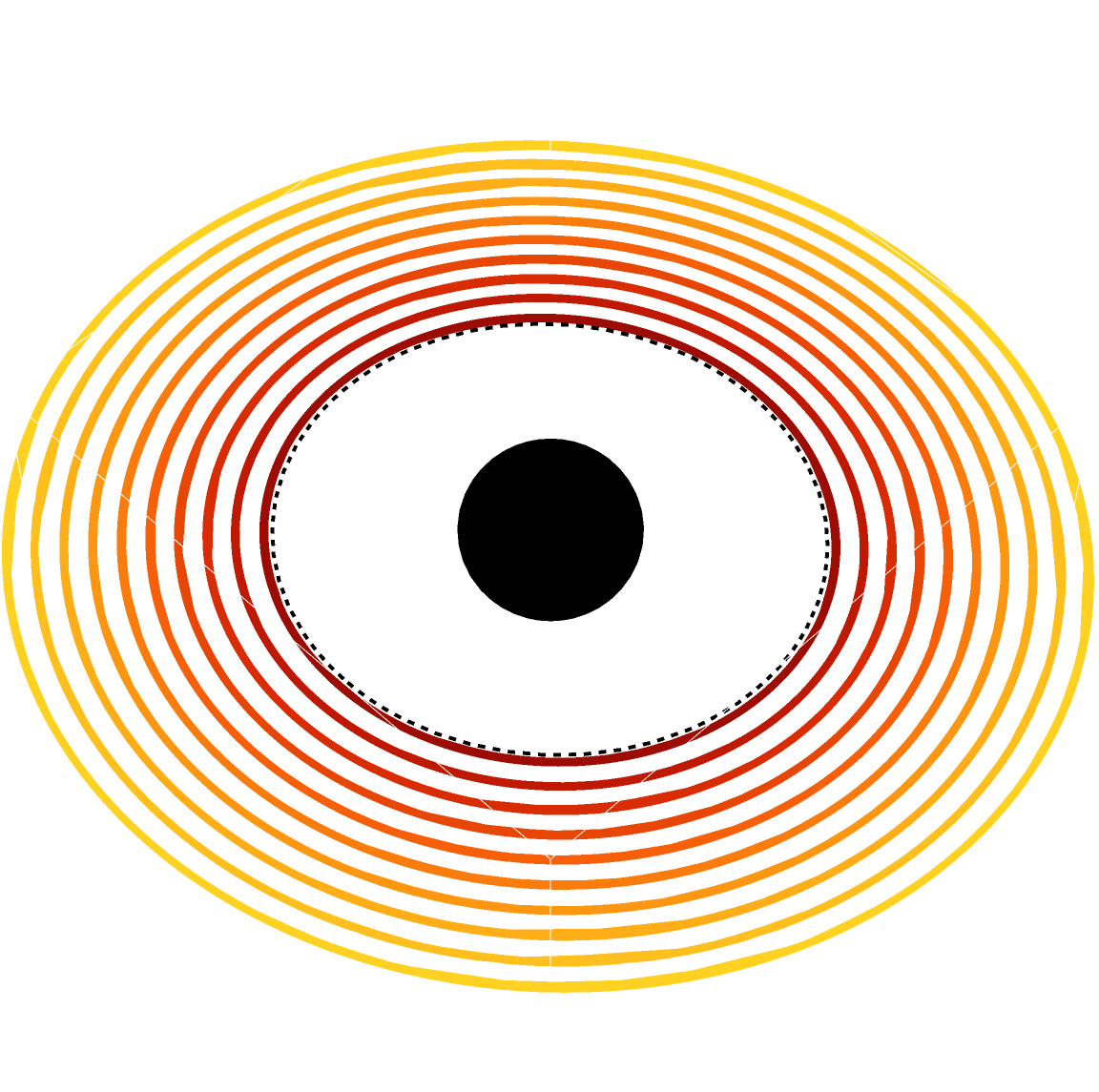}\label{}}
    \caption{\footnotesize{\it \textcolor{black}{Keplerian disk composed of neutral particles in the presence of CDM halos with density parameters $\bm{\kappa} = 0,\,10,\,20$ for curves (a), (b), and (c), respectively. The inclination of the disk is fixed at $\theta_0 = 1.51$ radians relative to the equatorial plane. The black dashed line marks the ISCO radius, while the central black sphere indicates the black hole event horizon.} }}
    \label{keplerian disk, different values of kappa}
\end{figure}
\textcolor{black}{The disk remains geometrically thin in all cases, but an increasing $\bm{\kappa}$ shifts the innermost stable circular orbit  outward, as well as the event horizon radius. This highlights the gravitational influence of the CDM halo on the inner disk structure and possible accretion dynamics.}




\textcolor{black}{Having established the Hamiltonian framework for charged particle dynamics and the formation of Keplerian disks, we now shift our focus to examining the role of the surrounding CDM halo in the evolution of the conserved quantities $L'$ and $E'$, under both attractive and repulsive Lorentz forces. Specifically, we investigate the motion of charged particles around a magnetized black hole embedded in a CDM halo, in order to assess the combined influence of the Lorentz force and dark matter on the ionization and stability of the Keplerian disk.}

\subsection{Attractive Lorentz force}
\begin{figure}[!ht]
    \centering
    \subfloat[]{\includegraphics[width=1\textwidth]{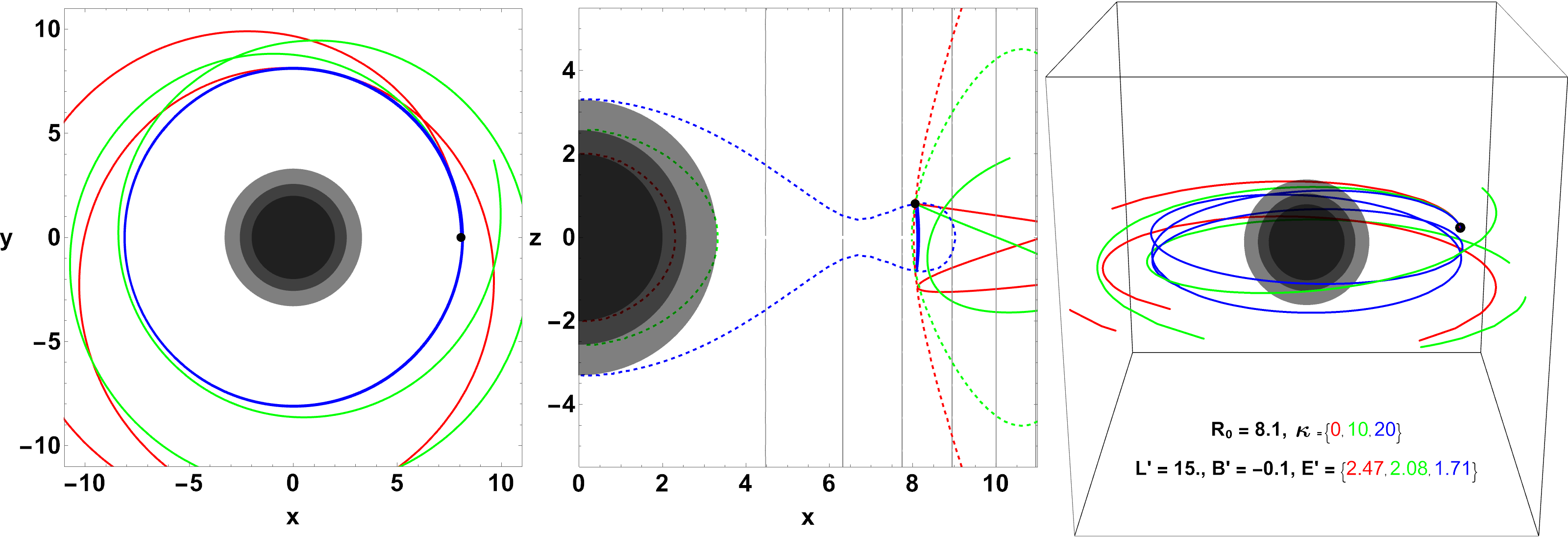}\label{isco for K=20 negative}}\\
    \subfloat[]{\includegraphics[width=1\textwidth]{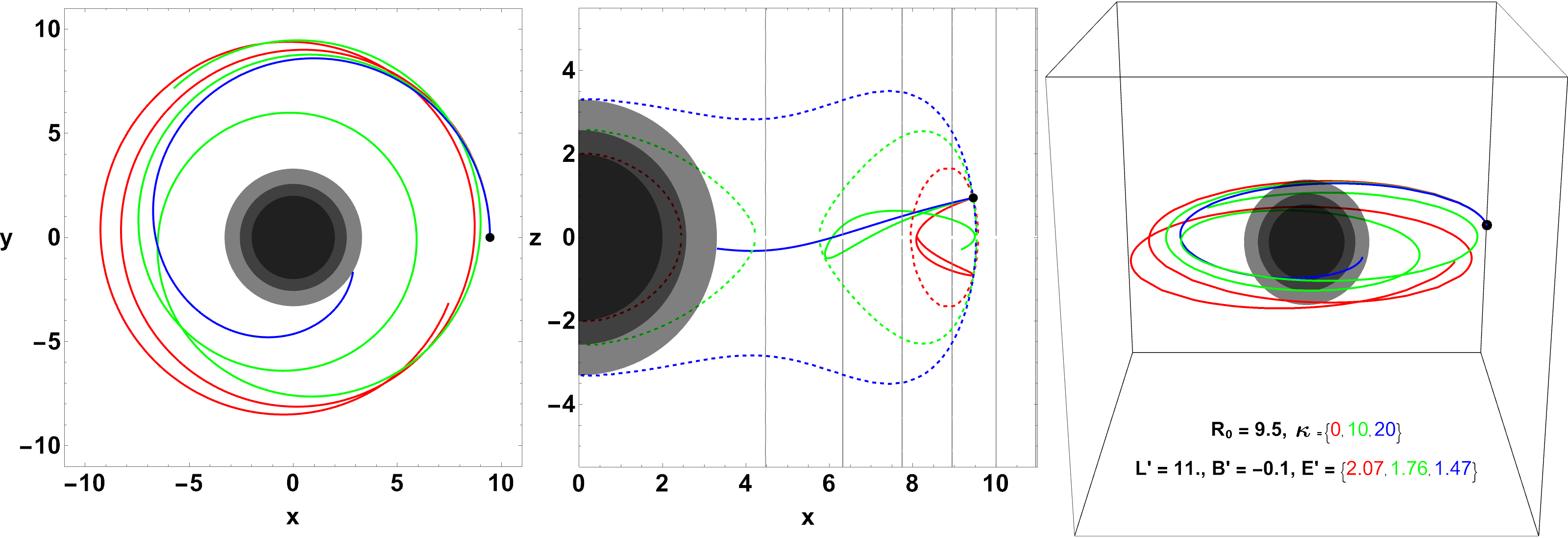}\label{trapped for negative}}
    \caption{\footnotesize{\it Illustration of charged particle trajectories for different values of $\bm{\kappa}$ around BH embedded in  a uniform magnetic field of strength $B'=-0.1$ and CDM halo within $\theta_0\approx\pi/2$. Parameters $R_0, L'$ and $E'$ present initial position, angular momentum and specific energy of charged particle respectively . The disks at the centers of the panels in the first and second rows represent the BH event horizon for varying $\bm{\kappa}$, while in the third row, the horizons are depicted as spherical objects. Panel (a) highlights the variation of the ISCO radius with increasing $\bm{\kappa}$, and panel (b) illustrates the effect of the CDM halo on particle dynamics. Notably, a trapped state appears for $\bm{\kappa}=20$, indicating the growing influence of CDM in confining particle motion.}}
    \label{figure 6}
\end{figure}

\begin{figure}[!ht]
    \centering
    \setlength{\tabcolsep}{2pt} 
    \renewcommand{\arraystretch}{1} 
    \begin{tabular}{c c c c}
        & \textbf{\(B'=-0.01\)} & \textbf{\(B'=-0.1\)} & \textbf{\(B'=-1\)} \\
        
        \adjustbox{valign=c}{\(\bm{\kappa}=0\)} &
        \adjustbox{valign=c}{\subfloat[]{\includegraphics[width=0.33\textwidth]{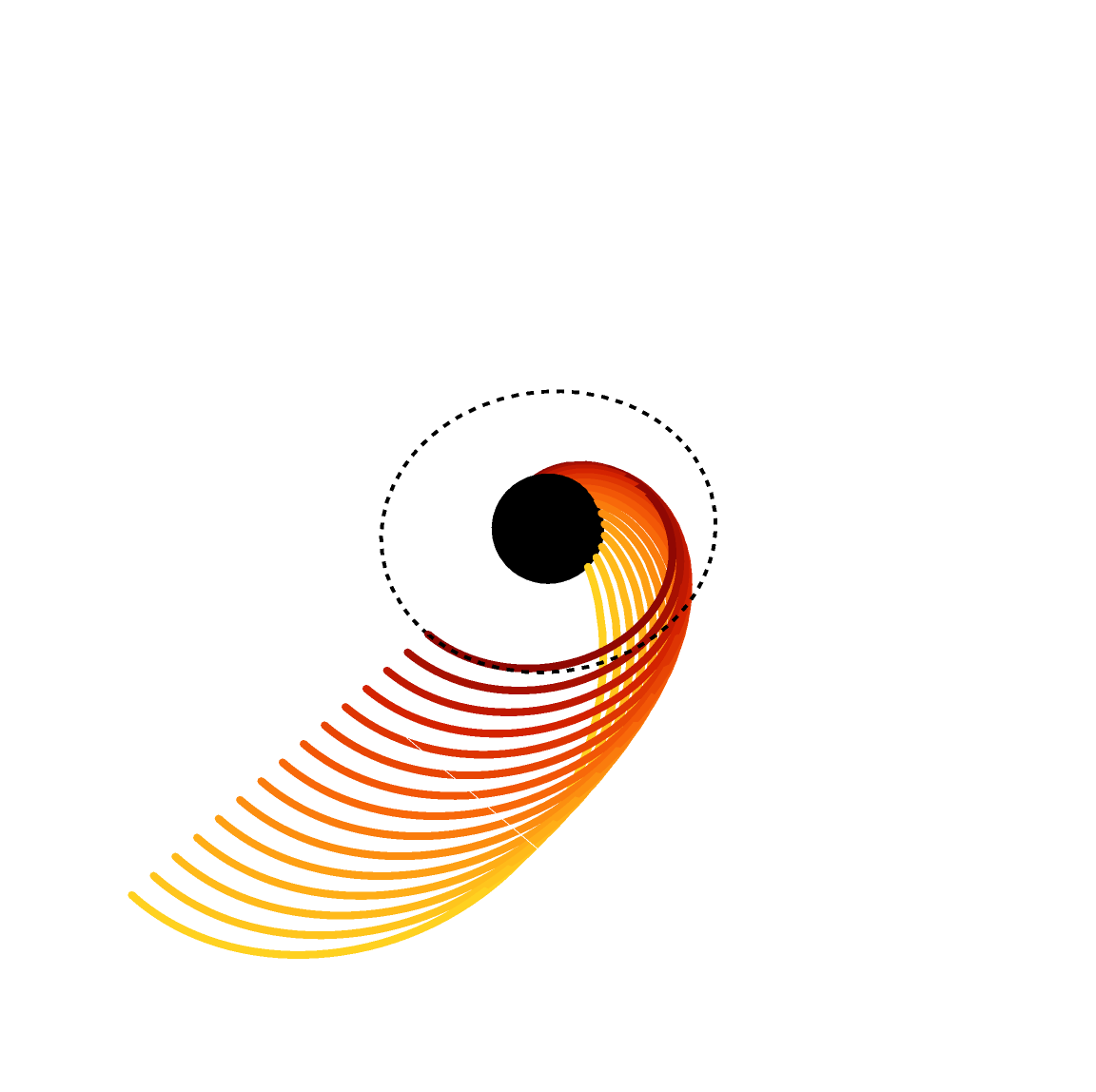}}} &
        \adjustbox{valign=c}{\subfloat[]{\includegraphics[width=0.33\textwidth]{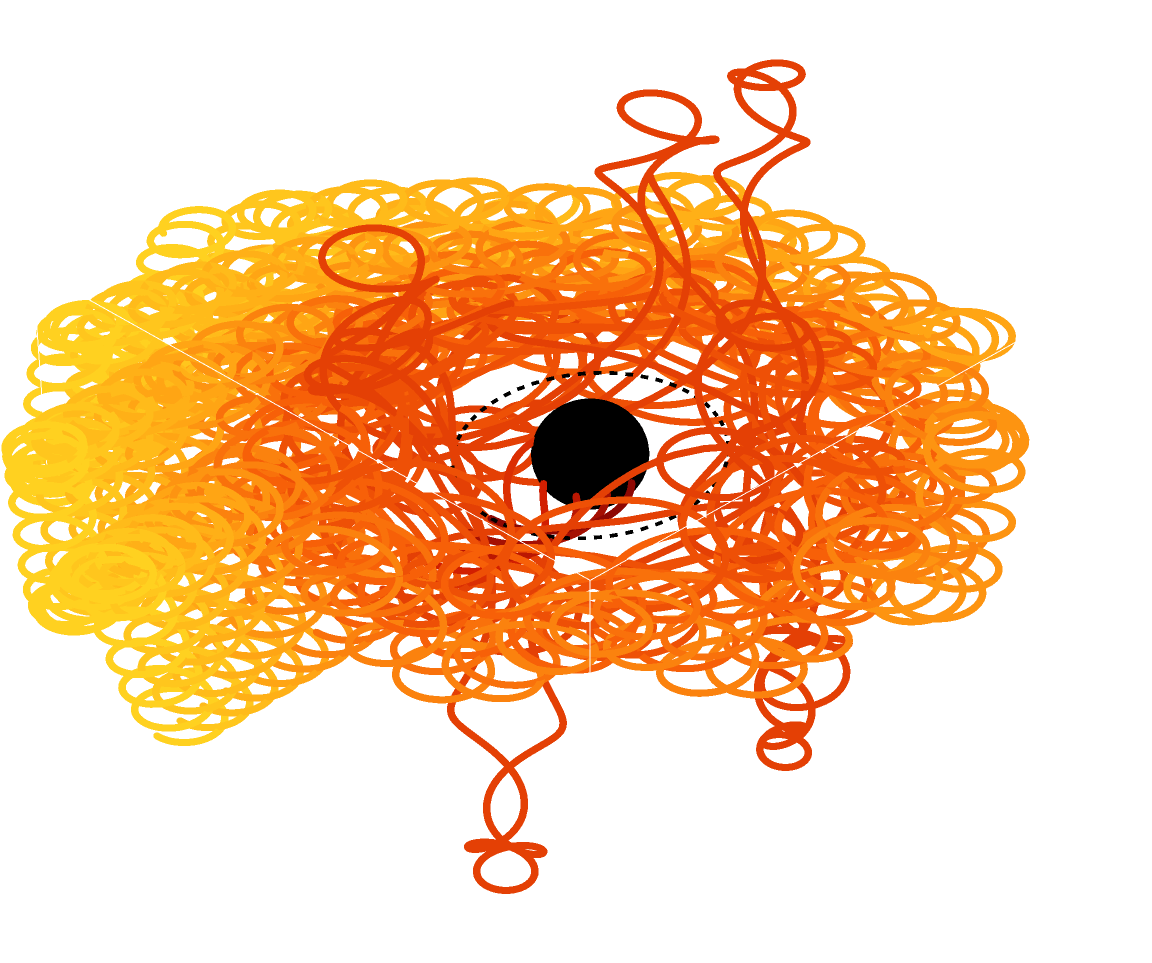}}} &
        \adjustbox{valign=c}{\subfloat[]{\includegraphics[width=0.33\textwidth]{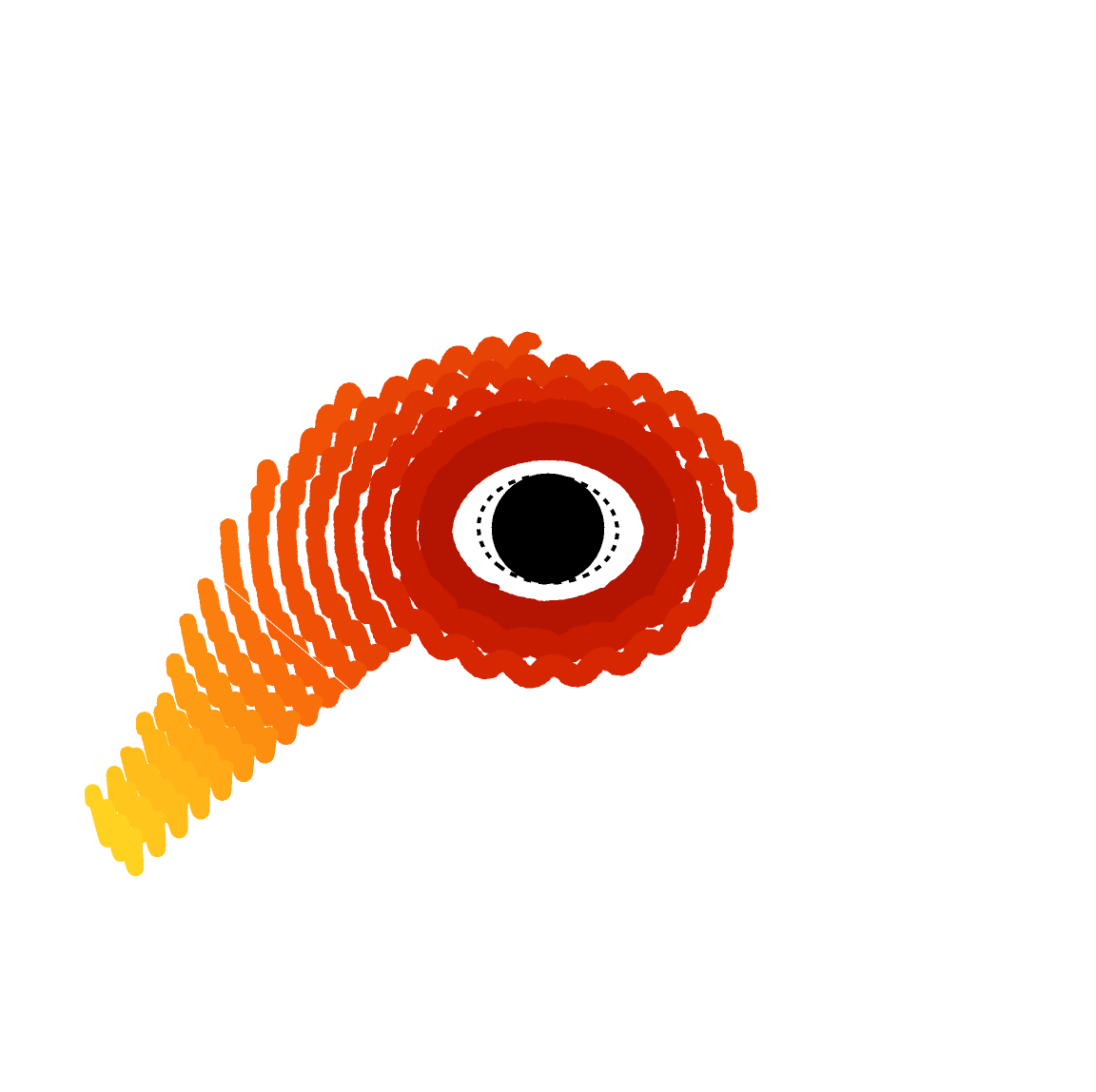}}} \\
        
        \adjustbox{valign=c}{\(\bm{\kappa}=10\)} &
        \adjustbox{valign=c}{\subfloat[]{\includegraphics[width=0.33\textwidth]{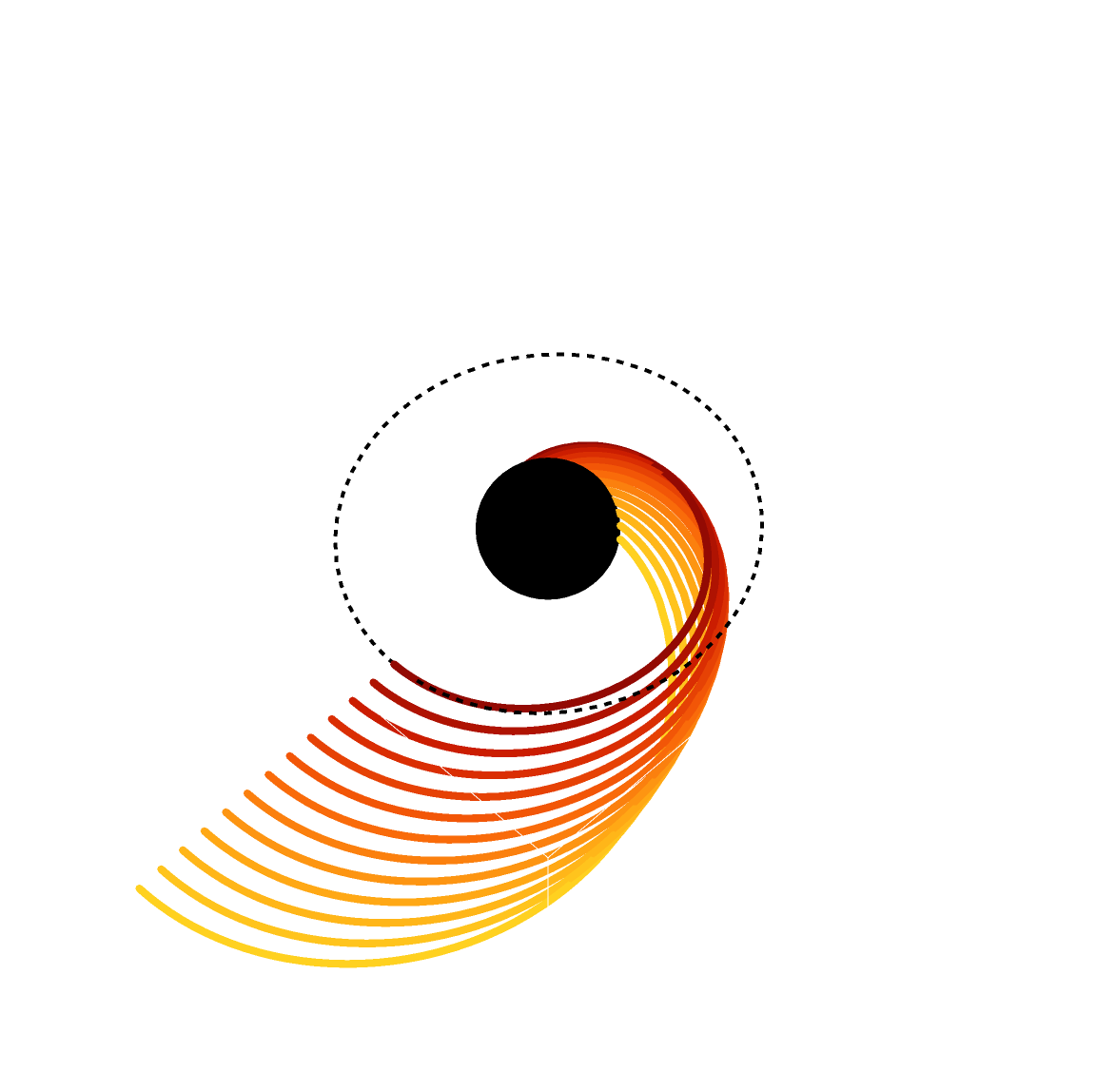}}} &
        \adjustbox{valign=c}{\subfloat[]{\includegraphics[width=0.33\textwidth]{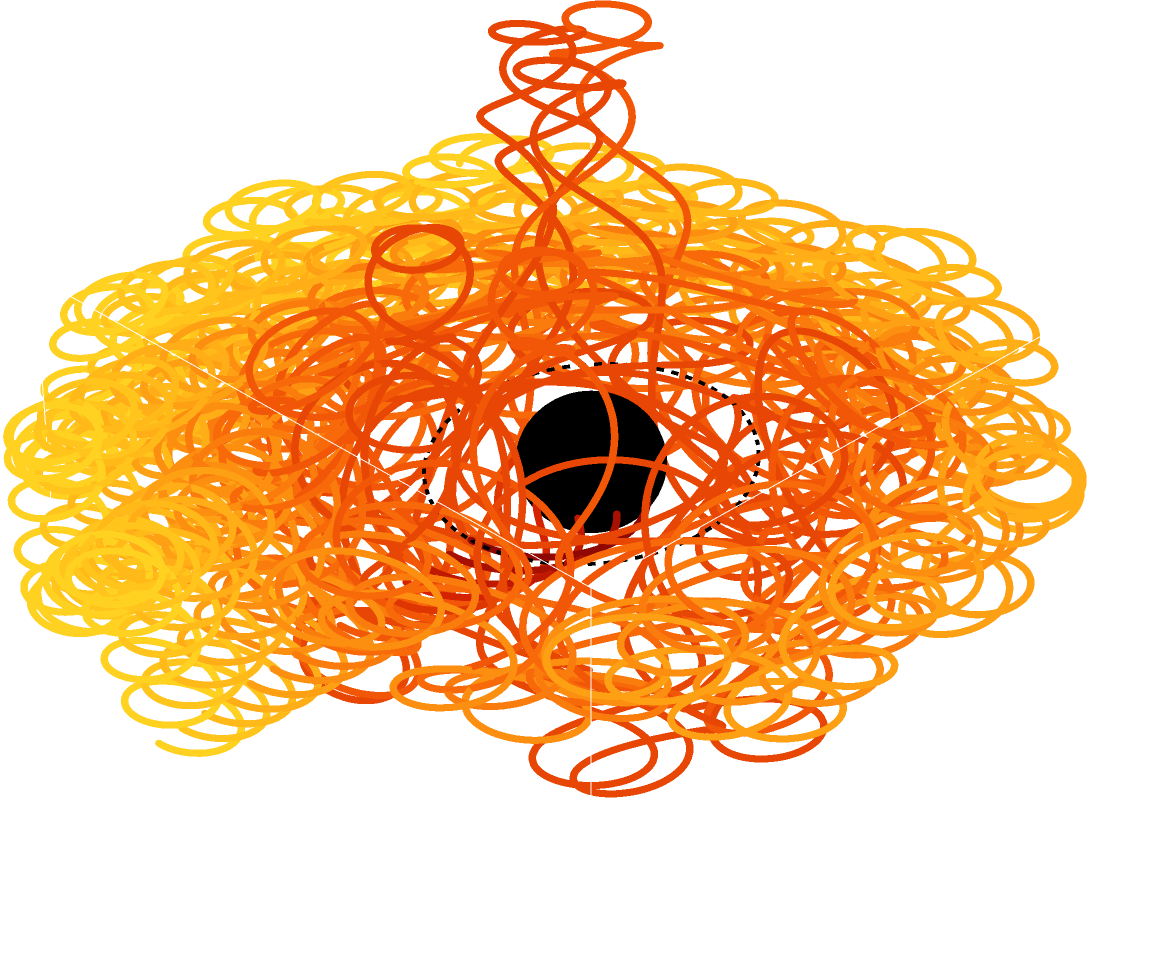}}} &
        \adjustbox{valign=c}{\subfloat[]{\includegraphics[width=0.33\textwidth]{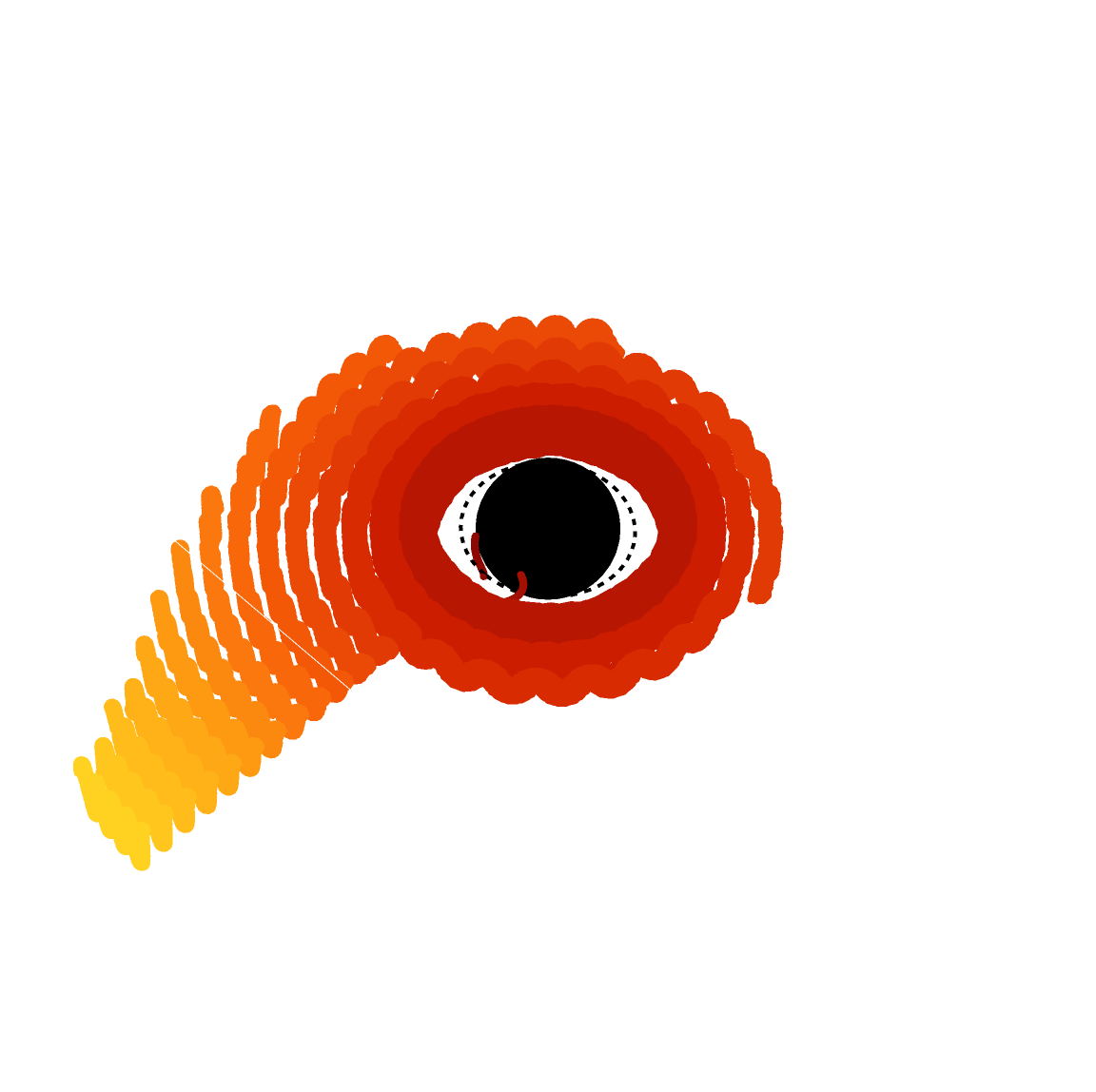}}} \\
        
        \adjustbox{valign=c}{\(\bm{\kappa}=20\)} &
        \adjustbox{valign=c}{\subfloat[]{\includegraphics[width=0.33\textwidth]{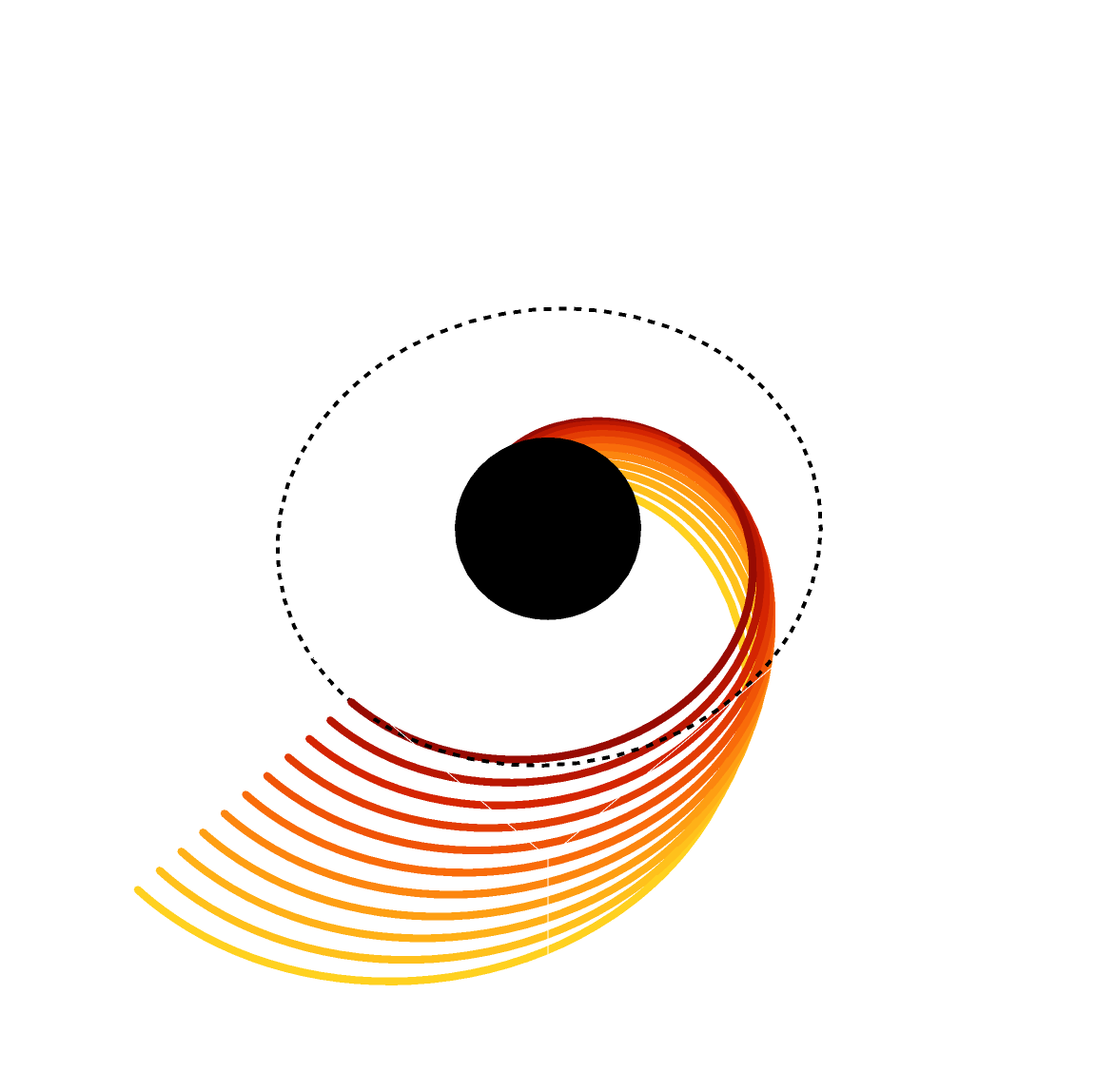}}} &
        \adjustbox{valign=c}{\subfloat[]{\includegraphics[width=0.33\textwidth]{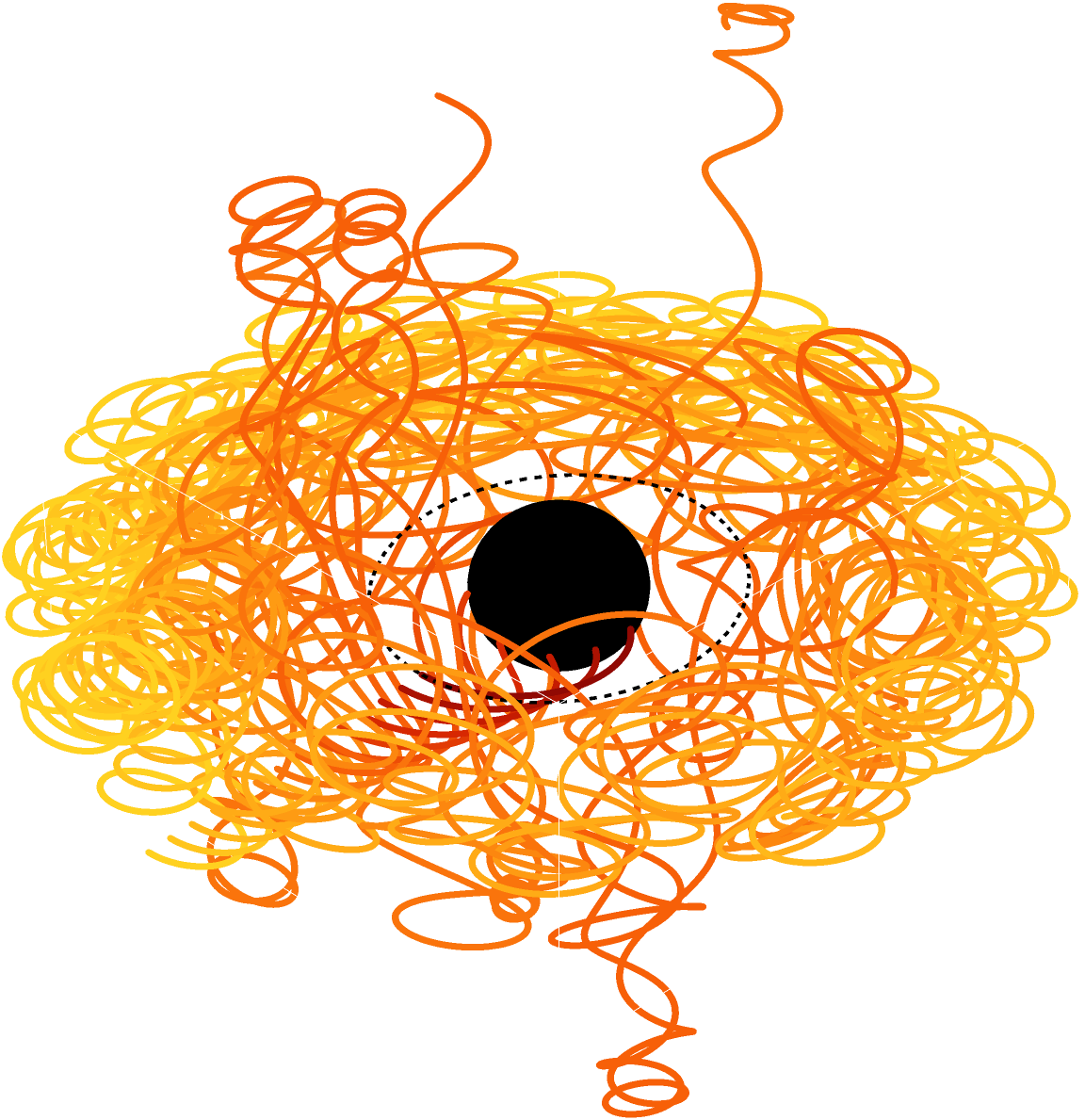}}} &
        \adjustbox{valign=c}{\subfloat[]{\includegraphics[width=0.33\textwidth]{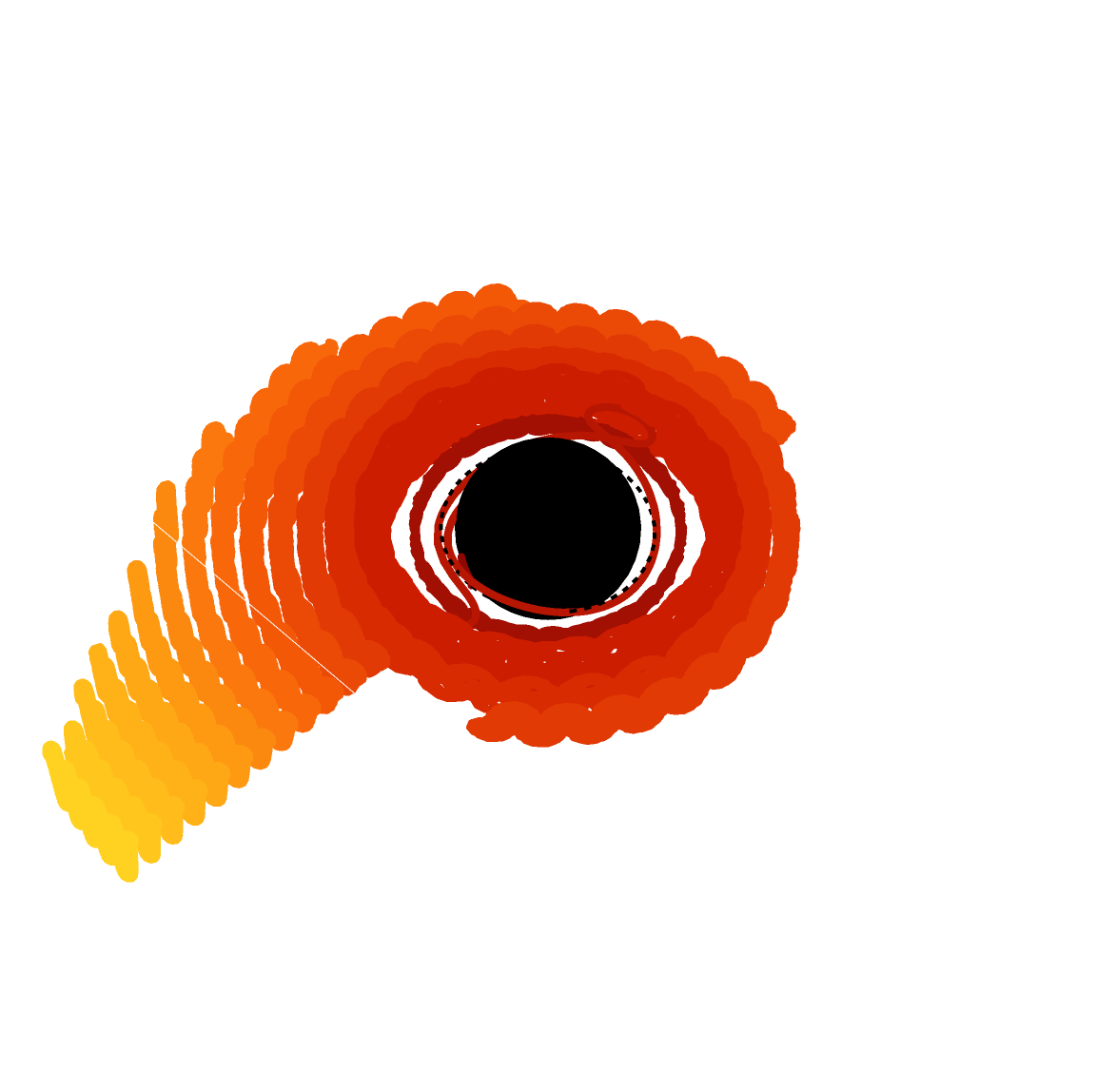}}} \\
    \end{tabular}
    \caption{\footnotesize{\it \textcolor{black}{Ionized Keplerian disk evolution under the influence of uniform magnetic field of strength ($B'=-0.01,-0.1,-1$) represented from first column to the third column respectively, also by considering the presence of CDM density ($\bm{\kappa}=0,10,20$), from first line to third line respectively. The inclination of the accretion disk settled to $\theta_0 =1.51$ to the equatorial plan. Dashed black line represent the isco, for spherically black object at the center represent the BH event horizon.}
    }}
    \label{IKD attractive}
\end{figure}

\textcolor{black}{In Fig.\ref{figure 6}, we present sample particle trajectories under an attractive Lorentz force configuration ($B'=-0.1$). The corresponding modifications to the Keplerian disk structure, arising from different strengths of the uniform magnetic field and CDM density, are depicted in Fig.\ref{IKD attractive}.}

From panels \ref{isco for K=20 negative}, one can notice that the $(x-y)$ plot displays the orbital trajectories of charged particles projected onto the equatorial plane. The color-coded trajectories linked to increasing values of the CDM density parameter $\bm{\kappa}=0$ (red), $10$ (green), and $20$ (blue), with all other parameters held constant: $R_0=8.1$, $L'=15$, and $B'=-0.1$. As $\bm{\kappa}$ increases, the orbits become more confined, indicating stronger gravitational effects due to the denser dark-matter halo. The red trajectory ($\bm{\kappa}=0$) shows the widest orbit, gradually transitioning to more tightly bound orbits for green and blue as the influence of CDM intensifies.

The $(x-z)$ plane shows the vertical (out-of-plane) extension of the particle motion. Equipotential contours (dashed lines) are included, which delineate the effective energy boundaries governing motion in the poloidal plane. The escape trajectories for ($\bm{\kappa}=0$) and $10$ are open, allowing the particles to exit along the $z-$axis, while for $\bm{\kappa}=20$, the particle appears trapped within the potential well.  The right panel offers a three-dimensional visualization of the trajectories for different $\bm{\kappa}$, clearly showing how the presence of CDM modifies the spatial extent and structure of the motion. The shrinking of the orbit's spatial envelope from red to blue confirms that CDM increases gravitational confinement of trajectories.
 
The \ref{trapped for negative} panels mirror the structure of \ref{isco for K=20 negative}, but under a different parameter set: $R_0=9.5$, $L'=11$, and the same $B'=-0.1$. The previously observed behavior in $(x-y)$ plane persist, but the initial orbits begin closer to the BH, and the confining contribution of CDM is more pronounced even at lower $\bm{\kappa}$.  The $(x-z)$ plane confirms the transition from open to closed energy surfaces as $\bm{\kappa}$ grows, with the particle at $\bm{\kappa}=20$ exhibiting a clear bound state within the polar confinement zone. Lastly, the $3D$ visualization reinforces the conclusion that CDM density plays a major role in spatially restricting particle motion. The orbits shrink progressively as $\bm{\kappa}$ increases, with tighter confinement around the BH.

\textcolor{black}{From the left column of Fig.\ref{IKD attractive}, we observe that in the case of an attractive weak Lorentz force ($B'=-0.01$), the IKD becomes unstable, leading to the complete destruction of the Keplerian disk, as all particle trajectories in the various CDM configurations ultimately spiral into the black hole.}

\textcolor{black}{The middle Fig.\ref{IKD attractive} column unveils that for a stronger magnetic field ($B'=-0.1$), the trajectories of charged particles exhibit pronounced chaotic behavior, particularly near the ISCO radius. This turbulence gives rise to a toroidal structure aligned with the magnetic field lines along the $z$-axis, extending in both upward and downward directions. At the inner edge of the ISCO, a large fraction of particles are rapidly driven toward the event horizon, leading to the disruption of the Keplerian disk. The role of CDM becomes evident: as $\bm{\kappa}$ increases, both the degree of chaotic motion and the size of the toroidal structure grow, while an even larger proportion of particles near the ISCO ultimately fall into the black hole.}

\textcolor{black}{ The right column of Fig.\ref{IKD attractive} associated with a stronger field ($B'=-1$) reveals that the attractive Lorentz force dominates over the CDM contribution for $\bm{\kappa}=0$ and $10$. When $\bm{\kappa}=20$, particles near the inner edge of the ISCO are driven rapidly toward the black hole, leading to the destruction of the Keplerian disk. The remaining trajectories oscillate vertically along the magnetic field lines, while simultaneously executing clockwise circular orbits around the black hole, consistent with the negative polarity of $B'$ ($B'<0$). In this regime, the Keplerian disk transforms from a thin structure into a thicker configuration as a consequence of the strong magnetization ($B'=-1$). It is important to emphasize that the particle trajectories deviate from the $z$-axis, and therefore do not strictly follow the uniform magnetic field lines..}


\textcolor{black}{In the following subsection, we turn to the case of a repulsive Lorentz force. Fig.\ref{figure 7} displays the trajectories of charged particles for $B'=0.1$, while the resulting modifications to the Keplerian disk structure, under different magnetic field strengths and CDM densities, are illustrated in Fig.\ref{IKD repulsive}.}
\subsection{Repulsive Lorentz force}
\begin{figure}[!ht]
    \centering
    \subfloat[]{\includegraphics[width=1\textwidth]{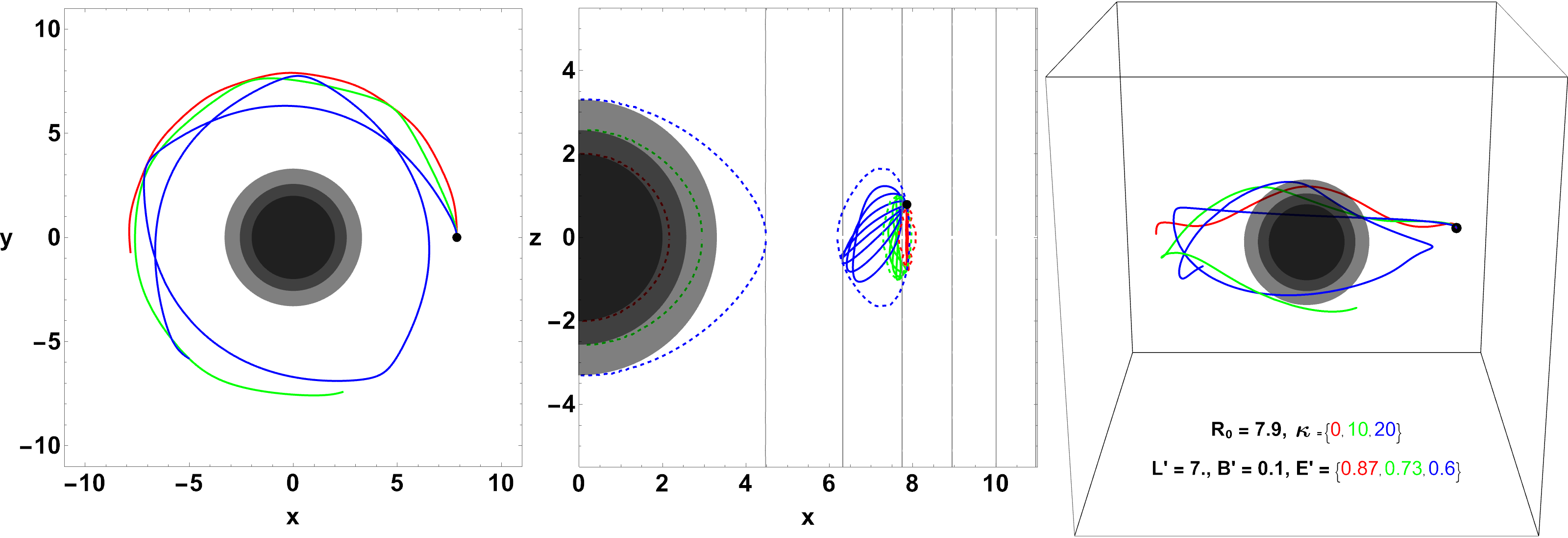} \label{isco for K=0 positive}}\\
    \subfloat[]{\includegraphics[width=1\textwidth]{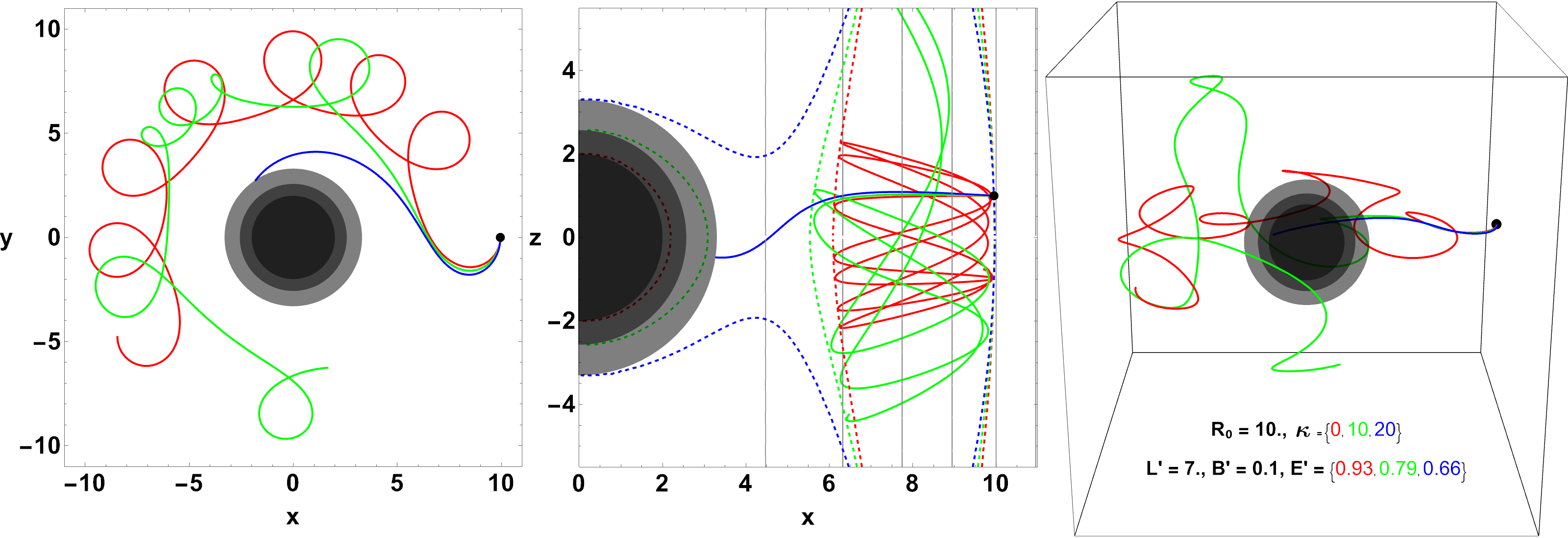}\label{trapped for positive}}
    \caption{\footnotesize{\it Charged particle trajectories for different values of $\bm{\kappa}$ around BH surrounded by a uniform magnetic field of strength $B'=0.1$ and CDM halo within $\theta_0\approx\pi/2$. Parameters $R_0, L'$, and $E'$ present the initial position, angular momentum, and specific energy of the charged particle, respectively. The disks at the centers of the panels in the first and second rows represent the BH event horizon for varying $\bm{\kappa}$, while in the third row, the horizons are depicted as spherical objects. Panel (a) isco located where $\bm \kappa =0$, for (b) charged particles trajectories for $\bm{\kappa}=20$ trapped toward BH, while for decreasing values of $\bm{\kappa}$ particles trajectories orbiting in a curled trajectories while it variate and become large where $\bm{\kappa}$ value close to zero.}}
    \label{figure 7}
\end{figure}

From Fig.\ref{figure 7} which illustrates the particle dynamics within a positive uniform magnetic field $B'=0.1$, with the same angular momentum $L'=7$ used for all trajectories,
 One can observe that the trajectories exhibit more extended and less confined motion, particularly in the $(x-z)$ planes and $3D$ views. Higher CDM values continue to influence the degree of curvature, effectively trapping particles. This effect is clearly illustrated in the case where $\bm{\kappa}=20$.
Furthermore, in Fig.\ref{trapped for positive}, the red and green trajectories (lower $\bm{\kappa}$) display complex looping structures in the $(x-y)$ plane, driven by magnetic effects, which are more pronounced than in the $B'<0$ depicted in Fig.\ref{figure 6}. 
Rigorously speaking, CDM, characterized by the parameter $\bm{\kappa}$, significantly perturbs the coordinate of the ISCO and alters $E'$ of charged particles, regardless of the magnetic field's sign. As shown in the figures, increasing $\bm{\kappa}$ enhances the likelihood of particles becoming trapped in bounded orbits. These trapped states are defined by inner and outer energy boundaries, which confine the motion to a finite radial region. However, at sufficiently high CDM density, the confinement may break down, causing particles to spiral inward toward the BH.

\begin{figure}[!ht]
    \centering
    \setlength{\tabcolsep}{1pt}
    \renewcommand{\arraystretch}{0.5} 
       \begin{tabular}{c c c c}
        & \textbf{\(B'=0.01\)} & \textbf{\(B'=0.1\)} & \textbf{\(B'=1\)} \\
        
        \adjustbox{valign=c}{\(\bm{\kappa}=0\)} &
        \adjustbox{valign=c}{\subfloat[]{\includegraphics[width=0.33\textwidth]{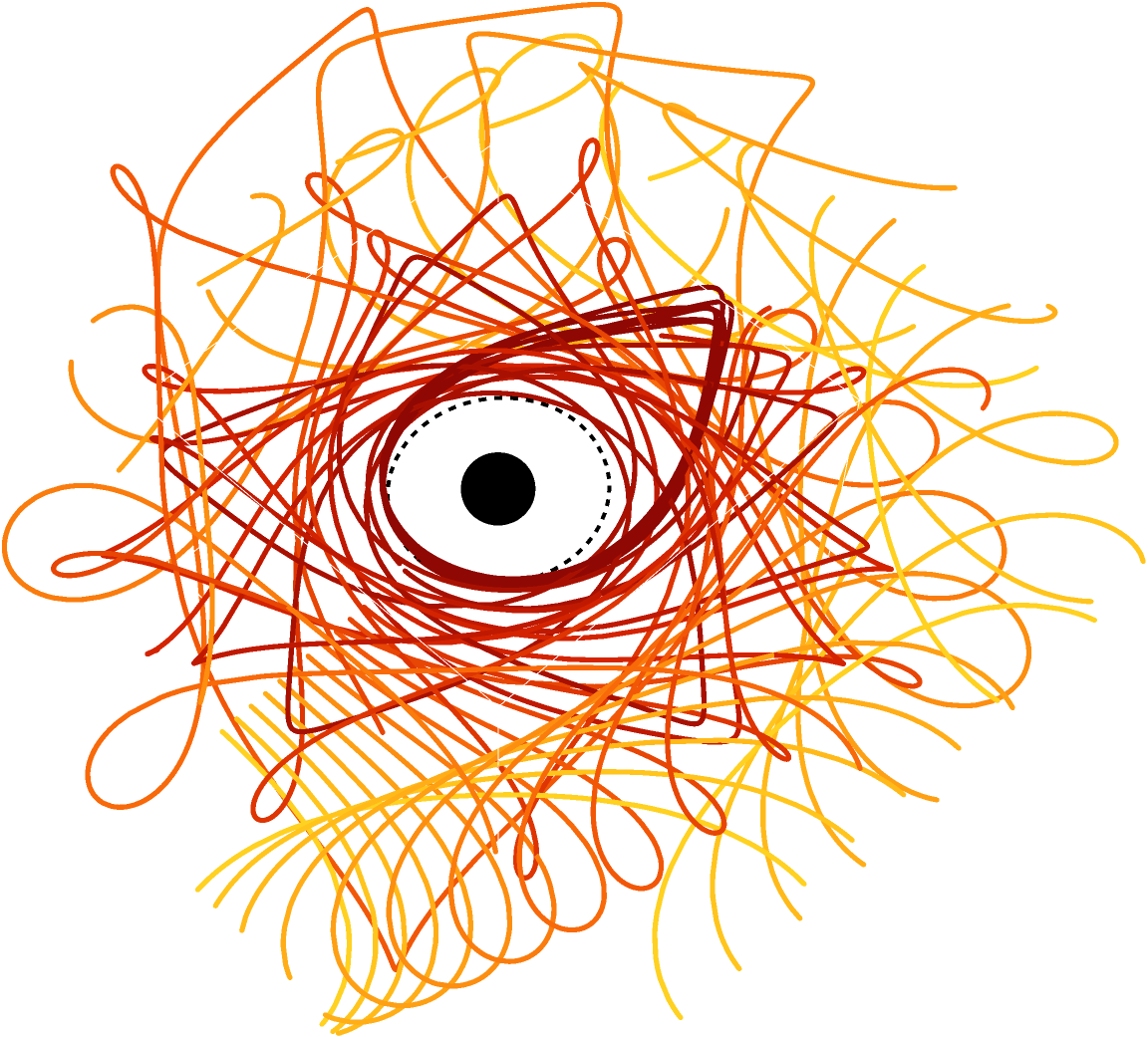}}} &
        \adjustbox{valign=c}{\subfloat[]{\includegraphics[width=0.33\textwidth]{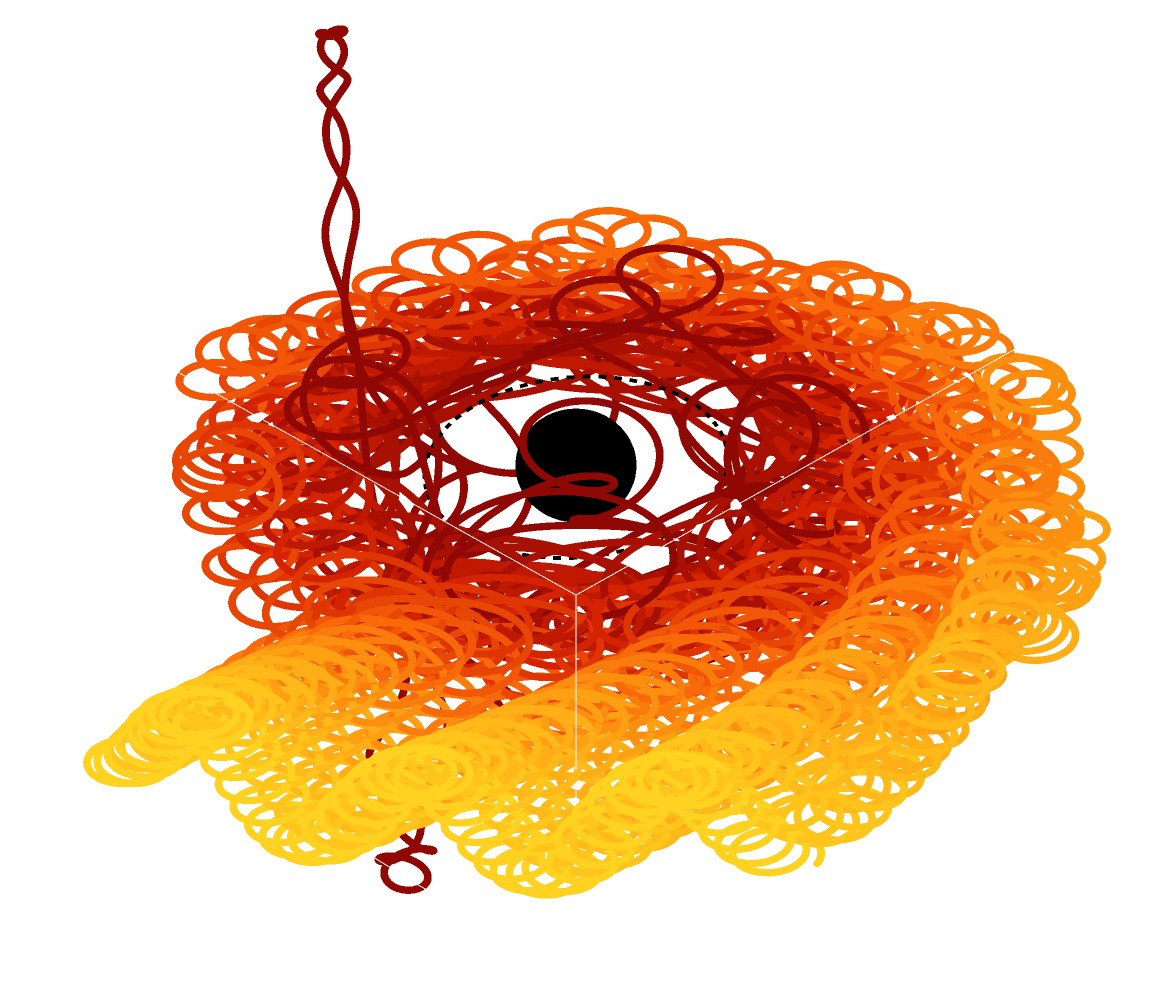}}} &
        \adjustbox{valign=c}{\subfloat[]{\includegraphics[width=0.33\textwidth]{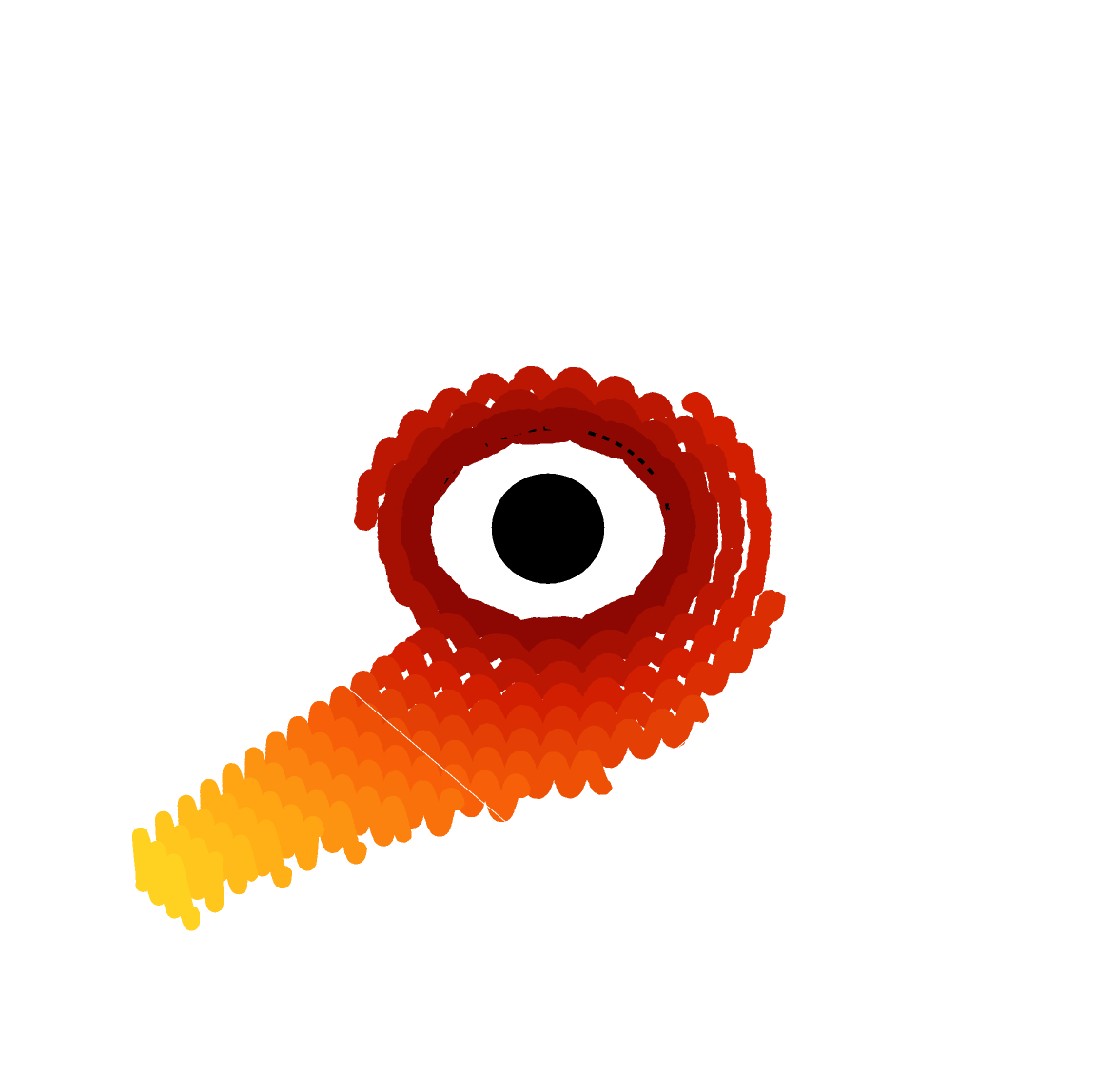}}} \\
        
        \adjustbox{valign=c}{\(\bm{\kappa}=10\)} &
        \adjustbox{valign=c}{\subfloat[]{\includegraphics[width=0.33\textwidth]{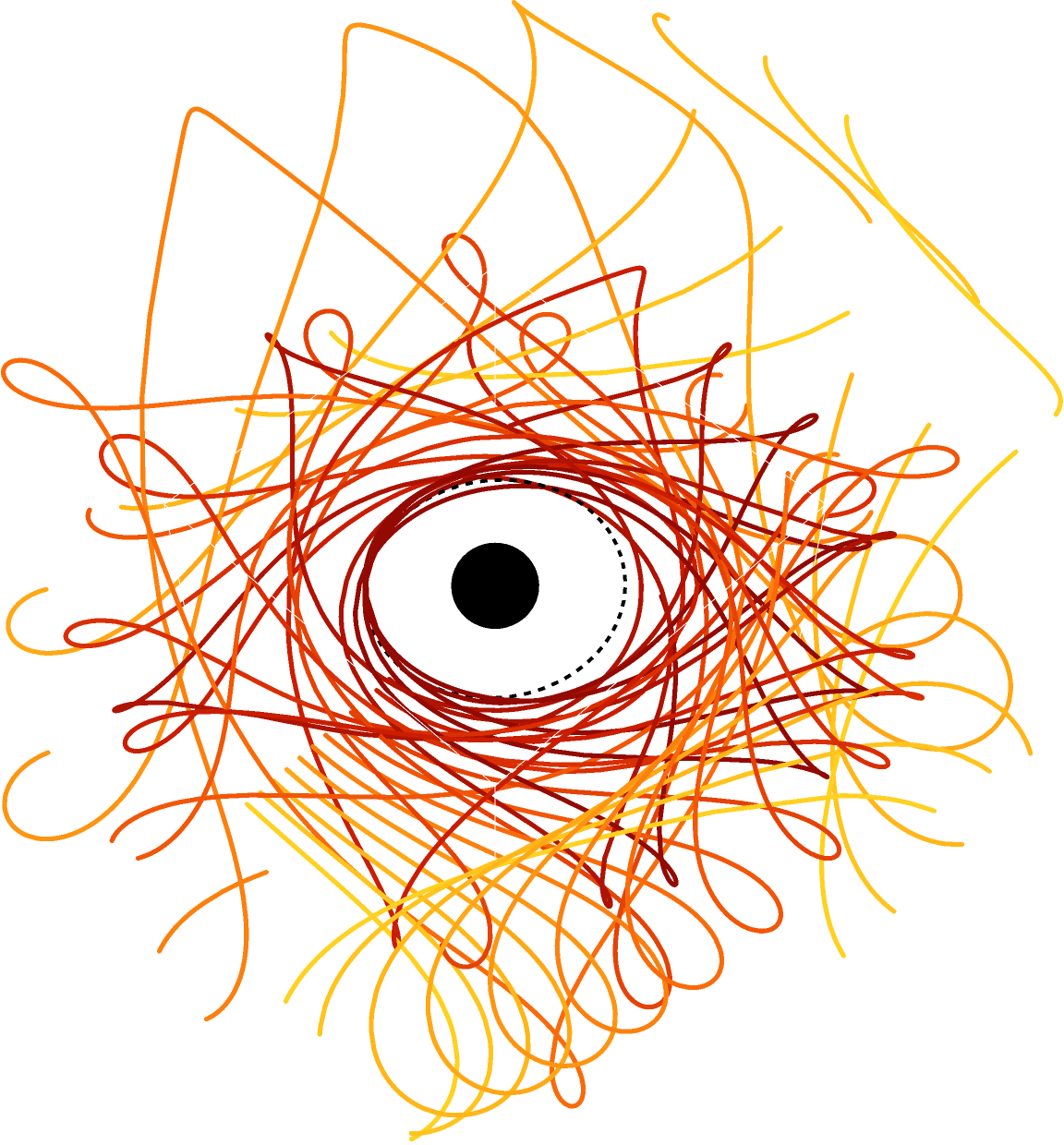}}} &
        \adjustbox{valign=c}{\subfloat[]{\includegraphics[width=0.33\textwidth]{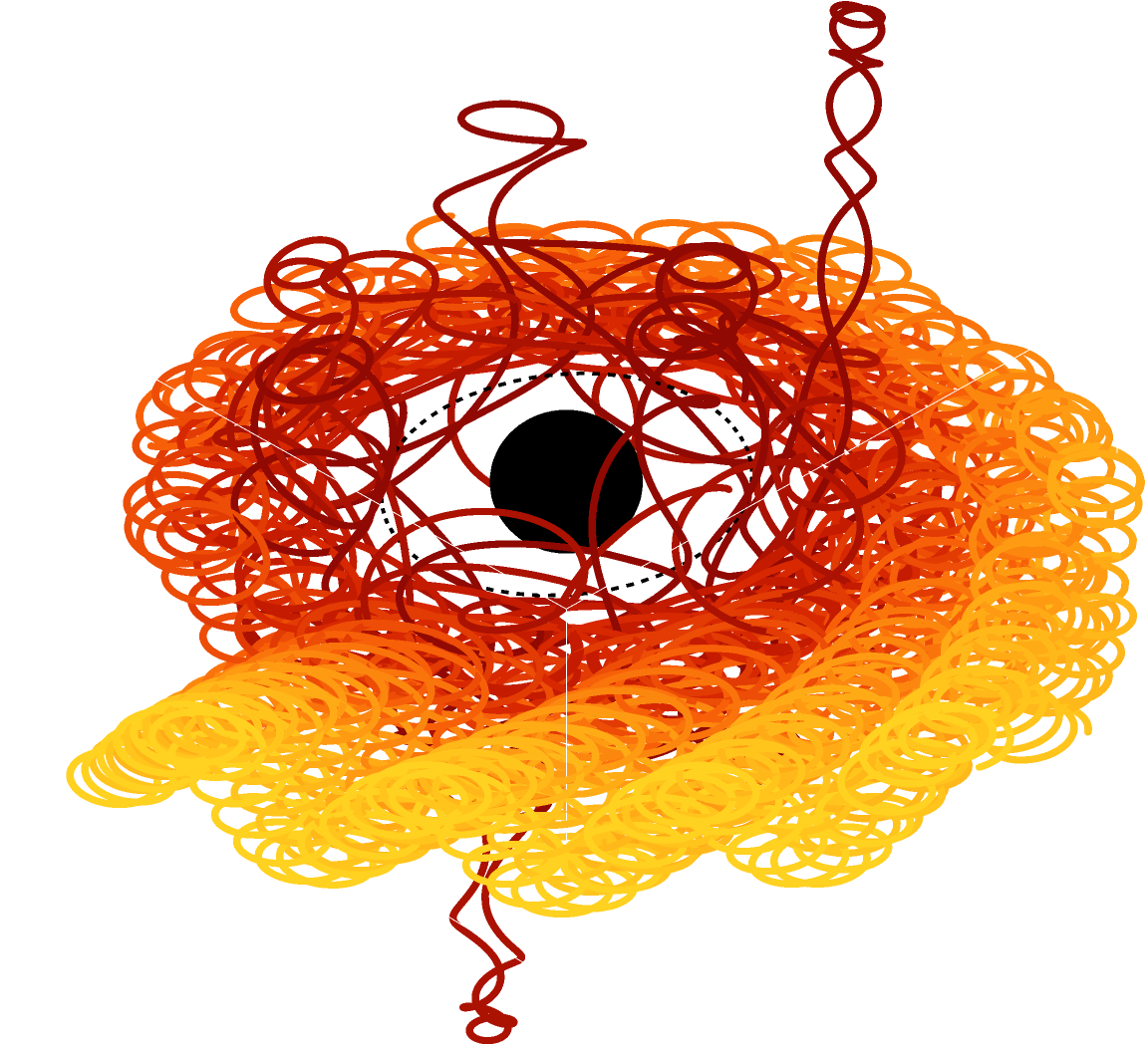}}} &
        \adjustbox{valign=c}{\subfloat[]{\includegraphics[width=0.33\textwidth]{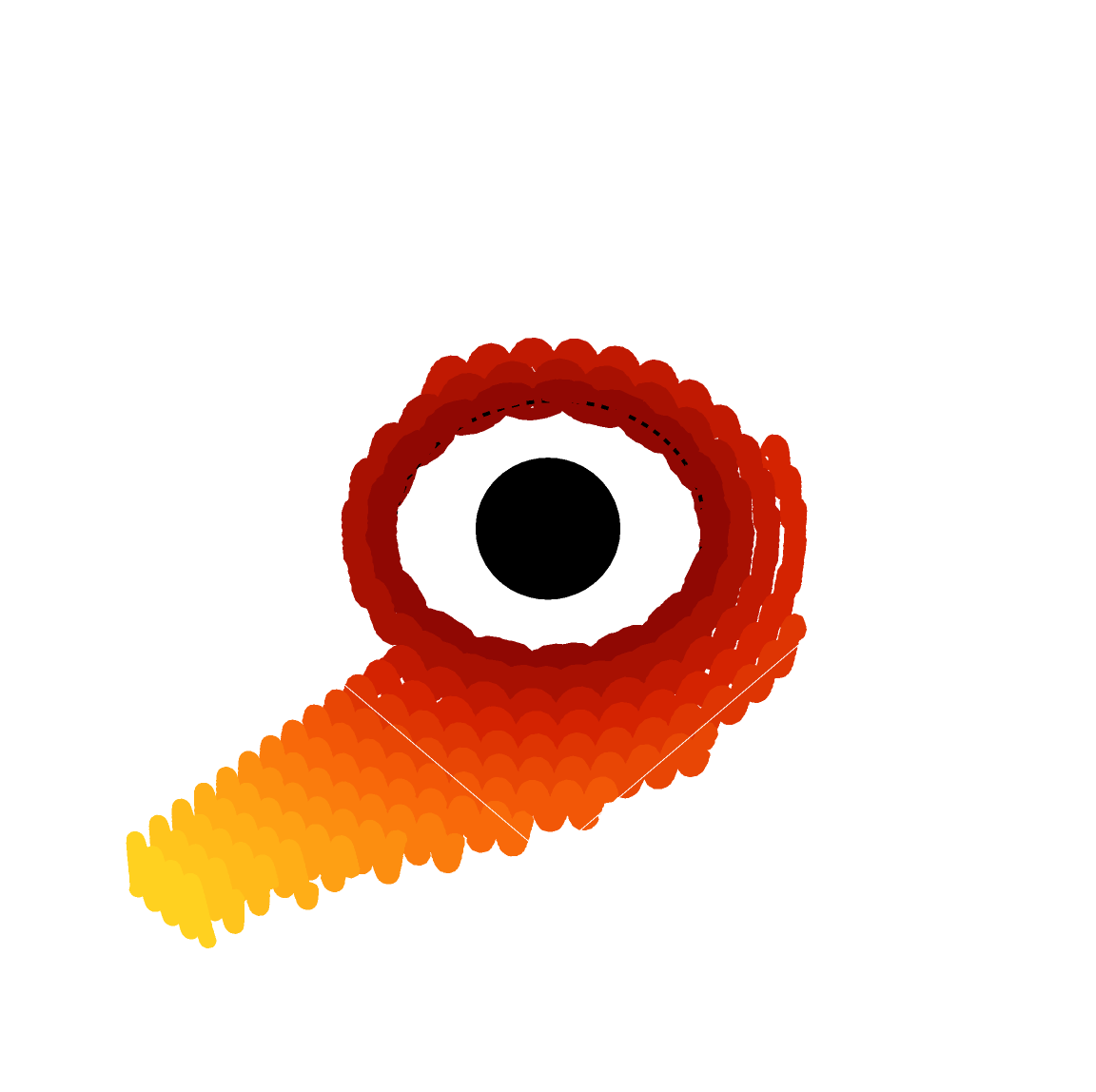}}} \\
        
        \adjustbox{valign=c}{\(\bm{\kappa}=20\)} &
        \adjustbox{valign=c}{\subfloat[]{\includegraphics[width=0.33\textwidth]{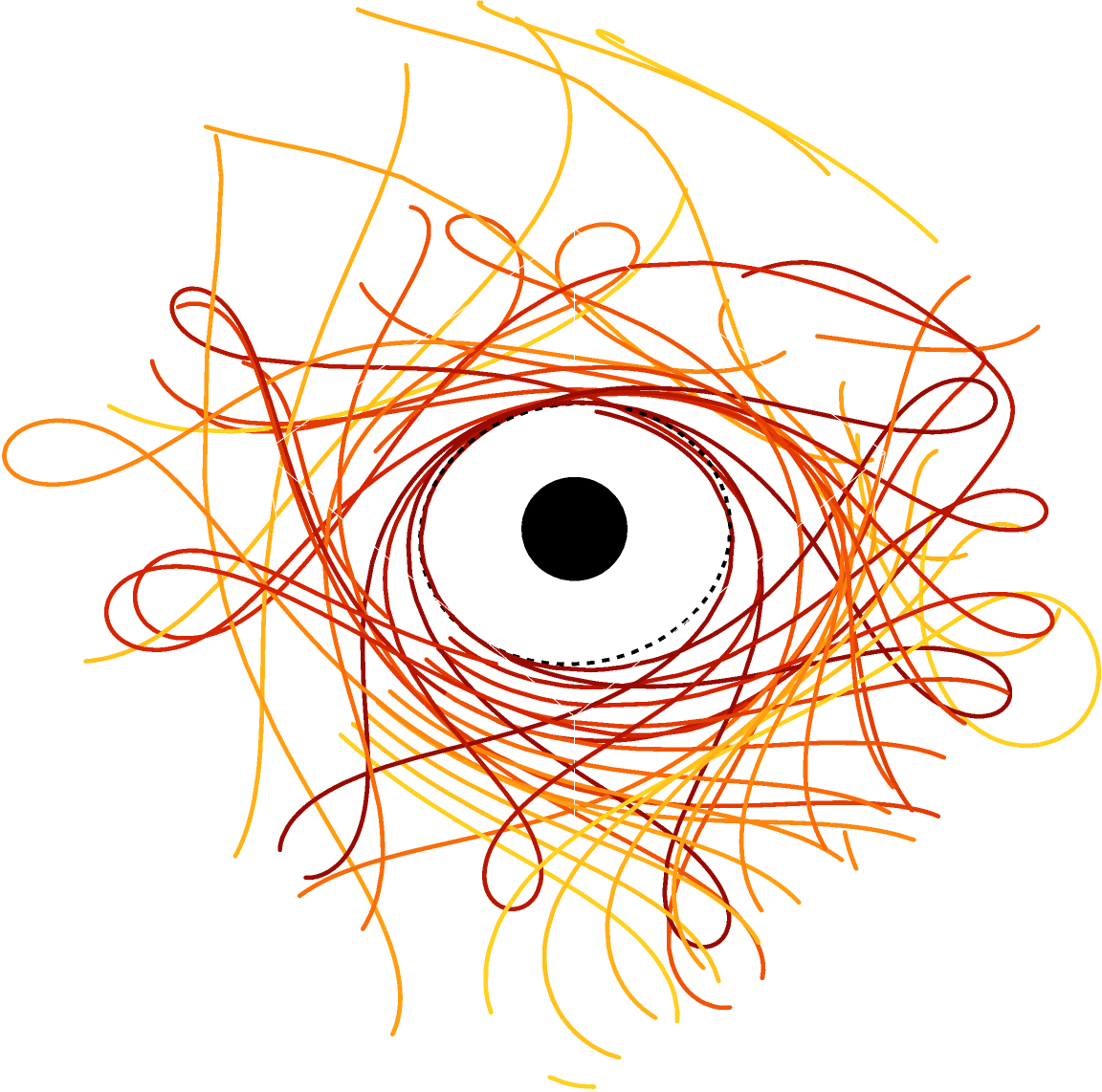}}} &
        \adjustbox{valign=c}{\subfloat[]{\includegraphics[width=0.33\textwidth]{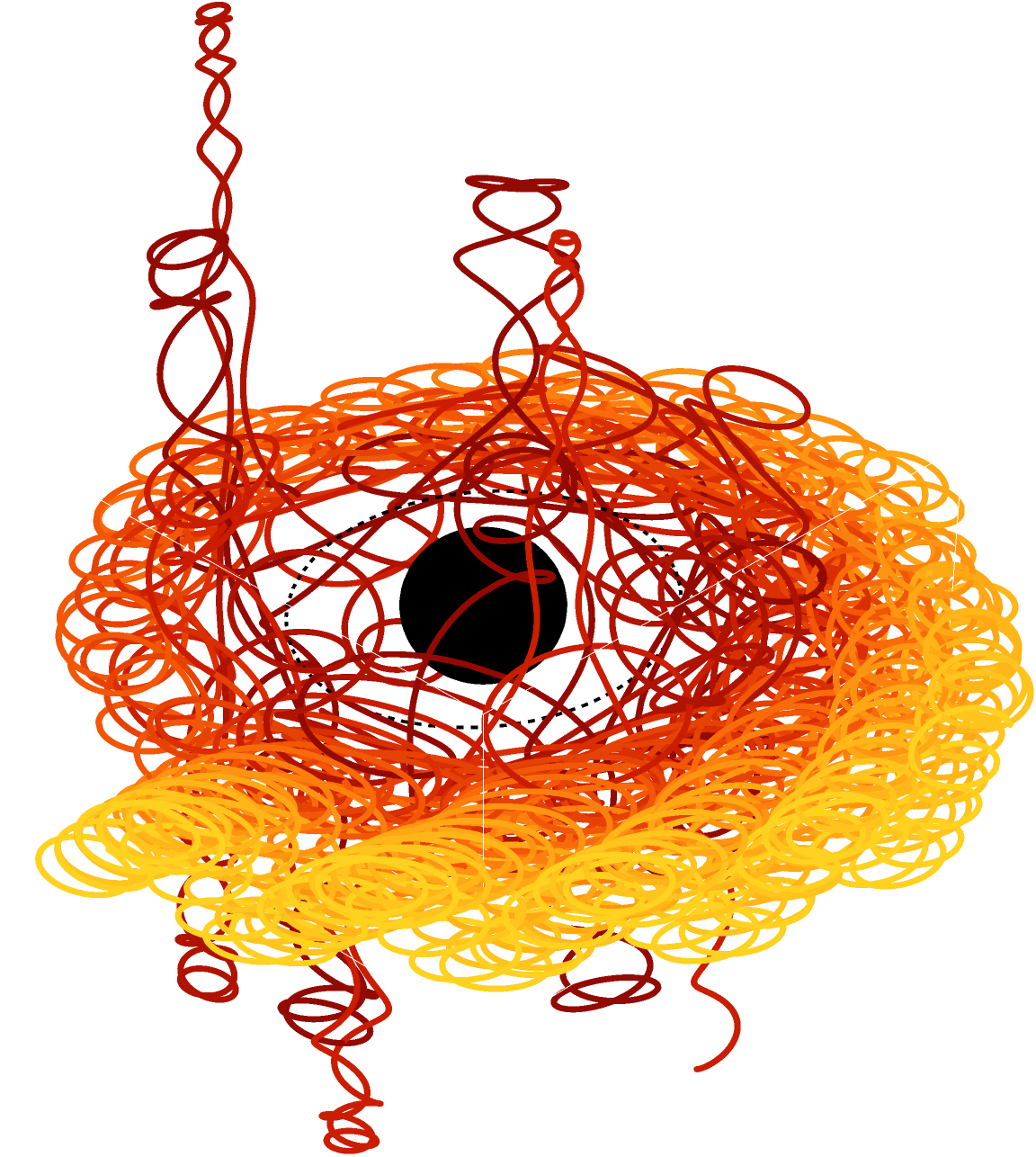}}} &
        \adjustbox{valign=c}{\subfloat[]{\includegraphics[width=0.33\textwidth]{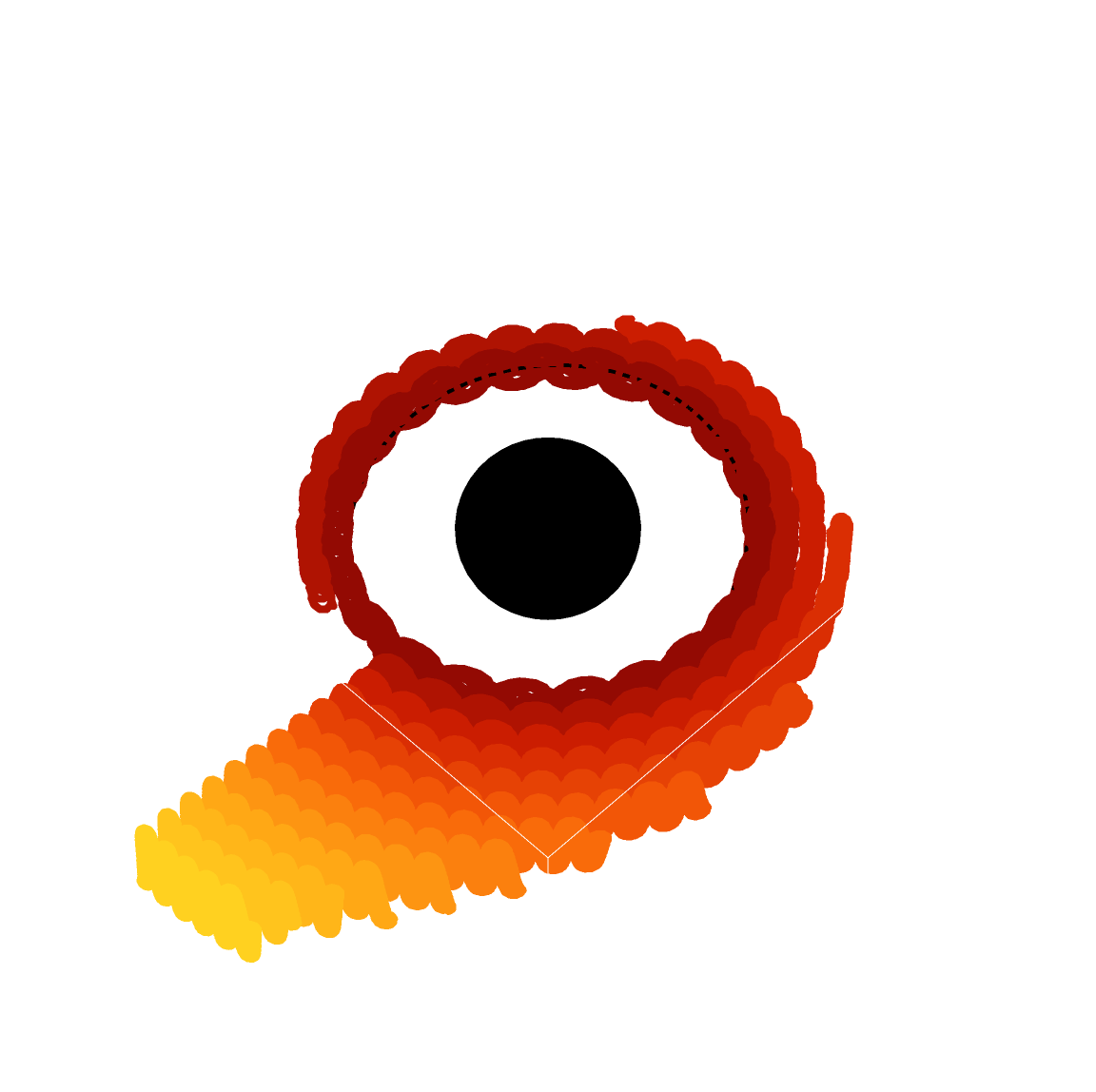}}} \\
    \end{tabular}

    \caption{\footnotesize{\it \textcolor{black}{Ionized Keplerian disk evolution under the influence of Uniform MF of strength ($B'=0.01,0.1,1$) represented from first column to the third column respectively, also by considering the presence of CDM density ($\bm{\kappa}=0,10,20$), from first line to third line respectively. The inclination of the accretion disk settled to $\theta_0 =1.51$ to the equatorial plane. The dashed black lines indicate the ISCO, while the central spherical region denotes the black hole event horizon.}}}
    \label{IKD repulsive}
\end{figure}

The Keplerian disk structure, induced by varying strengths of the uniform magnetic field and the surrounding CDM density, is illustrated in Fig.\ref{IKD repulsive}.
\textcolor{black}{In contrast to the attractive Lorentz force, the repulsive configuration ($B' > 0$) does not lead to the destruction of the Keplerian disk. For $B' = 0.01$, charged particles follow curled trajectories around the black hole, and as the CDM parameter $\bm{\kappa}$ increases, the degree of curvature in these orbits gradually decreases. Moreover, for $B' = 0.1$, the influence on the IKD resembles that of the $B' = -0.1$ case; however, unlike the attractive configuration, the Keplerian disk remains intact, and no disk destruction is observed. While in the final case of $B'=1$, the repulsive Lorentz force dominates across the entire range of CDM densities. Charged particles follow stable counter-clockwise oscillatory orbits around the BH. Unlike the $B'=-1$ configuration, the presence of CDM does not significantly affect the oscillatory motion of the particles.}

\textcolor{black}{This stability under strong repulsive fields naturally links to another key astrophysical phenomenon: the launching of relativistic jets. Jet particles (e.g., electrons, protons, and positrons) are a well-known feature correlated with the presence of strong $B'$ near BHs. These particles are typically ejected from the innermost regions of the accretion disk, composed of hot gas and dust spiraling around the BH. The mechanism driving their acceleration is primarily attributed to the intense repulsive Lorentz force acting on ionized particles, propelling them along magnetic field lines to relativistic speeds. Such outflows, or relativistic jets, provide crucial observational signatures of compact objects, offering insights into the structure of spacetime, the magnetic field configuration, and the energetic environment around the BH.
Beyond jet launching, magnetic fields can also induce complex and highly sensitive dynamical regimes in charged particle motion. In particular, chaotic behavior emerges for both attractive and repulsive Lorentz force configurations, typically in the range  $B'= 0.01$ to $B'= 0.1$  , and for $B'=-0.1$ with destruction of IKD in the inner edge of isco. In these cases, the destruction of the IKD may accompany the onset of chaos, which is manifested as particles oscillating irregularly up and down while still following the magnetic field lines. For stronger fields ($|B'|\geq 1$), the trajectories of charged particles are largely governed by the magnetic field itself, as discussed in~\cite{panis2019determination}. Interestingly, the transition to chaos is not determined by field strength alone but also depends on the initial inclination of the magnetic field with respect to the disk. As shown in~\cite{panis2019determination}, decreasing the inclination (from perpendicular to the disk down toward parallel alignment) progressively enhances chaotic behavior, which becomes most pronounced at intermediate values. At critical inclinations, particles are redirected inward, and in the limiting case of $\theta_0 \approx 0$ (magnetic field lines parallel to the disk), infall into the BH is strongly favored.
We now turn to the analysis of jet particle dynamics in the vicinity of a BH immersed in a strong uniform magnetic field and surrounded by a CDM halo. In this context, we consider an inclination angle $\theta_0 \approx 0.67$ (corresponding to $\pi/2 - 0.9$), which allows us to investigate chaotic particle trajectories aligned with the magnetic field lines in both upward and downward directions~\cite{panis2019determination}, under different CDM densities. Fig.\ref{jet particle K} illustrates the impact of a strong magnetic field ($B'=1$) on the trajectories of ionized particles for several discrete values of the CDM density parameter $\bm{\kappa}$. The scale radius $R_s$ is kept fixed throughout, ensuring that the variations observed are solely due to changes in CDM density.  
\begin{figure}[!ht]
    \centering
    \includegraphics[width=1\linewidth]{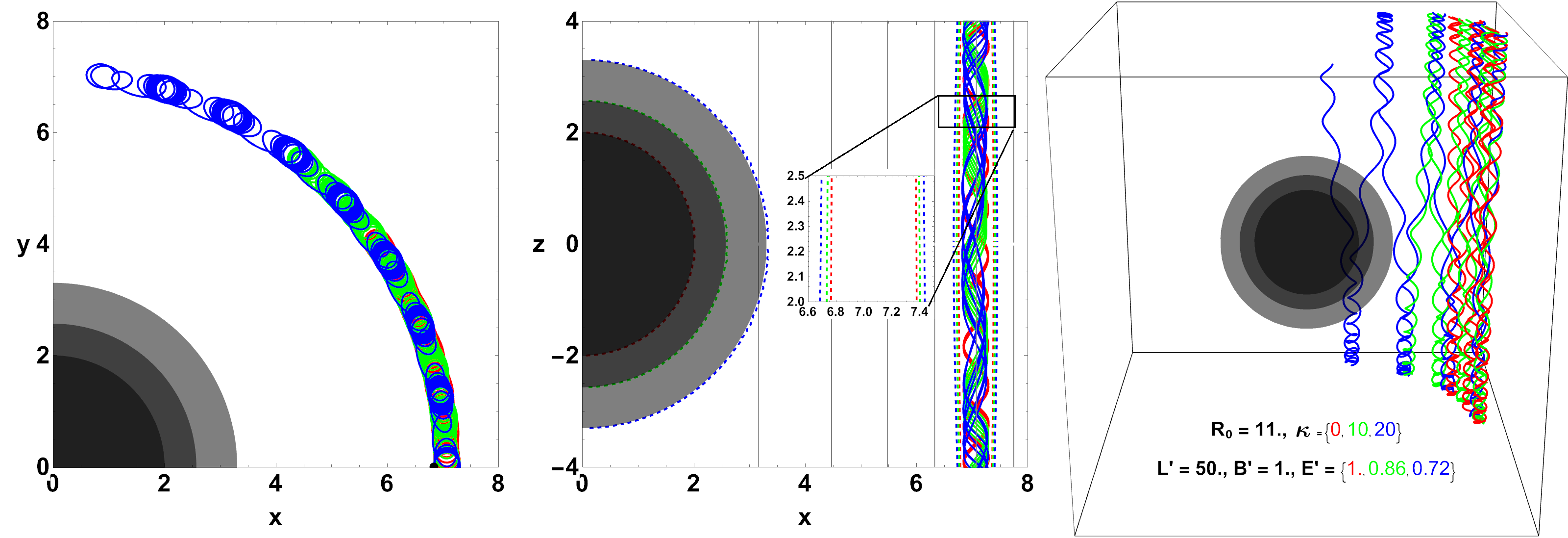}
    \caption{\footnotesize{\it 
    Trajectories of ionized charged particles for different values of the CDM density parameter $\bm{\kappa}$, around a BH immersed in a strong, positive uniform magnetic field of strength $B' = 1$ within the initial inclination $\theta_0=0.67$.
    }}
    \label{jet particle K}
\end{figure}
The leftmost panel displays the trajectories of charged particles near the BH for different CDM densities under a magnetic field strength of $B'=1$. The second column presents the corresponding specific energy $E'$ for each $\bm{\kappa}$ configuration, while the zoomed-in middle panels highlight the energetic boundaries of the particles, as determined by the effective potential (dashed lines). In both the left and middle columns, the central disks mark the positions of the BH horizons for the respective values of $\bm{\kappa}$. The rightmost column provides a $3D$ visualization of the system, offering clearer insight into the geometric configuration and the resulting particle dynamics.
\paragraph{} Our results indicate that increasing the CDM density parameter $\bm{\kappa}$ substantially modifies the structure of the confinement and acceleration zones, thereby reshaping particle trajectories and escape dynamics of jet-like particles. Specifically, most trajectories remain aligned with the $z$-axis. According to the escape condition defined in Eq.~\eqref{condition for positive B}, only the trajectory corresponding to $\bm{\kappa}=0$ (shown in red) carries sufficient energy to escape to infinity. As $\bm{\kappa}$ increases, the specific energy along particle trajectories decreases, demonstrating that the CDM halo exerts a dominant influence over the magnetic field in the near-horizon region. Moreover, the expansion of the energy boundaries with larger $\bm{\kappa}$ signals the emergence of a deeper potential well, resulting in stronger confinement of charged particles.}

\subsection{More on curled trajectories}
The conserved angular momentum associated with the axial symmetry of the charged particle's trajectory is expressed in the following expression
\begin{equation}
    \dot{\phi} = \frac{L'}{r^2} - B'.
\end{equation}
For $B'=0$, with $L'>0$ and $r>0$, the axial coordinate $\phi$ increases, indicating standard prograde motion. However, when $B'>0$, $\phi$ may decrease, implying a reversal in the azimuthal direction. This behavior results in curled or spirallike trajectories, as the Lorentz force modifies the particle's angular motion around the BH. 
The condition for the occurrence of curled trajectories can be derived from the angular momentum constraint, which must satisfy:
\begin{equation}
      L'(r;B')\geq B'r^2.
\end{equation}
This inequality leads to the corresponding condition on the specific energy required for such spiral motion, expressed as
\begin{equation}\label{open energy}
    E'(\bm{\kappa},R_s;B',L') > \sqrt{-\frac{2}{\sqrt{\frac{L'}{B'}}} + \left(1 + \frac{\sqrt{\frac{L'}{B'}}}{R_s}\right)^{-\frac{1}{\sqrt{\frac{L'}{B'}}}8\bm{\kappa}\pi}}.
 \end{equation}
Curled trajectories are clearly illustrated in Fig.\ref{figure 7}, where different initial radii are considered under the same angular momentum and magnetic field strength. As the DM parameter $\bm{\kappa}$ augmentation, the curvature of the trajectories becomes more significant, and the spatial extent of the curls gradually diminishes, indicating stronger gravitational confinement induced by the CDM halo.



The polarity of the Lorentz force plays a decisive role in shaping the coupling between the black hole and its accretion disk. By analyzing the dynamics of charged particles and their ionization pathways, one gains valuable insights into how CDM halos influence disk stability, angular momentum transport, and radiative efficiency. Such effects may leave imprints on observable signatures, including quasi-periodic oscillations (QPOs) and disk luminosity, thereby providing a potential avenue to probe the interplay between dark matter, magnetic fields, and black hole accretion.
\section{Harmonic oscillations of a charged particle in a magnetized BH  with CDM halos}\label{sec4}

A charged particle in equilibrium is sited at the ISCO, at a radius denoted by $r_{\text{ISCO}}$, immediately beyond the BH's event horizon. At this position, the particle reveals stable circular motion corresponding to the minimum of $V_\text{eff}$ in $\theta = \pi / 2$. When the particle experiences a very small displacement from this equilibrium point, it undergoes oscillatory motion around the potential minimum.

To describe such harmonic oscillations, we analyze small perturbations along both the radial and the latitudinal directions. These are represented by $r = r_{\text{ISCO}} + \delta$ and $\theta = \theta_{\text{ISCO}} + \delta$, corresponding respectively to horizontal (radial) and vertical (latitudinal) perturbations. The dynamics of these oscillations are regulated by the following equations
\begin{equation}
    \ddot{\delta}i + w_i^2\delta i = 0.
\end{equation}
Here, $i = (r, \theta)$, and the double dot denotes the second derivative with regard to the particle's proper time $\tau$. The quantities $\omega_r$ and $\omega_\theta$ represent the radial and latitudinal oscillation frequencies, respectively. In the context of a BH spacetime influenced by gravity, a magnetic field, and a surrounding DM halo, the corresponding frequencies can be computed using the following expression
\begin{align}
    &\omega_r^2 = \frac{\partial^2 V_{\text{eff}}}{\partial r^2}, \quad
&\omega_{\theta}^2 = \frac{1}{r^2 f(r)} \frac{\partial^2 V_{\text{eff}}}{\partial \theta^2}.
\end{align}
In addition, two other important frequencies arise: the Keplerian frequency $\omega_\phi$ and the Larmor frequency $\omega_L$ \cite{kolovs2015quasi}. The Keplerian frequency $w_\phi$ characterizes the angular (azimuthal) motion of the particle, while the Larmor frequency $w_L$ describes its circular motion induced by the uniform magnetic field. These frequencies are given by
\begin{align}
    &w_\phi=\frac{d\phi}{d\tau},&w_L=\frac{qB}{m}=2|B'|.
\end{align}
The Larmor frequency $w_L$ is unrelated to the radial coordinate, and its effect becomes increasingly dominant at extended range from the BH, where the uniform magnetic field plays a more significant role.

Finally, the radial, latitudinal, and Keplerian frequencies can be defined as follows \cite{kolovs2023charged}
\begin{eqnarray}\label{frequencies r}\nonumber
    w_r&=&\sqrt{\left(B'^2 + \frac{3L'^2}{r^4}\right) f(r) + \frac{\left(r^2 + (L' - B'r^2)^2\right) \left(f'(r)\right)^2}{r^2 f(r)} - \frac{\left(r^2 + (L' - B'r^2)^2\right) f''(r)}{2r^2}},\\
\\
\label{frequencies theta new}
\omega_\theta&=& \sqrt{\frac{2 L'^2}{r^4}-2 B'^2},
\\
\label{frequencies phi}
    \omega_\phi&=& \frac{L'}{r^2}-B'.
\end{eqnarray}

\begin{figure}[H]
        \centering
        \includegraphics[width=1\linewidth]{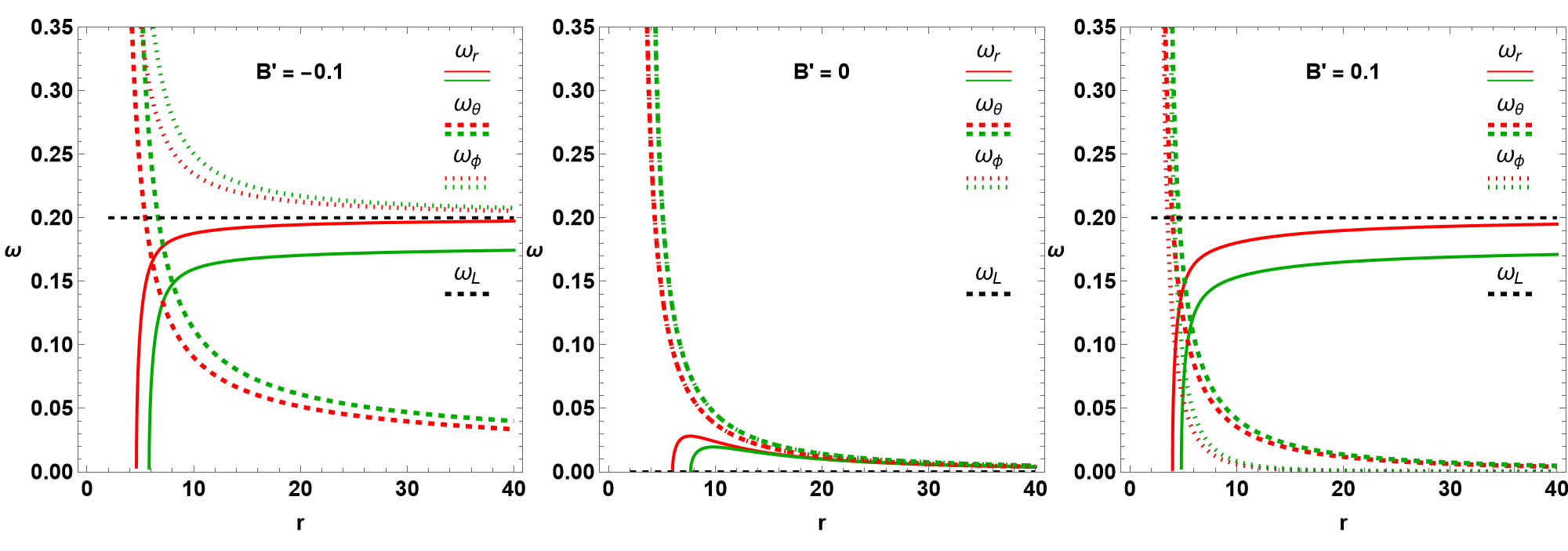}
       \caption{\footnotesize{\it Fundamental frequencies  $\omega_r$(Thick line), $\omega_\theta$(Dahsed line), $\omega_\phi$(Dotted line) and $\omega_L$(black dashed line), on function of radii where $\bm{\kappa}=0$ (Red)  $\bm{\kappa}=10$ (Green) within $R_s=1000$, for different values of $B'$. }}
       \label{figure oscillation}
    \end{figure}

In all expressions for the characteristic frequencies, $L'$ is set to $L'_+$, as defined in Eq.\eqref{extrema function}. Fig.\ref{figure oscillation} illustrates the behavior of the fundamental frequencies-radial ($\omega_r$), latitudinal ($\omega_\theta$), Keplerian ($\omega_\phi$), and Larmor ($\omega_L$) as functions of $r$.
 The ratios between these frequencies provide important insights into the dynamics and geometry of charged particle motion around a BH immersed in a uniform magnetic field and surrounded by a DM halo. In particular, these ratios can be used to estimate the shape and nature of the particle trajectories.
 
The comparison of these fundamental frequencies is presented in this figure for three distinct cases: $B' = 0$, and in the presence of both positive and negative magnetic fields ($B' = \pm 0.1$). Additionally, the figure explores how the charged particle orbits behave for a given CDM density parameter $\bm{\kappa}$.   
As previously discussed, gravitational effects dominate near the BH (at small radii), whereas the magnetic field becomes more influential at larger distances. This behavior is also reflected in the frequency profiles. 
The Red thick and dashed lines represent the frequencies in the presence of the magnetic field alone (i.e., $\bm{\kappa} = 0$), consistent with results in \cite{kolovs2015quasi}. In contrast, our case with $\bm{\kappa} > 0$ is illustrated by Green  (lines, dashed, dotted), showing the modifications induced by the dark-matter halo. 

According to Newtonian gravity, all characteristic frequencies are identical ($\omega_r = \omega_\theta = \omega_\phi$), which corresponds to charged particles following elliptical trajectories. This Newtonian behavior is plotted in the leftmost panel of Fig.\ref{figure oscillation}. 
The trajectory of charged particles is notably affected by the presence of a uniform magnetic field, regardless of its sign. For $B'>0$, the Keplerian frequency $\omega_\phi$ approaches the Larmor frequency $\omega_L$ (represented by the dashed black line in Fig.\ref{figure oscillation}) from above at large distances. In contrast, when $B'<0$, the frequencies $\omega_\phi$ and $\omega_\theta$ become nearly identical at large distances from the ISCO radius (where $\omega_r = 0$), and both asymptotically approach zero. 
The influence of the CDM halo is most significant in the inner region near $r_{\text{ISCO}}$, where it affects both $\omega_\theta$ and $\omega_\phi$. However, at large distances, the effect diminishes, and the frequency profiles converge, showing little difference between cases $\bm{\kappa} = 0$ and $\bm{\kappa} = 10$. \textcolor{black}{It should be noted that the equatorial motion remains stable, since Eqs.\eqref{frequencies r} and \eqref{frequencies theta new} stay positive for all considered values of $\bm{\kappa}$ and asymptotically approach zero, as illustrated in Fig.\ref{figure oscillation}.}

The fundamental frequencies can be translated into the corresponding frequencies as measured by distant observers located at spatial infinity. These observed frequencies are given by
\begin{equation}
 \nu_i = \frac{c^3}{2 \pi G M} \frac{\omega_i}{-g^{tt} E'} 
\end{equation}
where $i = \{r, \theta, \phi\}$, $G$ and $c$ denote the gravitational constant and the speed of light, respectively. The factor $(-g^{tt} E')^{-1}$ accounts for the gravitational redshift experienced by the particle.

\begin{figure}[!ht]
    \centering
    \begin{subfigure}[b]{\textwidth}
        \begin{minipage}{0.32\textwidth}
        \centering
        \includegraphics[width=1\linewidth]{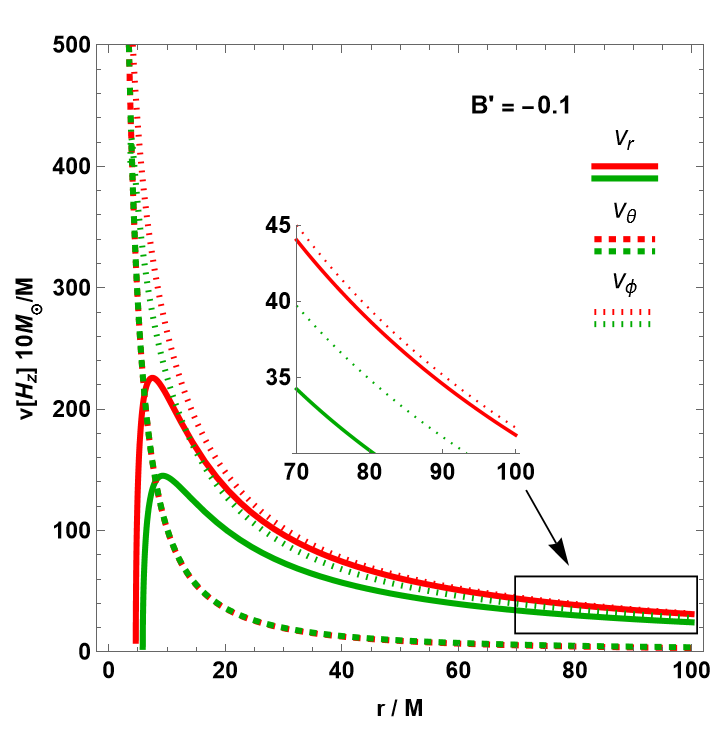}
    \end{minipage}
    \begin{minipage}{0.32\textwidth}
        \centering
        \includegraphics[width=1\linewidth]{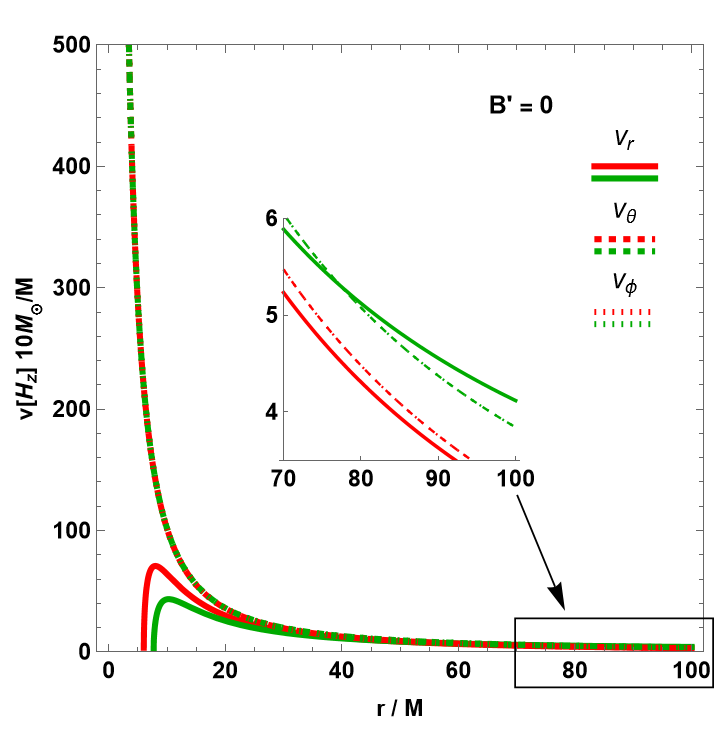}
    \end{minipage}
    \begin{minipage}{0.32\textwidth}
        \centering
        \includegraphics[width=1\linewidth]{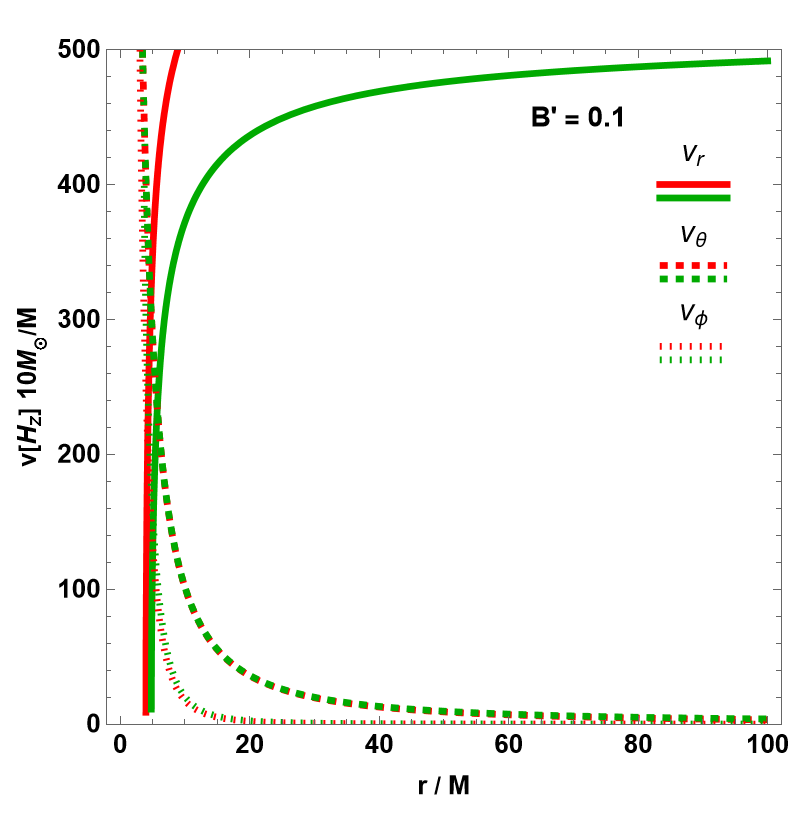}
    \end{minipage}
    \end{subfigure}
    \caption{\footnotesize{\it Fundamental frequencies $\nu_r(\text{Thick}), \nu_\theta (\text{Dashed}), \nu_\phi(\text{Dotted})$
    as measured by distant observers located at spatial infinity
     for different values of $\bm{\kappa}$, $0$(Red), $10$ (Green), for different values of $B'$.}  }
    \label{frequencies mesured at a distant}
\end{figure}

In Fig.\ref{frequencies mesured at a distant} we depict the outcome of the CDM parameter—density $\bm{\kappa}$ on the fundamental frequencies (radial, latitudinal, and Keplerian) of a charged particle orbiting a BH situated within a uniform magnetic field and DM halo as measured by distant observers located at spatial infinity. The behavior is shown for three different magnetic field strengths: $B' = 0, \pm 0.1$.

Green lines correspond to the case $\bm{\kappa} = 10$, while red lines represent the scenario with no DM contribution, i.e., $\bm{\kappa} = 0$. 
The presence of CDM leads to a reduction in the charged particle motion frequencies across all values of $B'$. In the case of $B'<0$, and for $\bm{\kappa} = 0$, the radial frequency $\nu_r$ and the Keplerian frequency $\nu_\phi$ converge at large distances. As $\bm{\kappa}$ increases, the gap between $\nu_r$ and $\nu_\phi$ becomes more pronounced. In addition, for $B' = 0$, and for $\bm{\kappa} = 10$, we observe an intersection between $\nu_r$ and $\nu_\phi$ at large distances, where the radial frequency lies above the Keplerian frequency.

\subsection{{Power spectral densities of ionized particles and jet particles}}
\begin{figure}[!ht]
    \centering
    \begin{subfigure}[b]{\textwidth}
    \begin{minipage}{0.32\textwidth}
        \includegraphics[width=1.1\textwidth]{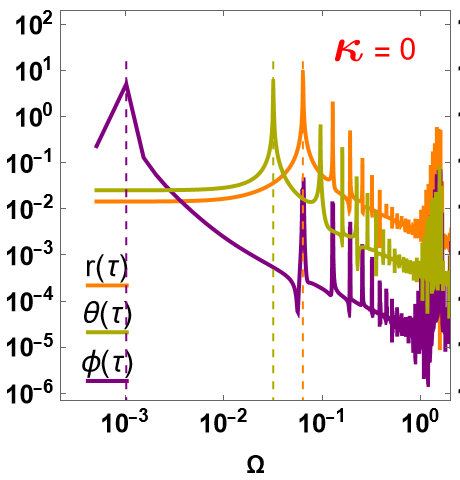}
        \caption{}
    \end{minipage}
        \begin{minipage}{0.32\textwidth}
            \includegraphics[width=1.1\textwidth]{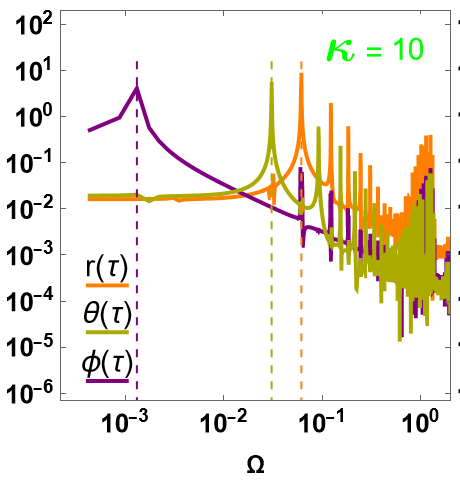}
        \caption{}
        \end{minipage}
        \begin{minipage}{0.32\textwidth}
            \includegraphics[width=1.1\textwidth]{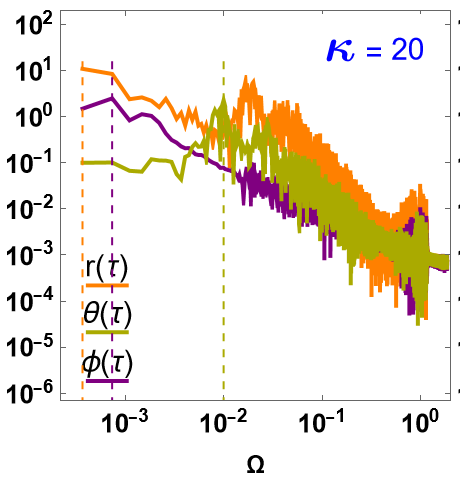}
        \caption{}
        \end{minipage}
    \end{subfigure}
    \caption{\footnotesize{\it Power spectral density of trajectories coordinates of charged particles $r(\tau)$, $\theta(\tau)$, $\phi(\tau)$, with correspond of trajectories plotted in Fig.\ref{jet particle K}. Peaks of the fundamental frequencies of particles' motion near the circular orbits are represented as dashed lines.}}
    \label{power spectra K}
\end{figure}

Throughout Fig.\ref{power spectra K}, we plot the power spectral densities (PSD) corresponding to the radial, latitudinal, and Keplerian coordinates. These spectra are derived from the trajectories of particles with charge orbiting a BH immersed in a strong uniform magnetic distribution and surrounded by a CDM halo \textcolor{black}{illustrated in Fig.\ref{jet particle K}}. The dashed lines refer to the peak positions of the fundamental frequencies of particle motion near circular orbits, as computed from Eqs.\eqref{frequencies r}–\eqref{frequencies phi}.%
The panel (a) corresponds to a particle with $E' \geq 1$, escaping to infinity, consistent with Eq.\eqref{condition for positive B}. As the CDM density parameter $\bm{\kappa}$ increases (panels b–c), the PSD structure changes significantly. For high $\bm{\kappa}$ (panel c), the radial frequency dominates while latitudinal and Keplerian components are suppressed, indicating stronger confinement and increased dynamical complexity. 

Furthermore, in contrast, for the case $\bm{\kappa} = 0$, the Keplerian component dominates at low frequencies, with radial and latitudinal components taking over at higher frequencies, especially the radial component. As $\bm{\kappa}$ increases, the frequency peaks of all trajectory coordinates tend to converge toward higher frequencies, indicating the onset of chaotic dynamics dominated by strong radial oscillations. 
This progression reveals how increasing $\bm{\kappa}$ enhances the chaotic character of particle motion. As seen in Fig.~\ref{jet particle K}, this behavior is associated with particles having specific energies below 1, which remain gravitationally bound and orbit close to the BH. Their corresponding PSDs exhibit more complex, broadband features, reflecting the increased dynamical instability introduced by the CDM halo.

By studying the fundamental frequencies—radial, latitudinal, and Keplerian—we have shown how both the magnetic field and CDM environment significantly impact the stability and structure of particle motion. In particular, the CDM halo modifies the confinement regions and enhances the chaotic behavior of bound orbits, especially as the density parameter $\bm{\kappa}$ increases. Building on this analysis, we now turn our attention to the concept of \textit{resonance radii}, which play a critical role in understanding QPOs. These special radii correspond to locations where characteristic frequency ratios (such as 3:2) occur, leading to resonant amplification of oscillatory modes in the accretion flow around compact objects.

\subsection{Resonance Radii}
HF QPOs are among the most intriguing phenomena observed in X-ray binaries (XBs), particularly in low-mass X-ray binaries (LMXBs) hosting stellar-mass BHs or neutron stars. Detected by X-ray observatories such as RXTE, NICER, and XMM-Newton, HF QPOs typically manifest as twin peaks in the Fourier power spectrum, corresponding to an upper and a lower frequency, denoted by $f_U$ and $f_L$, respectively \cite{Remillard:2006fc, van_der_Klis2006}. In microquasars, these twin peaks often appear in a near 3:2 frequency ratio, a feature that has been interpreted as a signature of resonance phenomena occurring in the inner regions of the accretion disk \cite{Abramowicz:2001bi, torok2005orbital, Motta:2013wwa}.

Generally, HF QPOs are understood to be accretion disk phenomena, closely linked to the fundamental frequencies of particle motion in the strong-field regime of gravity. They provide a powerful diagnostic tool for probing spacetime geometry in the proximity compact objects. Within the framework of our study, considering a BH immersed in a uniform magnetic field and enclosed by a CDM halo, this combination offers a rich and promising context for investigating the dynamical origin of HF QPOs. In particular, it allows us to explore how modifications to the effective potential, caused by magnetic and dark-matter interactions, can affect particle oscillations near the ISCO, where HF QPOs are believed to originate.

The presence of a CDM halo, in addition to altering the gravitational potential, can shift the ISCO radius and perturb the epicyclic frequencies of particles in the accretion disk \cite{Xu_2018, narzilloev2020dynamics}. These changes could have observable consequences for the frequencies of HF QPOs and offer a potential justification for deviations from the standard 3:2 ratio in some systems. Thus, incorporating both magnetic field effects and CDM contributions enhances our understanding of the complex physics governing QPO generation in LMXBs.

The observed QPOs in XBs are associated with fundamental frequencies of particle motion in the inner accretion disk: the radial frequency $v_r$, the latitudinal (vertical) frequency $v_\theta$, and the Keplerian (orbital) frequency $v_\phi$. Multiple theoretical models have been suggested to explain the origin of these QPOs, each characterized by different combinations of these frequencies. A detailed discussion of these models is provided below.

The model of {\it epicyclic resonance} (ER), originally proposed by Abramowicz and Kluzniak \cite{Abramowicz:2001bi}, interprets HF QPOs as a result of non-linear resonances between radial and vertical epicyclic oscillations of matter in the inner regions of the accretion disk around a compact object. In this framework, small perturbations around nearly circular and equatorial geodesics in a relativistic potential can lead to oscillations characterized by epicyclic frequencies.  When these frequencies enter into a rational ratio (most notably 3:2), nonlinear resonances can be triggered. The observed twin peaks of the HF QPOs are then interpreted as arising from these resonances. Several variants of the ER model exist, depending on the identification of the upper and lower frequencies $f_U$ and $f_L$ with combinations of $v_r$ and $v_\theta$, as summarized in Tab.\ref{tab:qpo_models}.

One widely studied model is the {\it relativistic precession} (RP) model, which attributes QPOs to relativistic corrections in particle orbits near the BH. This model does not rely on strong magnetic fields or resonance radii. Instead, it considers blobs of plasma within the accretion disk following slightly eccentric and tilted geodesics. In this framework, QPOs arise from small perturbations in the motion of these blobs, governed by the spacetime's intrinsic frequencies. In such an RP model, the upper HF QPO frequency is identified via the Keplerian frequency in the innermost regions of the disk: $f_U = v_\phi$. The lower HF QPO, $f_L$, is associated with the periastron precession of slightly eccentric orbits, defined as $f_L = v_\phi - v_r$, where $v_r$ is the radial epicyclic frequency. Since $v_r < v_\phi$, this difference captures the orbital precession.
Additionally, horizontal branch oscillations (HBOs) are attributed to nodal precession, described by $v_\phi - v_\theta$, which reflects vertical oscillations due to frame dragging in rotating BH configurations. However, in our case, the frame dragging is absent. This implies $v_\theta = v_\phi$. \textcolor{black}{However, assuming a uniform magnetic field, the equality condition $v_\theta = v_\phi$ is no longer valid, and we must instead consider $v_\theta \neq v_\phi$.}
As a result, the frequency of periastron precession $f_L = v_\phi - v_r$ becomes the dominant contributor to the lower QPO of the HF over a wide range of parameters \cite{stella1999correlations,hazarika2025signatures}.

Another relevant framework is the {\it Warped Disk} (WD) model, which addresses QPO generation in geometrically warped accretion disks. Although most accretion disk models assume equatorial symmetry, both theoretical and observational evidence support the existence of warped disks around stellar-mass and supermassive BHs \cite{tremaine2014dynamics}. The WD model involves resonant interactions between radial and vertical motions, incorporating both gravity (g-mode) and pressure (p-mode) oscillations. Horizontal resonances appear in the g- and p-modes, while vertical resonances are exclusive to the g-mode oscillations \cite{hazarika2025signatures}.

In summary, both the RP and WD models interpret QPOs as a result of oscillatory behavior within the accretion disk. Concerning our study, featuring a BH embedded in a magnetic and dark-matter environment, these frequency relations and their perturbations are crucial for identifying possible resonance conditions, especially in relation to the observed 3:2 HF QPO frequency ratio.

Lastly, the {\it tidal disruption} (TD) model interprets HF QPOs as resulting from the tidal disruption of clumps of matter (blobs or inhomogeneities) in the accretion disk by the gravitational field of the BH \cite{Germana:2009ce,Kostic:2009hp}. In this scenario, the orbital motion and deformation of these clumps generate periodic modulations in the X-ray flux. Concerning $f_U$ and $f_L$ of such a model, the observed modulation is produced by a combination of $v_\phi$ and $v_r$, which reflects the rate of deformation or compression due to the tidal force. 
This model has the advantage of offering a more physically intuitive mechanism: actual matter structures are being distorted by gravity and magnetic effects, causing quasi-periodic emission patterns. The model is particularly useful in systems where strong radial motion is expected, such as near the innermost regions of thick accretion disks or in the presence of non-axisymmetric perturbations.
\begin{table}[!ht]
\centering
\begin{tabular}{|c|l|c|c|}
\hline
\textbf{Model} & \textbf{Label} & \textbf{Upper Frequency $f_U(r)$} & \textbf{Lower Frequency $f_L(r)$} \\
\hline
\multirow{6}{*}{{\it Epicyclic Resonance} (ER)} 
& ER0 & $v_\theta$ & $v_r$ \\
& ER1 & $v_\theta$ & $v_\theta - v_r$ \\
& ER2 & $v_\theta - v_r$ & $v_r$ \\
& ER3 & $v_\theta + v_r$ & $v_\theta$ \\
& ER4 & $v_\theta + v_r$ & $v_\theta - v_r$ \\
& ER5 & $v_r$ & $v_\theta - v_r$ \\
\hline
\multirow{3}{*}{{\it Relativistic Precession} (RP)} 
& RP0 & $v_\phi$ & $v_\phi - v_r$ \\
& RP1 & $v_\theta$ & $v_\phi - v_r$ \\
& RP2 & $v_\phi$ & $v_\theta - v_r$ \\
\hline
{\it Tidal Disruption} (TD) & TD & $v_\phi + v_r$ & $v_\phi$ \\
\hline
{\it Warped Disk} (WD) & WD & $2 v_\phi - v_r$ & $2(v_\phi - v_r)$ \\
\hline
\end{tabular}
\caption{{\it  Frequency identification schemes for various QPO models. Each model defines the upper and lower observed frequencies $f_U(r)$ and $f_L(r)$ in terms of fundamental frequencies of particle motion}\cite{stuchlik2016models,shahzadi2021epicyclic}.}
\label{tab:qpo_models}
\end{table}

\begin{figure}[H]
    \centering
    \subfloat[]{\includegraphics[width=0.33\textwidth]{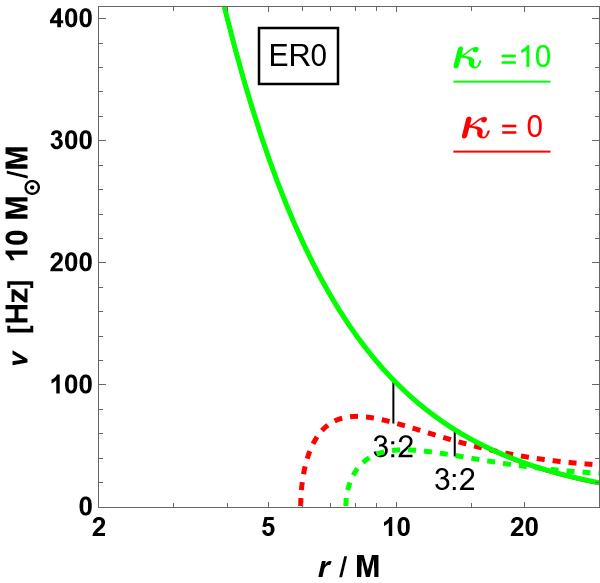}}
    \subfloat[]{\includegraphics[width=0.33\textwidth]{fig15a.png}}
    \subfloat[]{\includegraphics[width=0.33\textwidth]{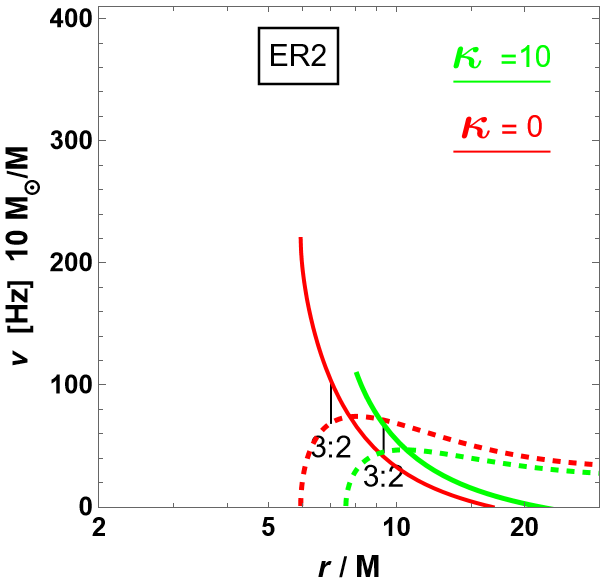}} \\
    \subfloat[]{\includegraphics[width=0.33\textwidth]{fig15a.png}}
    \subfloat[]{\includegraphics[width=0.33\textwidth]{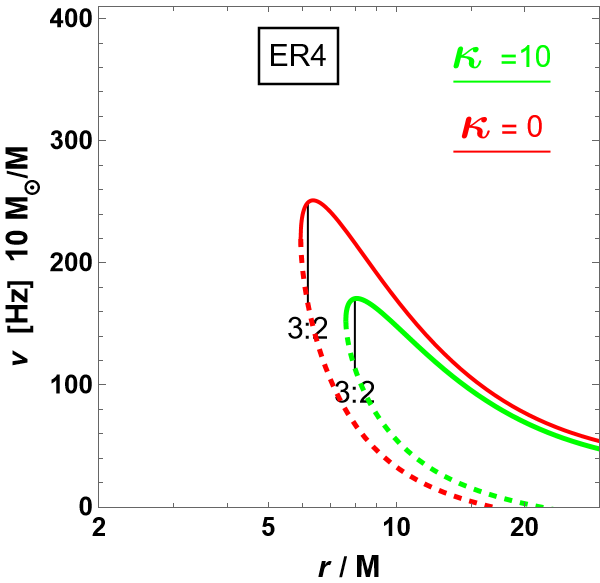}}
    \subfloat[]{\includegraphics[width=0.33\textwidth]{fig15a.png}} \\
    \subfloat[]{\includegraphics[width=0.33\textwidth]{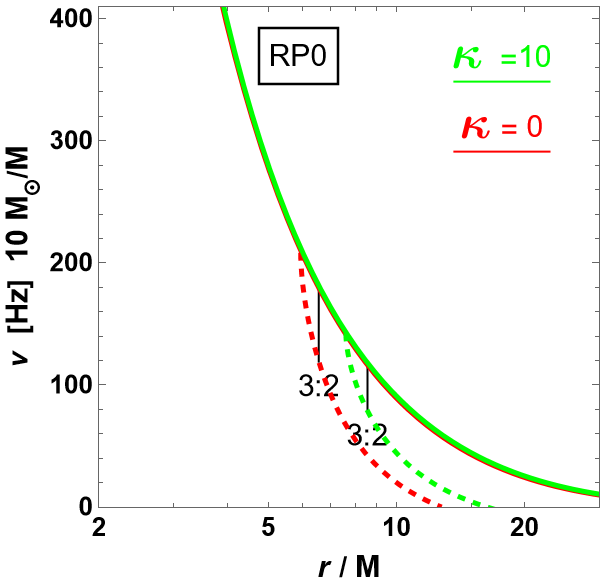}}
    \subfloat[]{\includegraphics[width=0.33\textwidth]{fig15a.png}}
    \subfloat[]{\includegraphics[width=0.33\textwidth]{fig15a.png}} \\
    \subfloat[]{\includegraphics[width=0.33\textwidth]{fig15a.png}}
    \subfloat[]{\includegraphics[width=0.33\textwidth]{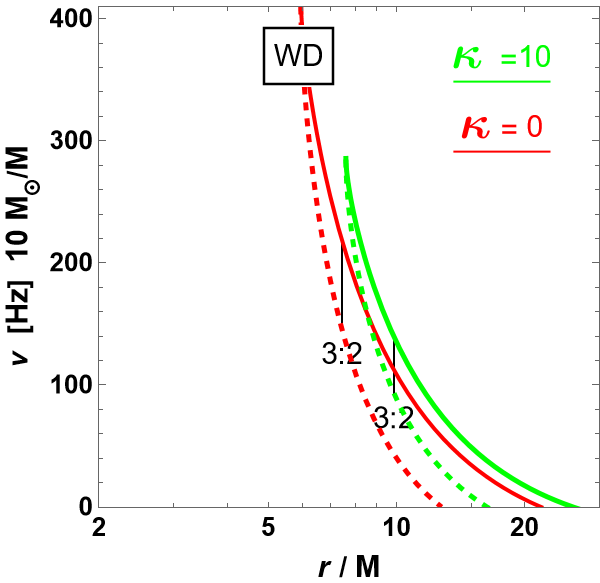}}
    \caption{\footnotesize{\it Radial profiles of lower $\nu_L(r)$ and upper $\nu_U(r)$ frequencies for several HF QPOs models, whit different values of density profile $\bm{\kappa}$ within $R_s=1000$ on week magnetic field of strength $B'=0.01$ in all cases, The Displacement $f_U(r):f_L(r)=$ 3:2 of resonance radii also plotted}}
    \label{resonance K}
\end{figure}

The twin peaks of $f_U$:$f_L$ of HF QPOs are in function of the BH mass $M$, $B'$ strength, and the parameter that controls the DM halo $\bm{\kappa}$. Fig.\ref{resonance K} represents the radial profiles of $f_U(r)$ and $f_L(r)$ for various HF QPOs models such as epical resonance ER, RP, TD, and WD.
From Fig.\ref{resonance K}, one can notice that the locations of the 3:2 resonance radii vary significantly across different models. Among them, the ER0 model exhibits the largest resonance radius, located at $r = 9.83589$ for $\bm{\kappa} = 0$, and shifting outward to $r = 13.7114$ for $\bm{\kappa} = 10$. In contrast, the ER4 model yields the smallest resonance radius, with $r =6.19465$ for $\bm{\kappa} = 0$, and $r = 7.98434$ when $\bm{\kappa} = 10$. 
For more details, \textcolor{black}{We provide Tabs.\ref{tab:resonance radii} and \ref{tab:resonance radii nega}, which list the resonance radii for the different models, evaluated at various values of the CDM density with $B'=0.01$ and $B'=-0.01$, respectively. These tables highlight how both the polarity of the magnetic field and the CDM density parameter $\bm{\kappa}$ influence the location of the resonance radii.}

\begin{table}[H]
\centering
\begin{tabular}{| l S S S S S|}
\hline
\textbf{Model} & {\boldmath$\kappa = 0$} & {\boldmath$\kappa = 5$} & {\boldmath$\kappa = 10$} & {\boldmath$\kappa = 15$} & {\boldmath$\kappa = 20$} \\
\hline
ER0 & 9.83589 &11.6557 & 13.7114  & 15.9826 & 18.4565 \\
ER1 & 6.66194 & 7.6306& 8.74095 &10.0102 & 11.4554  \\
ER2 & 7.0205 & 8.09117 & 9.32746  & 10.7485  &  12.3714 \\
ER3 & 7.78482 & 9.08374 & 10.6015 &12.3589 & 14.3709 \\
ER4 & 6.19465 & 7.03426  & 7.98434  &9.05739  & 10.2665 \\
ER5 & 8.86429 & 10.4525 & 12.2827  &14.3491 & 16.6405 \\
RP0 & 6.57077&7.49789  & 8.54967   &9.73796 & 11.0739 \\
RP1 & 6.42045 &  7.28662 &8.25678  & 9.33831& 10.5381 \\
RP2 & 6.89008& 7.98011&9.27671  & 10.8297 & 12.7011 \\
TD  & 7.44865& 8.58149 & 9.86633  &11.3099  & 12.9161 \\
WD  & 7.44865& 8.58149 & 9.86633   & 11.3099 & 12.9161  \\
\hline
\end{tabular}
\caption{{\it Resonance Radius Variation Across Different Models, Each model defines the location of the resonance radius for both cases: without DM ($\bm\kappa = 0$) and with DM ($\bm\kappa = 5, 10, 15, 20$) within $B'=0.01$.}}
\label{tab:resonance radii}
\end{table}
\begin{table}[H]
\centering
\begin{tabular}{| l S S S S S|}
\hline
\textbf{Model} & {\boldmath$\kappa = 0$} & {\boldmath$\kappa = 5$} & {\boldmath$\kappa = 10$} & {\boldmath$\kappa = 15$} & {\boldmath$\kappa = 20$} \\
\hline
ER0 & 9.88508 &11.7483 & 13.8734   & 16.2467  & 18.8619  \\
ER1 & 6.6675   & 7.6402& 8.75741  & 10.0381  & 11.5022   \\
ER2 & 7.02814 & 8.10476 & 9.35143  & 10.7902   &  12.4424 \\
ER3 & 7.77096  & 9.0579 & 10.5543 &12.2754  & 14.2288 \\
ER4 & 6.1982  & 7.0401  & 7.99388  &9.07288   & 10.2914  \\
ER5 & 8.89306   & 10.5075 & 12.3828   &14.5207 & 16.9179  \\
RP0 & 6.777 & 7.80631  & 9.00915   & 10.4185 & 12.0738 \\
RP1 & 7.06335  &  8.27752 &9.81008   & 11.8534& 14.9408 \\
RP2 & 6.49154  & 7.38565& 8.39284  & 9.52241 & 10.7831  \\
TD  & 8.27355& 9.89312 & 11.9388  &14.5394   & 17.8355 \\
WD  & 8.27355& 9.89312 & 11.9388    & 14.5394  & 17.8355  \\
\hline
\end{tabular}
\caption{{\it Resonance Radius Variation Across Different Models, Each model defines the location of the resonance radius for both cases: without DM ($\bm\kappa = 0$) and with DM ($\bm\kappa = 5, 10, 15, 20$) within $B'=-0.01$.}}
\label{tab:resonance radii nega}
\end{table}

These results demonstrate that the presence of CDM leads to a systematic outward shift of the resonance radii, reflecting the influence of the halo's density on the gravitational potential near the BH. \textcolor{black}{The resonance radii are strongly influenced by the polarity of the Lorentz force. For $B'<0$, the radii are shifted outward more prominently than in the case $B'>0$, across all models except ER3 and RP2, where this behavior remains consistent for all values of the CDM density parameter $\bm{\kappa}$. In addition, the TD and WD models yield identical resonance radii for every considered value of $\bm{\kappa}$, independent of the magnetic field polarity ($B'=\pm0.01$).}
 
In \cite{shahzadi2024testing}, the authors investigated the resonance radii corresponding to the 3:2 frequency ratio while incorporating the effects of CDM in rotating black hole spacetimes. Their results demonstrated that the rotation of a Kerr black hole shifts the resonance radii inward, drawing them closer to the event horizon. In contrast, our study, which focuses on a non-rotating black hole embedded in a CDM halo,  shows the opposite trend: increasing the CDM density pushes the resonance radii outward, moving them farther away from the black hole.

Next, to explore some astrophysical implications, it's essential to assess whether the models examined can reproduce the HF QPOs observed in real astrophysical systems. 

\section{On some astrophysical estimations of  HF QPOs }\label{sec5}

\paragraph{} In this section, we perform a comparative examination between theoretical model predictions and observational data from well-known microquasars \textcolor{black}{and supermassive BHs in active galactic nuclei}, serving as a testbed to assess the viability of our framework in the presence of a CDM halo.

One of the notable effects attributed to CDM is the enhancement of a BH's gravitational potential, which may contribute to explaining HF QPOs observed in accreting systems \cite{torok2005orbital,shahzadi2021epicyclic}. To investigate this, we fit the HF QPO data of four microquasars—GRS 1915+105, H1743-322, XTE 1550-564, and GRO 1655-40—summarized in Table~\ref{astrodata}, \textcolor{black}{and in several supermassive BHs in AGNs represented in Table~\ref{astrodata AGNs}}, under the assumption of non-rotating BHs \cite{kolovs2015quasi,stuchlik2022geodesic}. \textcolor{black}{The magnetic field strengths considered in this study are chosen to reflect realistic astrophysical conditions. For electrons (protons), the required values are $B_{e^-}\approx 0.1,\mathrm{mGs}$ ($B_{p^+}\approx 0.2,\mathrm{Gs}$), which are comparable to typical values in the heliosphere and at Earth’s surface, respectively. For iron atoms, the estimated strength is $B_{\mathrm{Fe}}\approx 10,\mathrm{Gs}$, consistent with estimates of Earth’s magnetic field in its core~\cite{kolovs2015quasi}. Since stronger magnetic fields tend to enhance chaotic behavior in charged particle dynamics, our analysis focuses primarily on moderate field strengths, namely $B'=0,\,\pm 0.01$. 
In what follows, we systematically analyze all representative models summarized in Tab.\ref{tab:qpo_models}.
}
\begin{table}[!ht]
\centering
\begin{tabular}{|c|c|c|c|c|}
\hline
 & {GRS 1915+105} & {H1743-322} & {XTE 1550-564} & {GRO 1655-40} \\
\hline
$M \; [M_{\odot}]$ & {$9.6-18.4$} & ${9.25-12.86}$ & ${8.5-9.7}$ & ${6.03-6.57}$ \\
\hline
$\nu_U \; [\text{Hz}]$ & ${165-171}$ & ${237-243}$ & ${273-279}$ & ${447-453}$ \\
\hline
$\nu_L \; [\text{Hz}]$ & ${108-118}$ & ${158-174}$ & ${179-189}$ & ${295-305}$ \\
\hline
\end{tabular}
\caption{\footnotesize{\it Observed twin HF QPOs data for the considered microquasars, including their masses and corresponding upper and lower frequencies 
\cite{remillard2006x,molla2017estimation,ingram2014solutions}.}}
\label{astrodata}
\end{table}
\begin{table}[h!]
\centering
\begin{tabular}{|c l c c c |}
\hline
Number & Name & BH Spin & $\log M_{\text{BH}}$ & $f_{\text{UP}}$ (Hz) \\
& & & $(M_{\odot})$ & \\
\hline
1 & RE J1034+396 & 0.998 & $6.0^{+1.0}_{-3.49}$ & $2.7 \times 10^{-4}$ \\
2 & 1H0707-495 & $>$0.976 & $6.36^{+0.24}_{-0.06}$ & $2.6 \times 10^{-4}$ \\
3 & MCG-06-30-15 & $>$0.917 & $6.20^{+0.09}_{-0.12}$ & $2.73 \times 10^{-4}$ \\
4 & Mrk 766 & $>$0.92 & $6.82^{+0.05}_{-0.06}$ & $1.55 \times 10^{-4}$ \\
5 & ESO 113-G010 & 0.998 & $6.85^{+0.15}_{-0.24}$ & $1.24 \times 10^{-4}$ \\
6 & ESO 113-G010b & 0.998 & $6.85^{+0.15}_{-0.24}$ & $6.79 \times 10^{-5}$ \\
7 & 1H0419-577 & $>$0.98 & $8.11^{+0.50}_{-0.50}$ & $2.0 \times 10^{-6}$ \\
8 & ASASSN-14li & $>$0.7 & $6.23^{+0.35}_{-0.35}$ & $7.7 \times 10^{-3}$ \\
9 & TON S 180 & $<$0.4 & $6.85^{+0.5}_{-0.5}$ & $5.56 \times 10^{-6}$ \\
10 & RXJ 0437.4-4711 & ... & $7.77^{+0.5}_{-0.5}$ & $1.27 \times 10^{-5}$ \\
11 & XMMU J134736.6 & ... & $6.99^{+0.46}_{-0.20}$ & $1.16 \times 10^{-5}$ \\
& +173403 & & & \\
12 & MS 2254.9-3712 & ... & $6.6^{+0.39}_{-0.60}$ & $1.5 \times 10^{-4}$ \\
13 & Sw J164449.3 & ... & $7.0^{+0.30}_{-0.35}$ & $5.01 \times 10^{-3}$ \\
& +573451 & & & \\
\hline
\end{tabular}
\caption{Observations of QPOs around Supermassive BHs in active galactic nuclei \cite{stuchlik2022geodesic,smith2021confrontation}}
\label{astrodata AGNs}
\end{table}

\begin{figure}[!ht]
    \centering
    \subfloat[]{\includegraphics[width=0.3\textwidth,height=4.8cm]{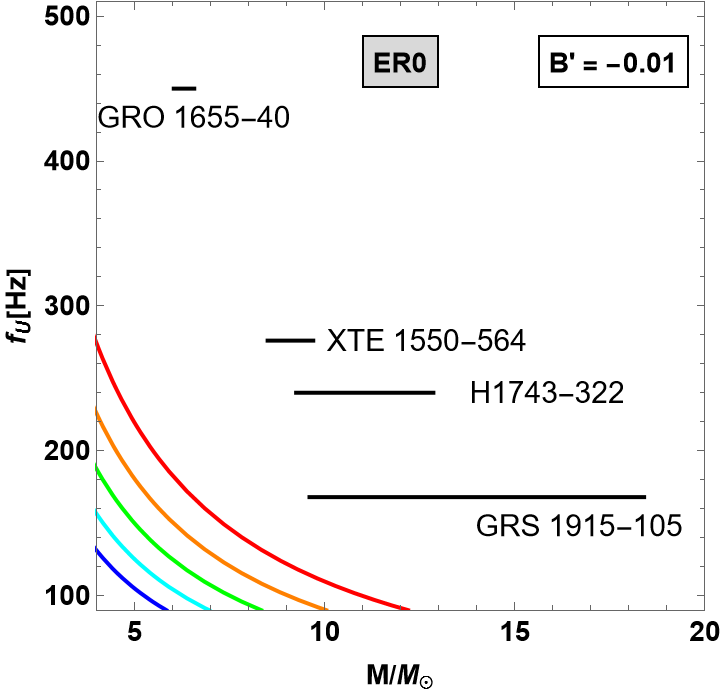}}
    \subfloat[]{\includegraphics[width=0.3\textwidth,height=4.8cm]{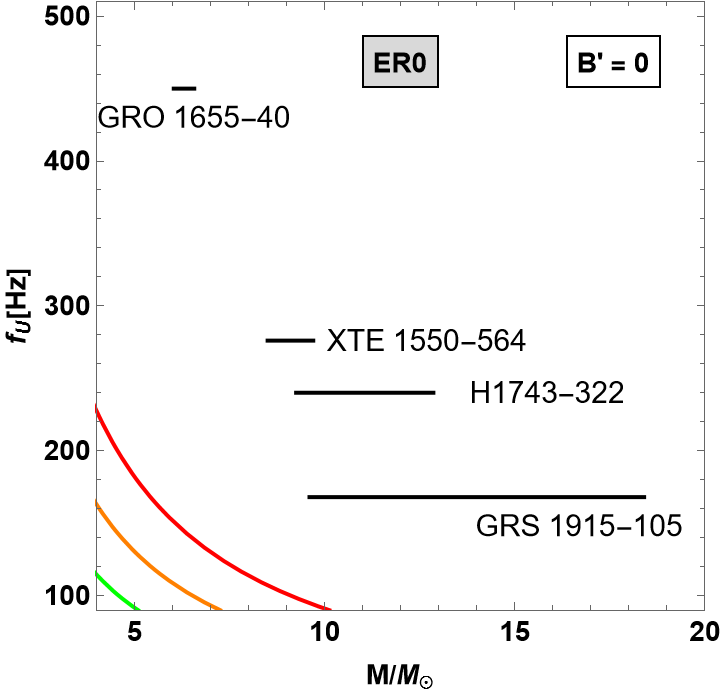}}
    \subfloat[]{\includegraphics[width=0.3\textwidth,height=4.8cm]{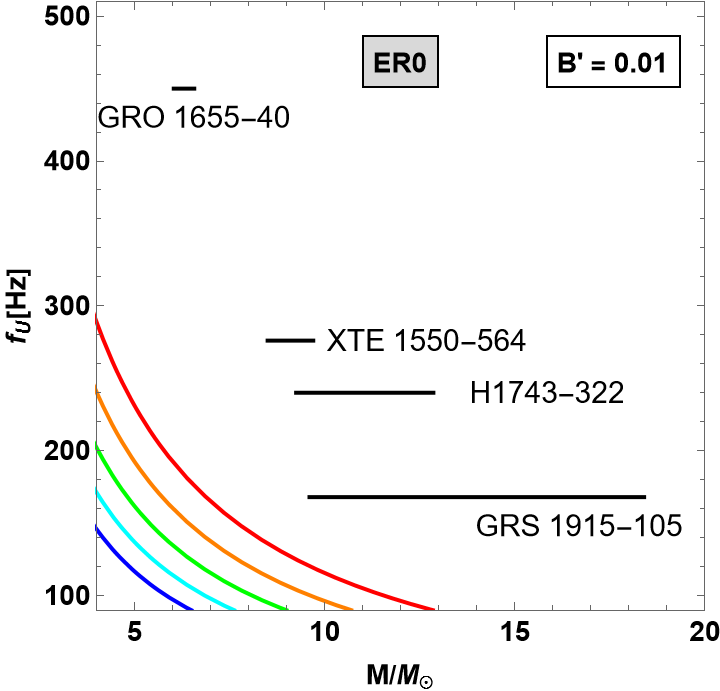}}\\
     \subfloat[]{\includegraphics[width=0.3\textwidth,height=4.8cm]{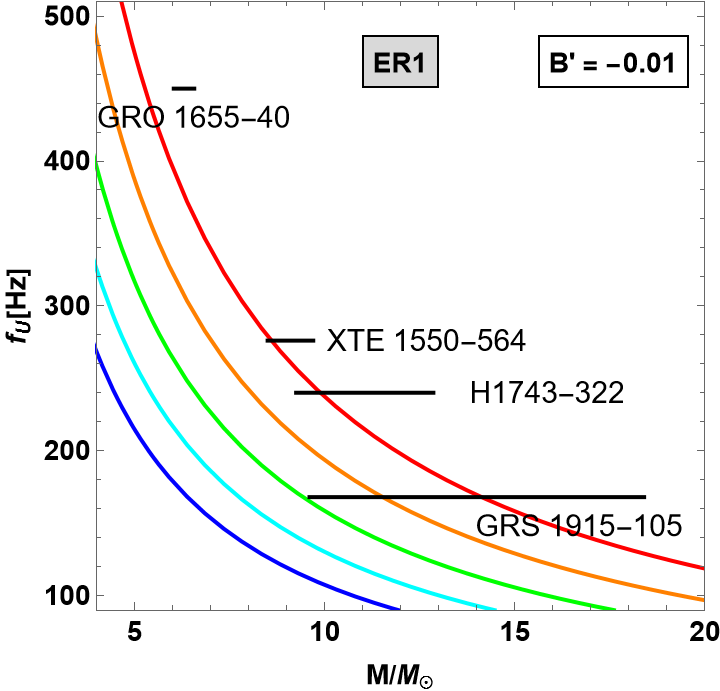}}
    \subfloat[]{\includegraphics[width=0.3\textwidth,height=4.8cm]{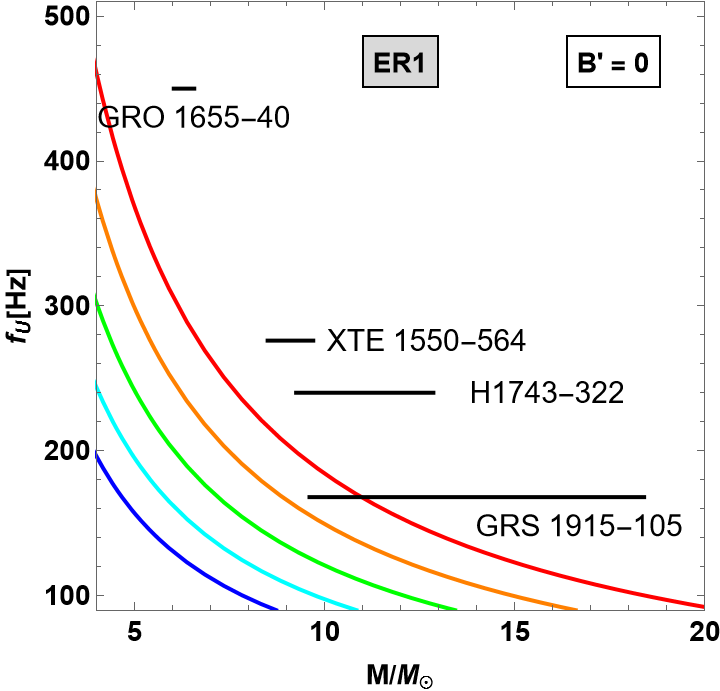}}
    \subfloat[]{\includegraphics[width=0.3\textwidth,height=4.8cm]{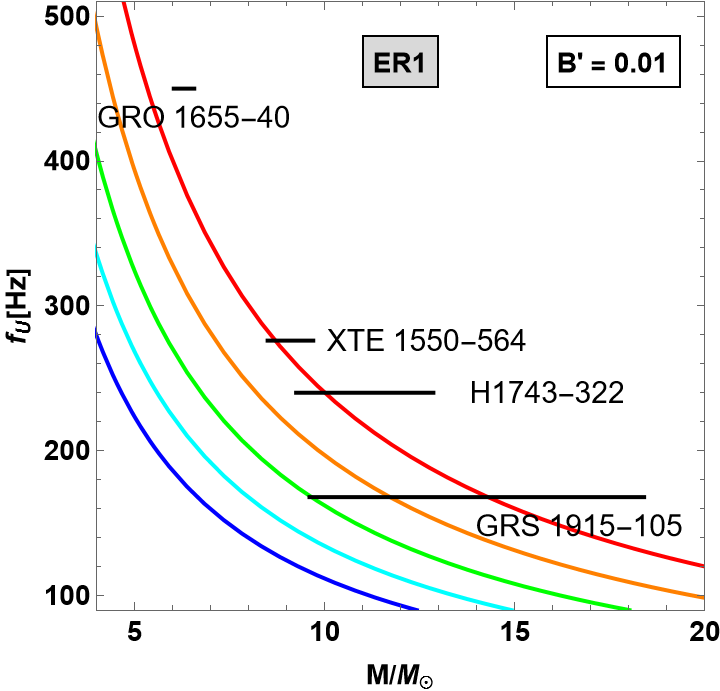}}\\
     \subfloat[]{\includegraphics[width=0.3\textwidth,height=4.8cm]{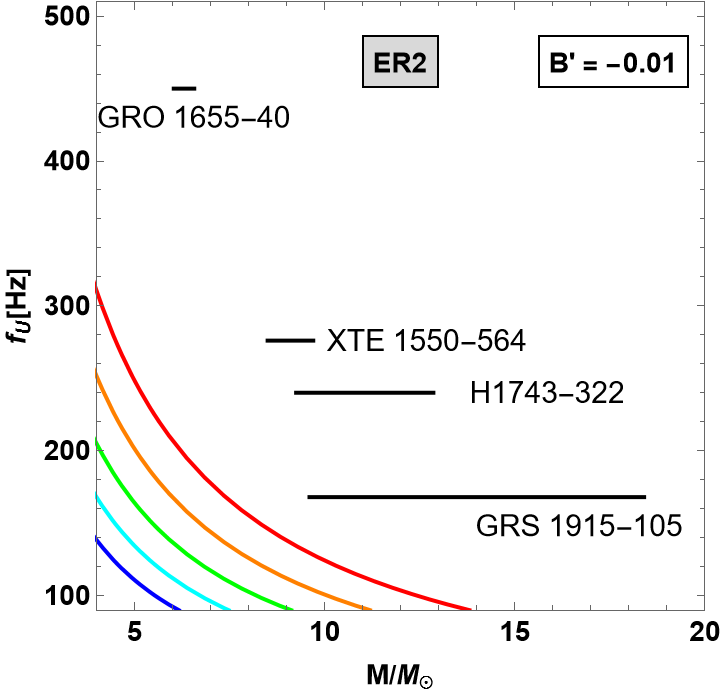}}
    \subfloat[]{\includegraphics[width=0.3\textwidth,height=4.8cm]{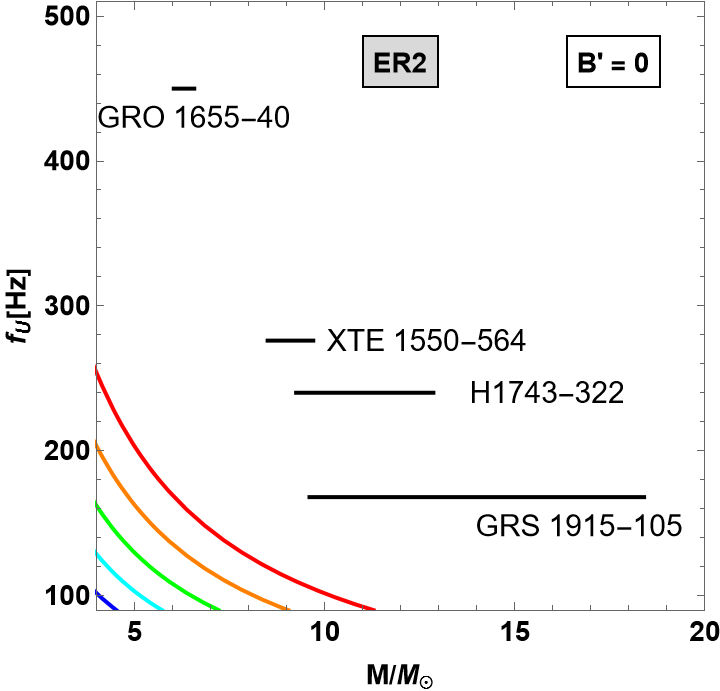}}
    \subfloat[]{\includegraphics[width=0.3\textwidth,height=4.8cm]{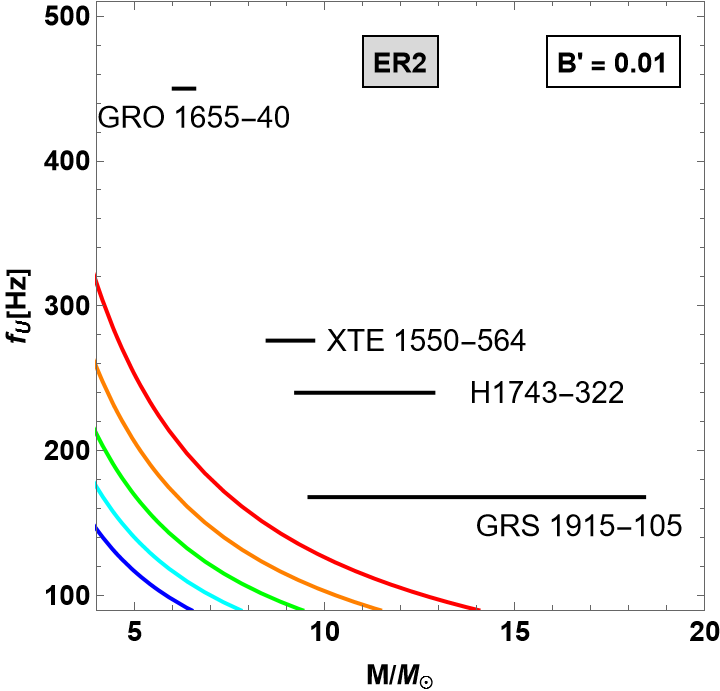}}\\
     \subfloat[]{\includegraphics[width=0.3\textwidth,height=4.8cm]{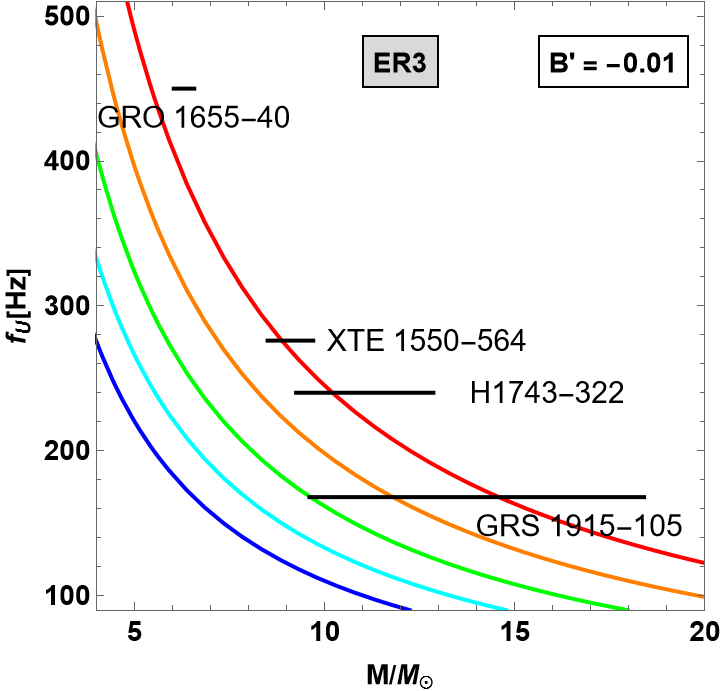}}
    \subfloat[]{\includegraphics[width=0.3\textwidth,height=4.8cm]{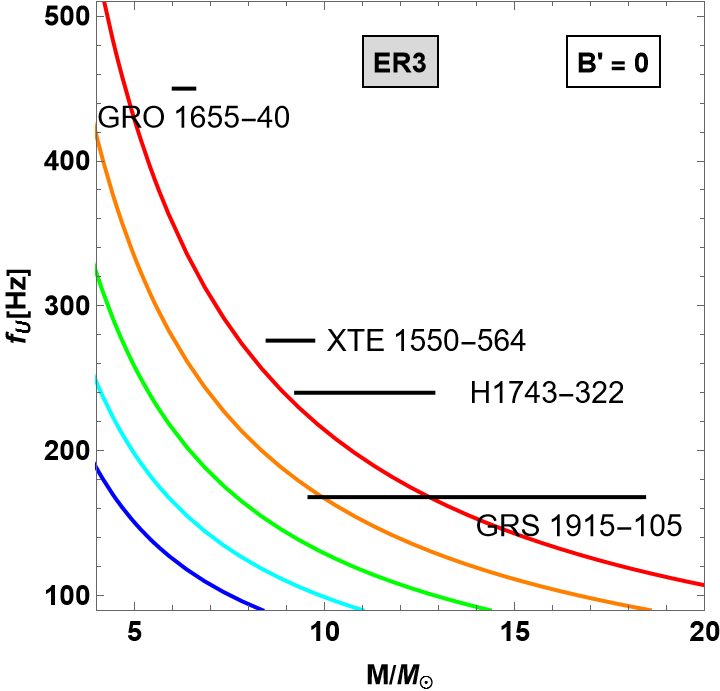}}
    \subfloat[]{\includegraphics[width=0.3\textwidth,height=4.8cm]{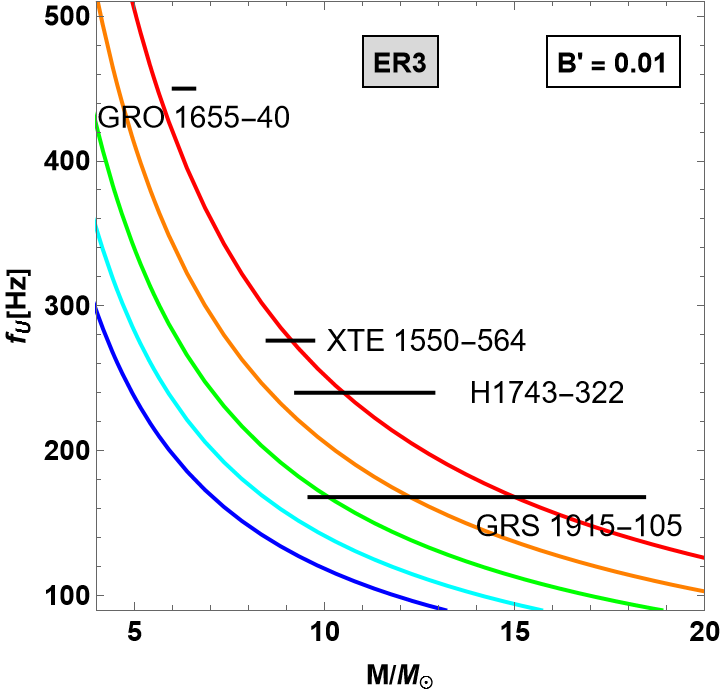}}\\
    \vspace{0.1em}
   \begin{tikzpicture}
  \matrix[column sep=0.3cm, row sep=0.2cm] {
    \draw[red, thick] (0,0) -- ++(1,0); & \node {$\bm{\kappa}$ = 0,}; &
    \draw[orange, thick] (0,0) -- ++(1,0); & \node {$\bm{\kappa}$ = 5,}; &
    \draw[green, thick] (0,0) -- ++(1,0); & \node {$\bm{\kappa}$ = 10,};& 
    \draw[cyan, thick] (0,0) -- ++(1,0); & \node {$\bm{\kappa}$ = 15,}; &
    \draw[blue, thick] (0,0) -- ++(1,0); & \node {$\bm{\kappa}$ = 20}; & & \\
  };
\end{tikzpicture}
    \caption{\footnotesize{\it Fitting the observed HF QPOs frequencies in different models \textcolor{black}{ER0, ER1, ER2, ER3} of  the upper oscillation frequency $f_U$ at the resonance radii $3:2$ for different values of $\bm{\kappa}$ within $R_s=1000$ for different cases of magnetic field strength $B'$ \textcolor{black}{of microquasars of Tab.\ref{astrodata}} }}
    \label{HFQPO Rs=1000 ER0-ER3}
\end{figure}

From Fig.\ref{HFQPO Rs=1000 ER0-ER3} \textcolor{black}{to Fig.\ref{AGNs RP2-WD}}, we perform a fitting of the observed HF QPO  $f_U$ through distinct theoretical models, evaluated at the 3:2 resonance radii for distinct values of the CDM density parameter $\bm{\kappa}$. The analysis is carried out across multiple intensities of $B'$, allowing us to assess the influence of both CDM and magnetic fields on the predicted QPO frequencies.

\begin{figure}[!ht]
    \centering
    \subfloat[]{\includegraphics[width=0.3\textwidth,height=4.8cm]{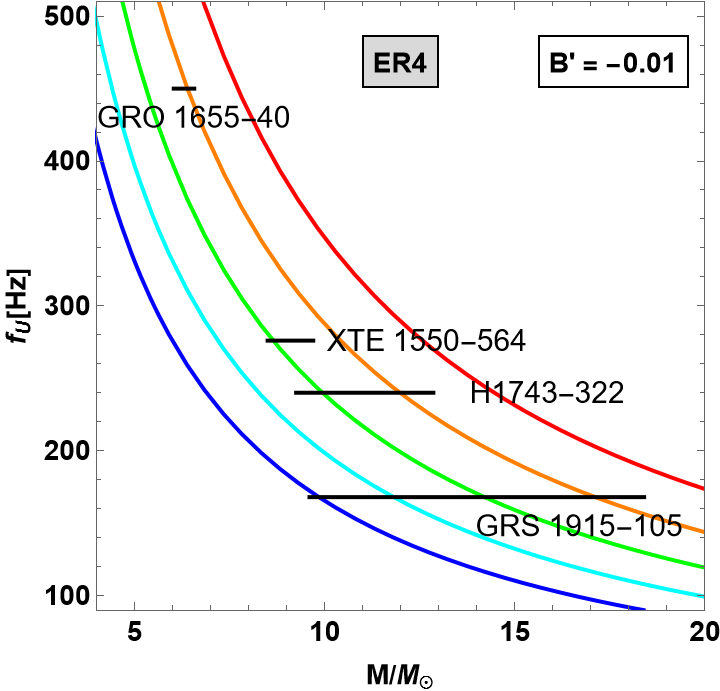}}
    \subfloat[]{\includegraphics[width=0.3\textwidth,height=4.8cm]{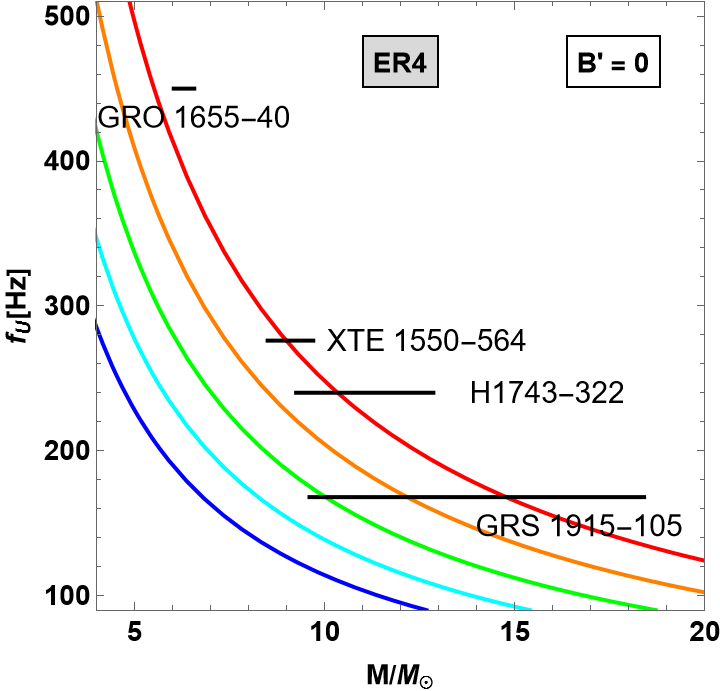}}
    \subfloat[]{\includegraphics[width=0.3\textwidth,height=4.8cm]{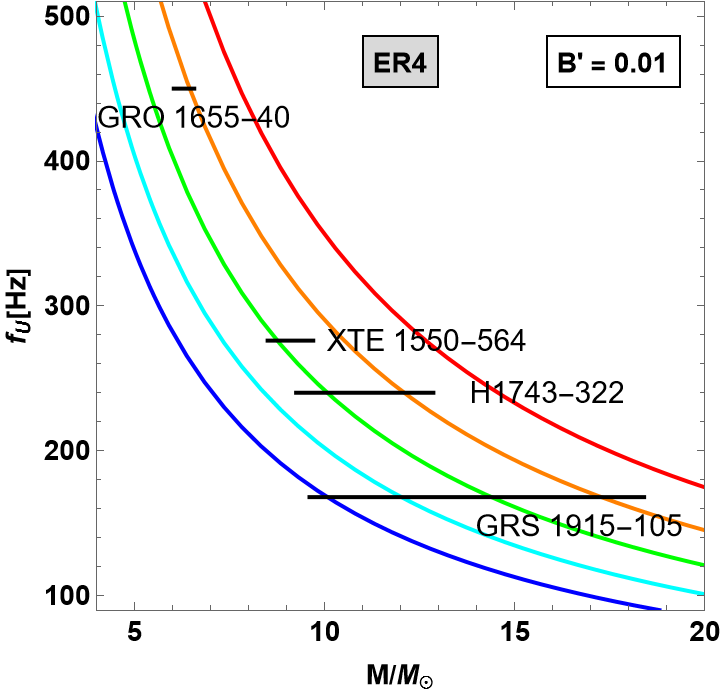}}\\
     \subfloat[]{\includegraphics[width=0.3\textwidth,height=4.8cm]{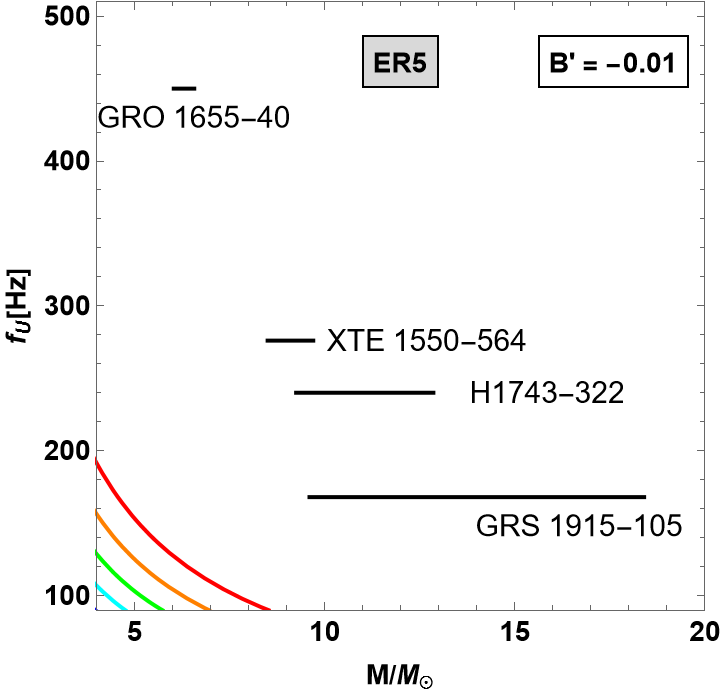}}
    \subfloat[]{\includegraphics[width=0.3\textwidth,height=4.8cm]{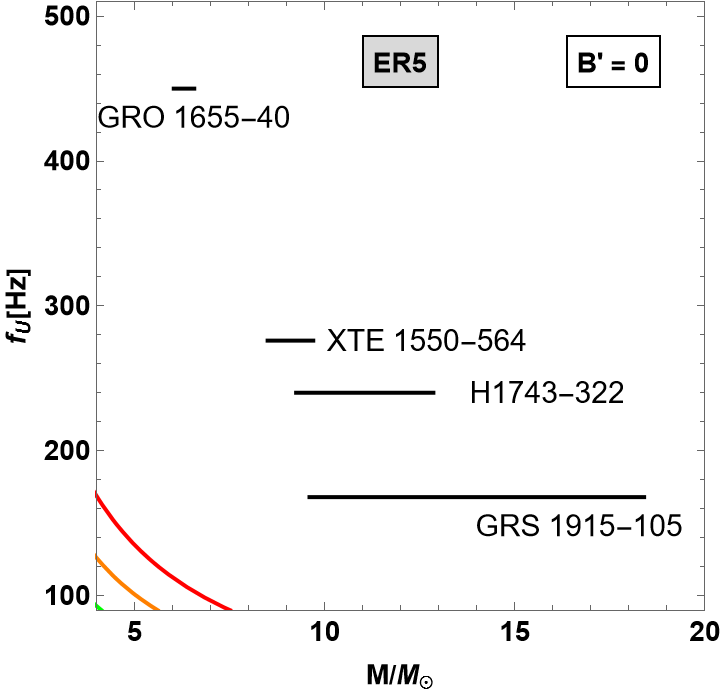}}
    \subfloat[]{\includegraphics[width=0.3\textwidth,height=4.8cm]{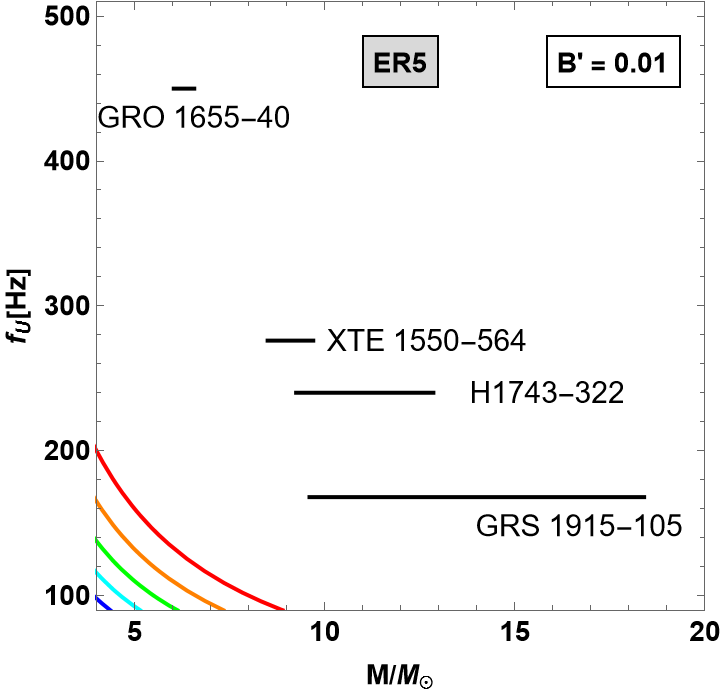}}\\
     \subfloat[]{\includegraphics[width=0.3\textwidth,height=4.8cm]{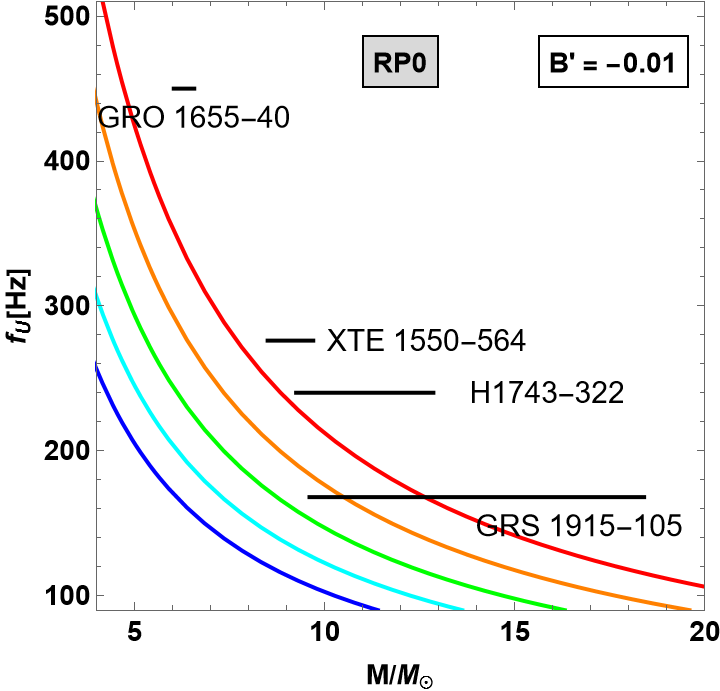}}
    \subfloat[]{\includegraphics[width=0.3\textwidth,height=4.8cm]{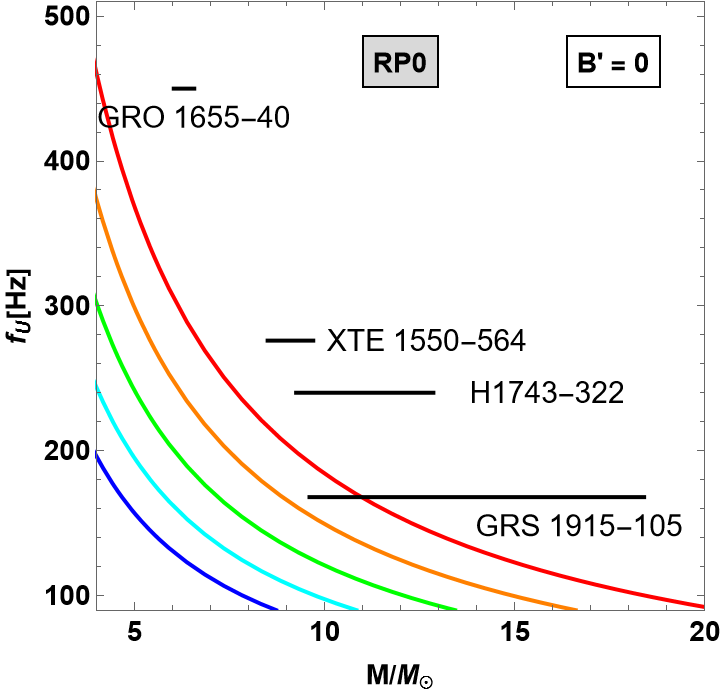}}
    \subfloat[]{\includegraphics[width=0.3\textwidth,height=4.8cm]{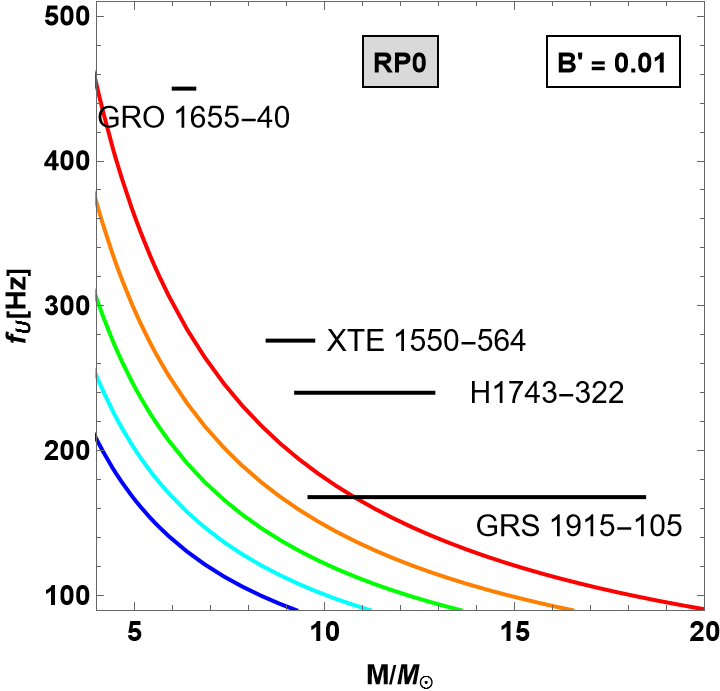}}\\
     \subfloat[]{\includegraphics[width=0.3\textwidth,height=4.8cm]{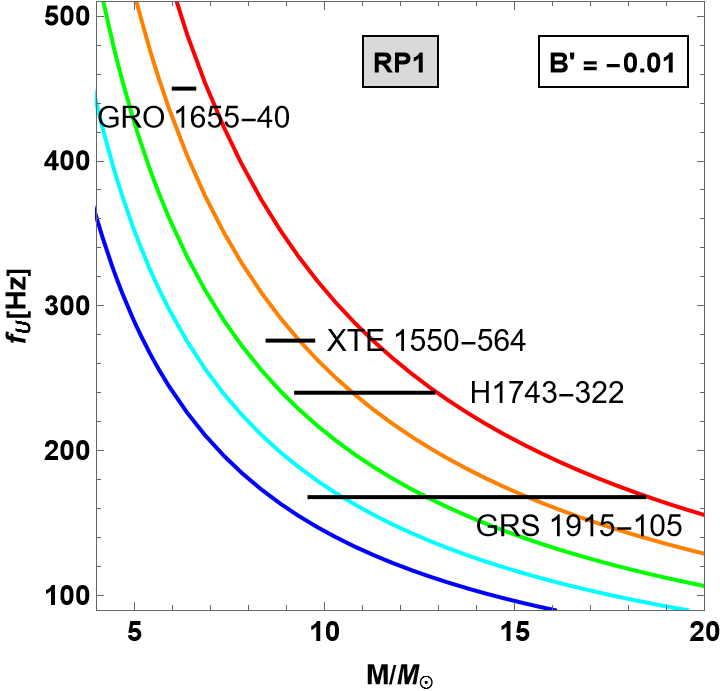}}
    \subfloat[]{\includegraphics[width=0.3\textwidth,height=4.8cm]{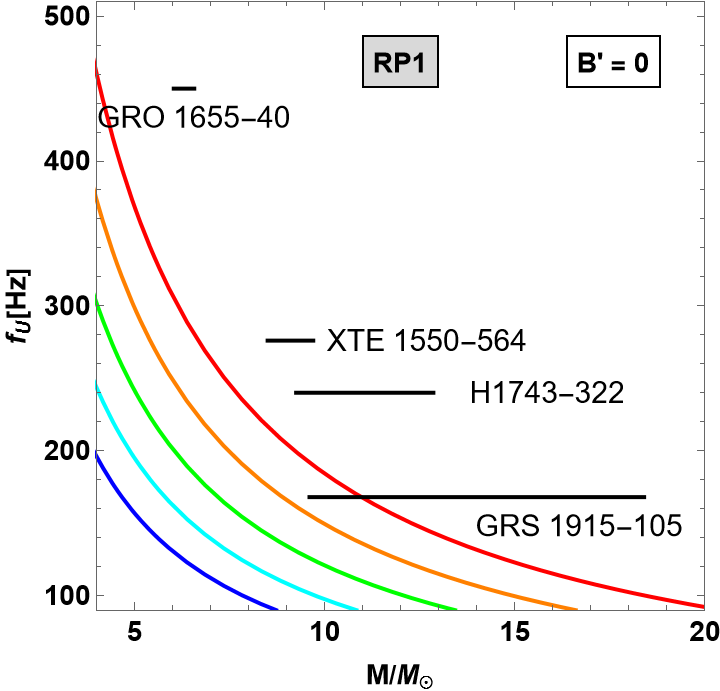}}
    \subfloat[]{\includegraphics[width=0.3\textwidth,height=4.8cm]{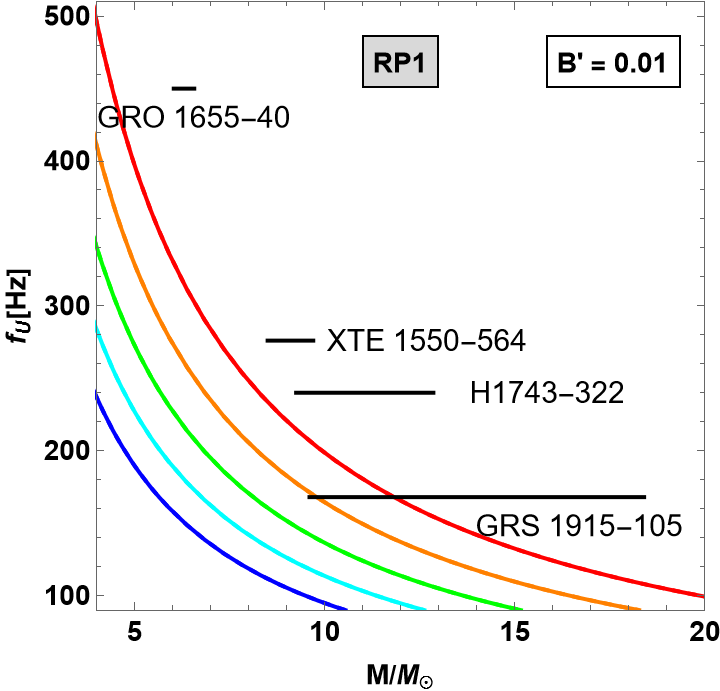}}\\
    \vspace{0.1em}
   \begin{tikzpicture}
  \matrix[column sep=0.3cm, row sep=0.2cm] {
    \draw[red, thick] (0,0) -- ++(1,0); & \node {$\bm{\kappa}$ = 0,}; &
    \draw[orange, thick] (0,0) -- ++(1,0); & \node {$\bm{\kappa}$ = 5,}; &
    \draw[green, thick] (0,0) -- ++(1,0); & \node {$\bm{\kappa}$ = 10,};& 
    \draw[cyan, thick] (0,0) -- ++(1,0); & \node {$\bm{\kappa}$ = 15,}; &
    \draw[blue, thick] (0,0) -- ++(1,0); & \node {$\bm{\kappa}$ = 20}; & & \\
  };
\end{tikzpicture}
    \caption{\footnotesize{\it Fitting the observed HF QPOs frequencies in different models \textcolor{black}{ER4, ER5, RP0, RP1} of  the upper oscillation frequency $f_U$ at the resonance radii $3:2$ for different values of $\bm{\kappa}$ within $R_s=1000$ for different cases of magnetic field strength $B'$ \textcolor{black}{of microquasars of Tab.\ref{astrodata}} }}
    \label{HFQPO Rs=1000 ER4-RP1}
\end{figure}

\begin{figure}[!ht]
    \centering
    \subfloat[]{\includegraphics[width=0.3\textwidth,height=4.8cm]{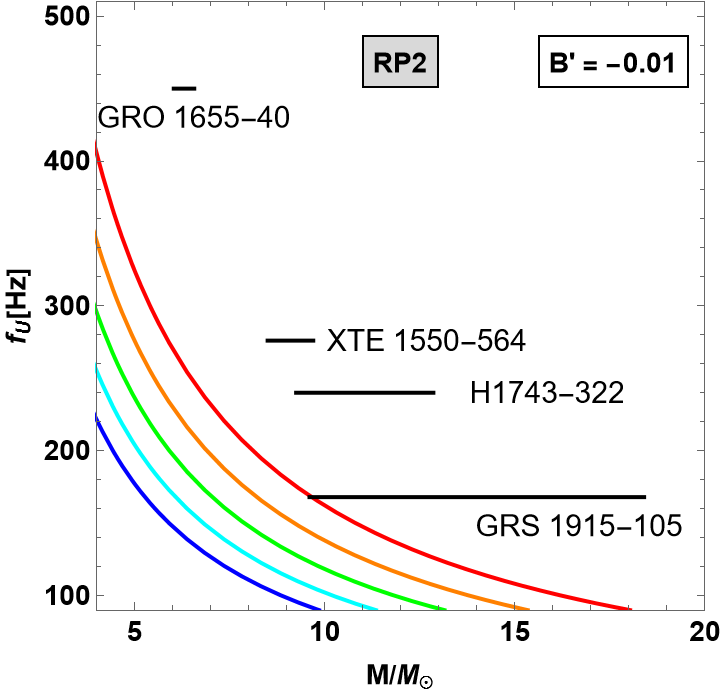}}
    \subfloat[]{\includegraphics[width=0.3\textwidth,height=4.8cm]{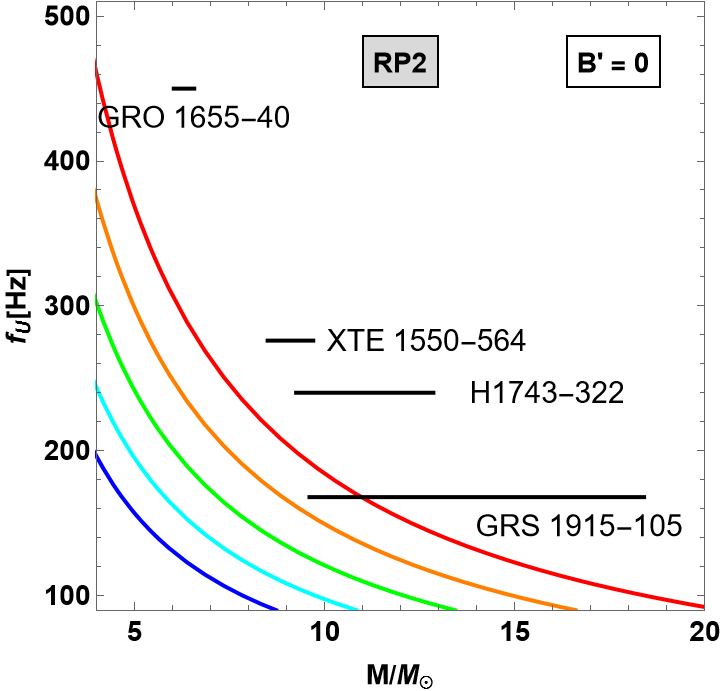}}
    \subfloat[]{\includegraphics[width=0.3\textwidth,height=4.8cm]{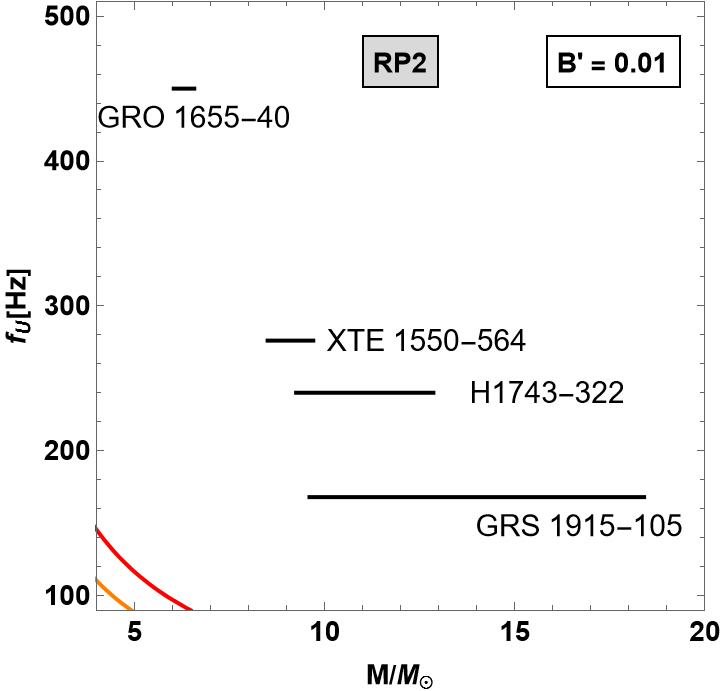}}\\
     \subfloat[]{\includegraphics[width=0.3\textwidth,height=4.8cm]{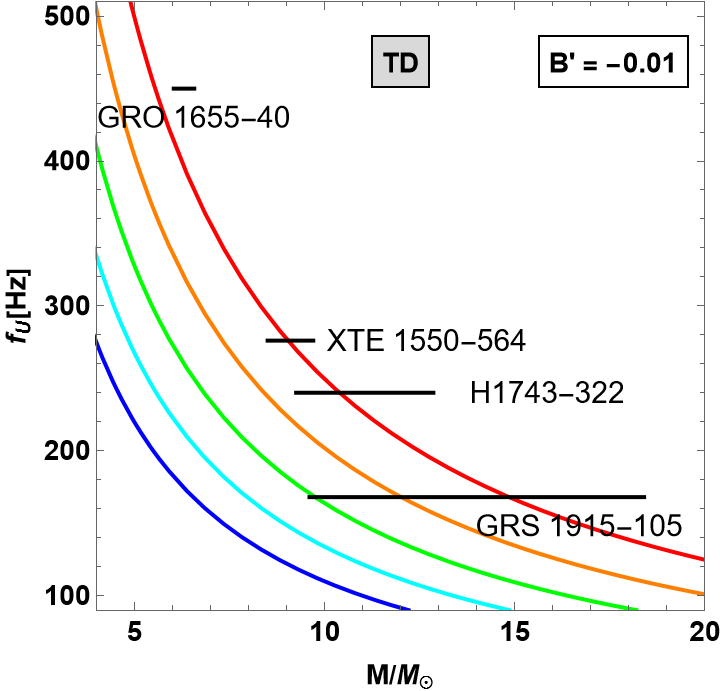}}
    \subfloat[]{\includegraphics[width=0.3\textwidth,height=4.8cm]{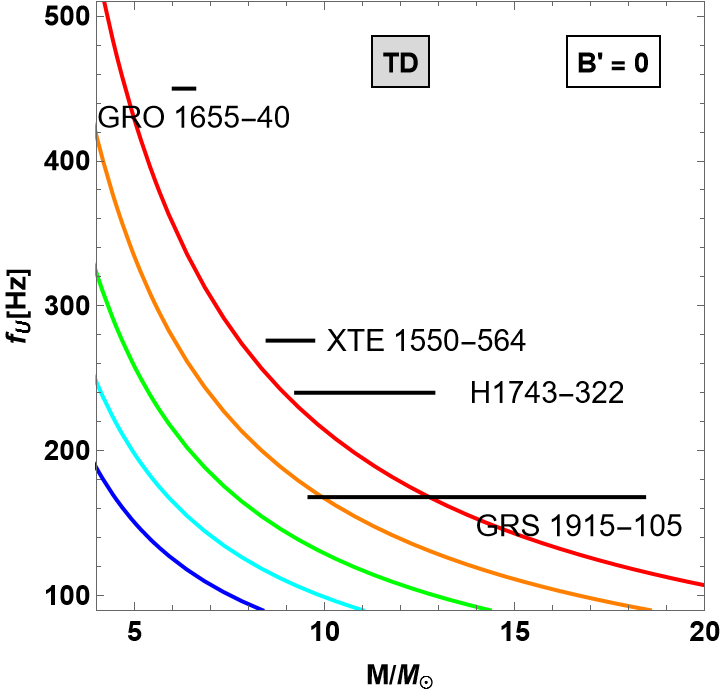}}
    \subfloat[]{\includegraphics[width=0.3\textwidth,height=4.8cm]{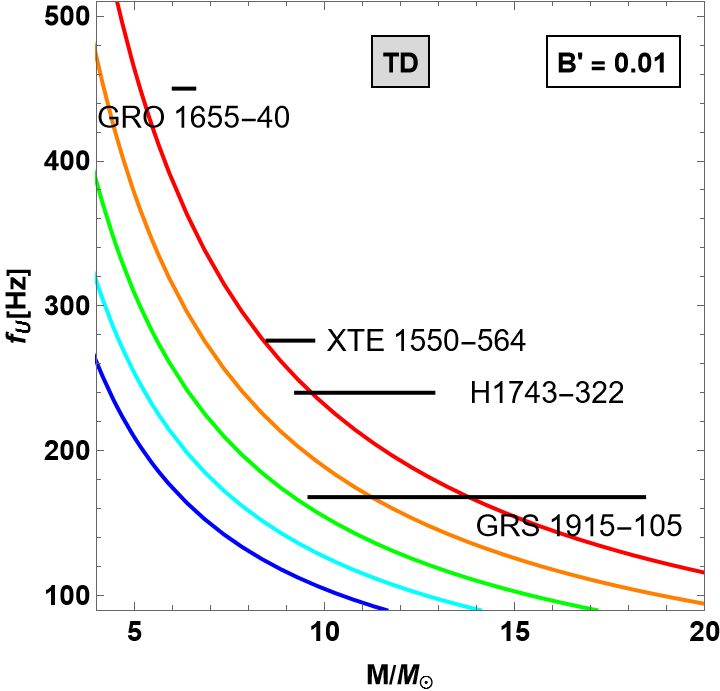}}\\
     \subfloat[]{\includegraphics[width=0.3\textwidth,height=4.8cm]{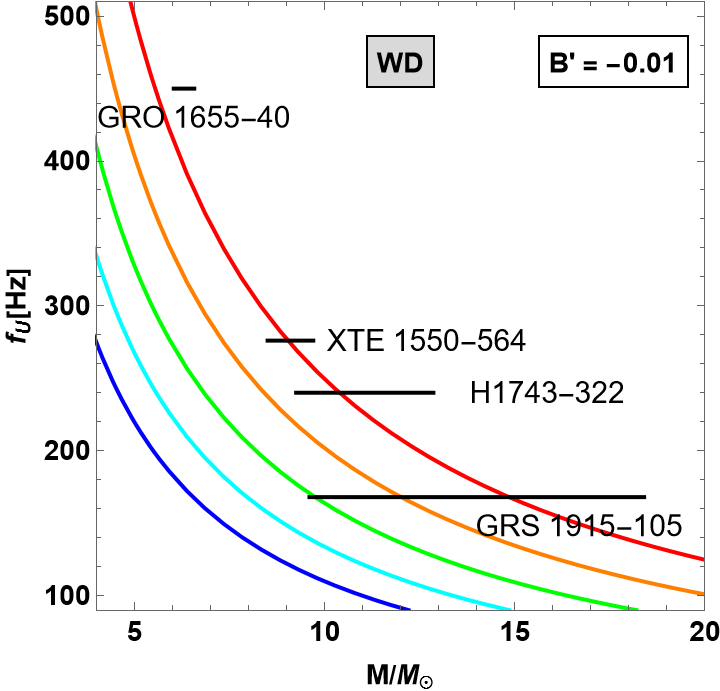}}
    \subfloat[]{\includegraphics[width=0.3\textwidth,height=4.8cm]{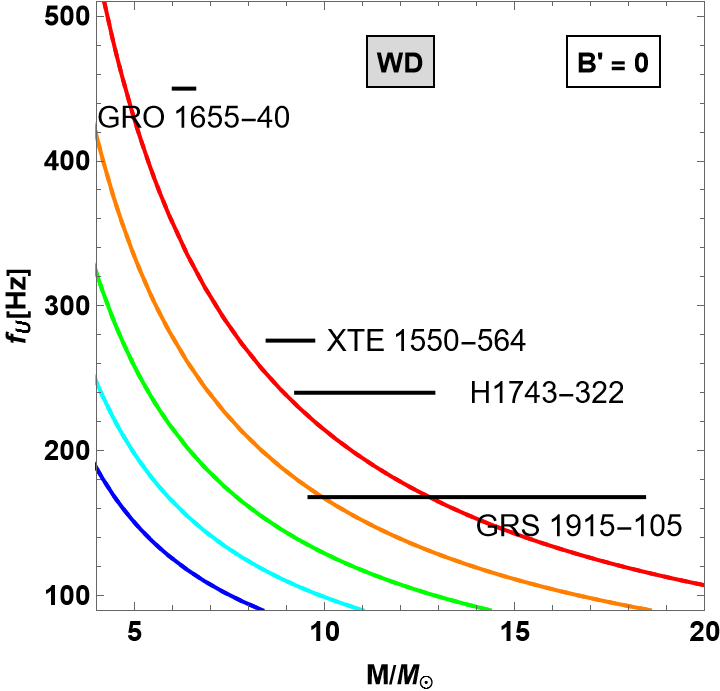}}
    \subfloat[]{\includegraphics[width=0.3\textwidth,height=4.8cm]{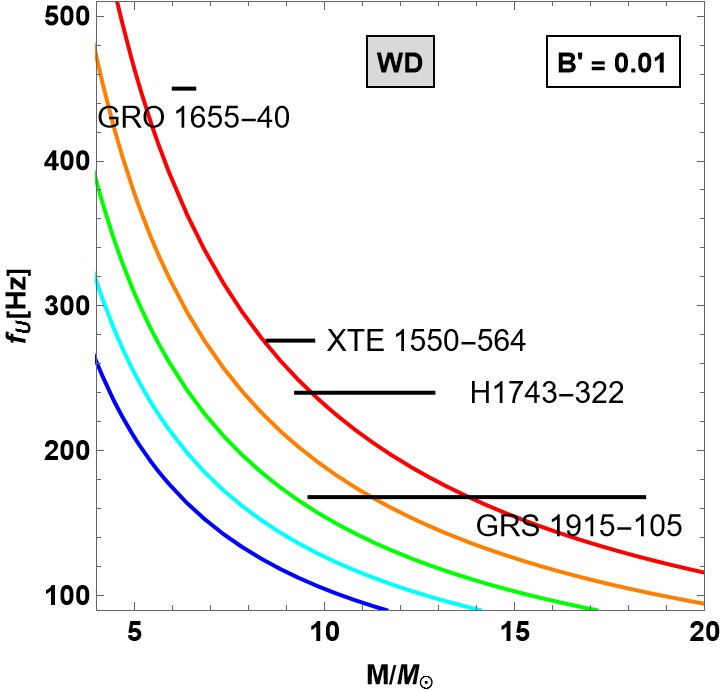}}\\
    \vspace{0.1em}
   \begin{tikzpicture}
  \matrix[column sep=0.3cm, row sep=0.2cm] {
    \draw[red, thick] (0,0) -- ++(1,0); & \node {$\bm{\kappa}$ = 0,}; &
    \draw[orange, thick] (0,0) -- ++(1,0); & \node {$\bm{\kappa}$ = 5,}; &
    \draw[green, thick] (0,0) -- ++(1,0); & \node {$\bm{\kappa}$ = 10,};& 
    \draw[cyan, thick] (0,0) -- ++(1,0); & \node {$\bm{\kappa}$ = 15,}; &
    \draw[blue, thick] (0,0) -- ++(1,0); & \node {$\bm{\kappa}$ = 20}; & & \\
  };
\end{tikzpicture}
    \caption{\footnotesize{\it Fitting the observed HF QPOs frequencies in different models \textcolor{black}{RP2, TD, WD} of  the upper oscillation frequency $f_U$ at the resonance radii $3:2$ for different values of $\bm{\kappa}$ within $R_s=1000$ for different cases of magnetic field strength $B'$ \textcolor{black}{of microquasars of Tab.\ref{astrodata}} }}
    \label{HFQPO Rs=1000 RP2-WD}
\end{figure}

\textcolor{black}{In particular, Figs.~\ref{HFQPO Rs=1000 ER0-ER3}, \ref{HFQPO Rs=1000 ER4-RP1}, and \ref{HFQPO Rs=1000 RP2-WD} illustrate the fitting of the upper oscillation frequency $f_U$ at the $3:2$ resonance radii of microquasars, considering different magnetic field strengths and CDM densities $\bm{\kappa}$.}

\textcolor{black}{ For a weak magnetic field ($B'=0.01$), several models—namely ER0, ER2, ER5, RP1, and RP2—fail to reproduce any microquasar data. In contrast, ER1 $\equiv$ ER3 and TD $\equiv$ WD succeed in fitting XTE 1550-564 and H1743-322 in the absence of CDM ($\bm{\kappa}=0$). For GRS 1915+105, these same models also provide fits at low CDM densities: ER1 $\equiv$ ER3 for $\bm{\kappa}=\{0,\,5,\,10\}$ and TD $\equiv$ WD for $\bm{\kappa}=\{0,\,5\}$. The RP0 model is more restrictive, as it fits only GRS 1915+105 without CDM. Notably, ER4 demonstrates the most robust performance, successfully fitting multiple microquasars over a wider range of CDM densities, $\bm{\kappa}=\{5,\,10,\,15,\,20\}$. For a negative magnetic field strength ($B'=-0.01$), most models behave similarly to the $B'=0.01$ case, with only minor differences between the two field polarities. However, the Relativistic Precession models show a marked sensitivity to the sign of $B'$. Specifically, RP0 fits GRS 1915+105 at a lower CDM density ($\bm{\kappa}=5$), whereas at $B'=0.01$ it succeeds only in the absence of CDM. More strikingly, RP1 and RP2—which fail to reproduce any microquasar data for $B'=0.01$—become viable under $B'=-0.01$: RP1 provides a good match to most of the microquasar data, while RP2 successfully fits GRS 1915+105 at $\bm{\kappa}=0$. Consistently, ER4 remains the most reliable model, offering successful fits to the majority of the data in the attractive Lorentz force regime. While in the absence of a magnetic field ($B'=0$), models ER0, ER2, and ER5 reduce to equivalent forms but remain inconsistent with the observational data across all sources. In contrast, the group ER1 $\equiv$ RP0 $\equiv$ RP1 $\equiv$ RP2 is able to reproduce the QPOs of GRS 1915+105, though only in the case $\kappa = 0$, i.e., without dark matter. When a weak CDM density is introduced ($\kappa = 5$), models ER3, TD, and WD yield good fits for GRS 1915+105. Notably, among all models in the vanishing magnetic field limit, ER4 consistently provides the best agreement with the majority of the observational data.}

\textcolor{black}{Summarizing, across all magnetic field strengths considered, ER4 emerges as the most robust model, consistently fitting the majority of the microquasars data. We should note that TD and WD models are equal in all cases of $B'$.}

\begin{figure}[!ht]
    \centering
    \subfloat[]{\includegraphics[width=0.3\textwidth,height=4.8cm]{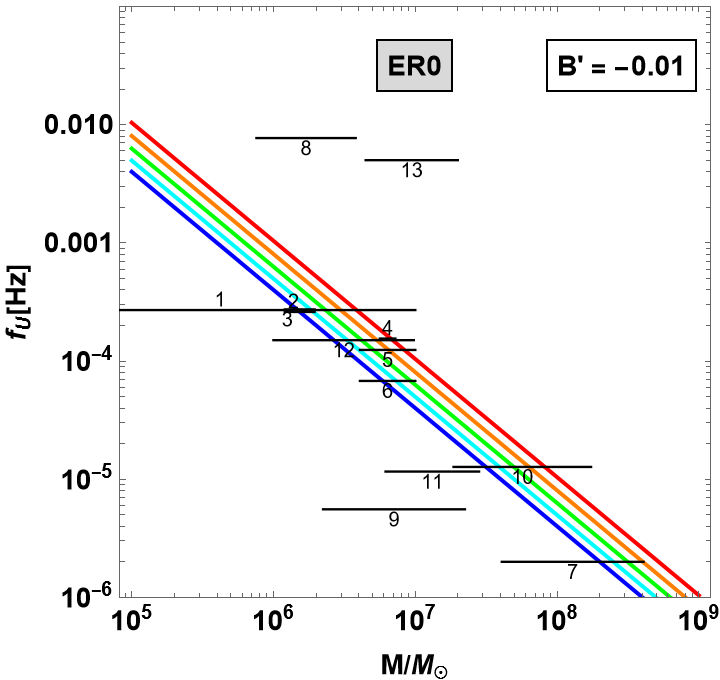}}
    \subfloat[]{\includegraphics[width=0.3\textwidth,height=4.8cm]{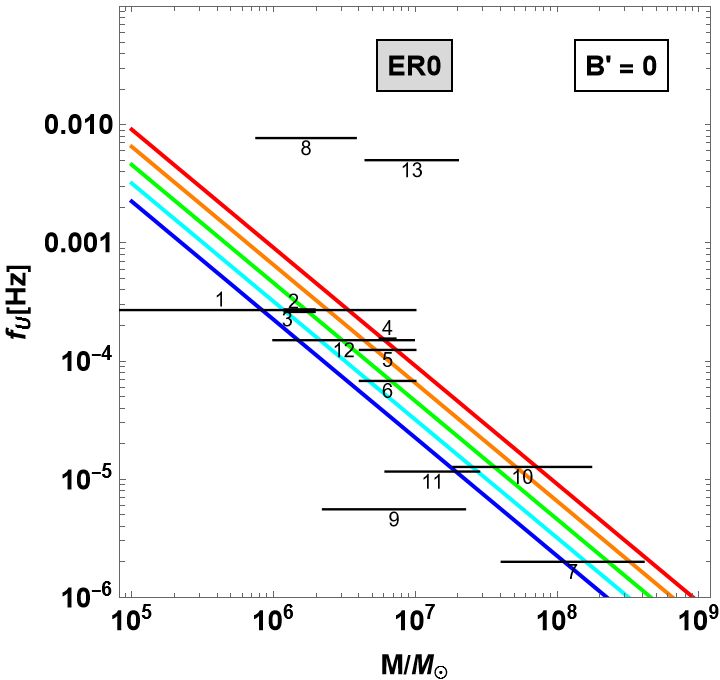}}
    \subfloat[]{\includegraphics[width=0.3\textwidth,height=4.8cm]{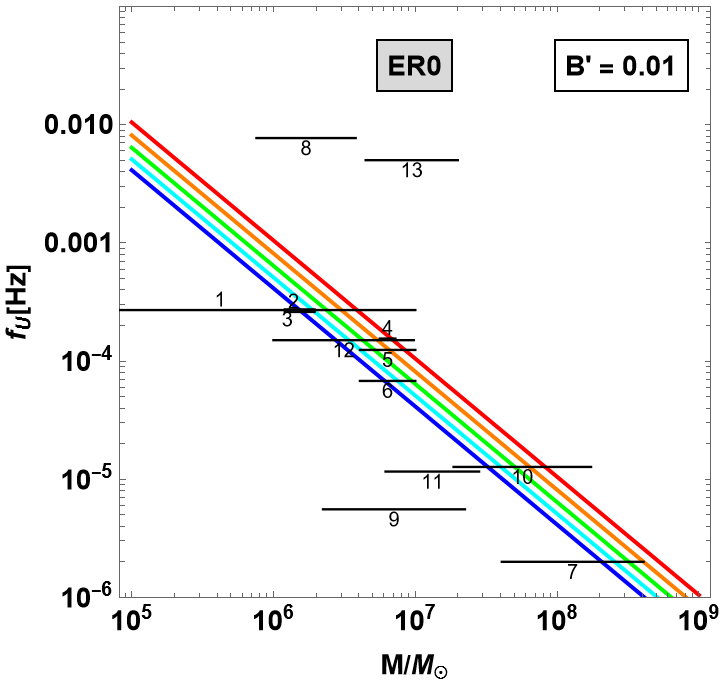}}\\
     \subfloat[]{\includegraphics[width=0.3\textwidth,height=4.8cm]{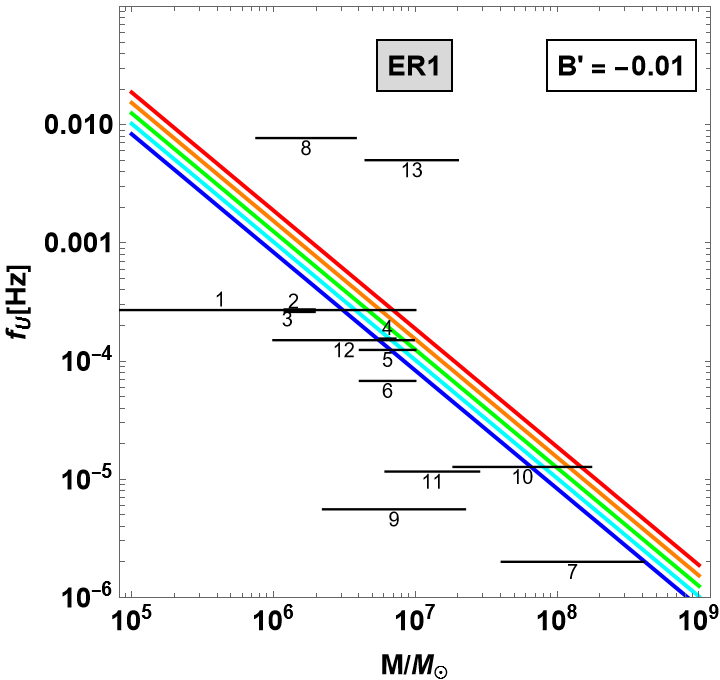}}
    \subfloat[]{\includegraphics[width=0.3\textwidth,height=4.8cm]{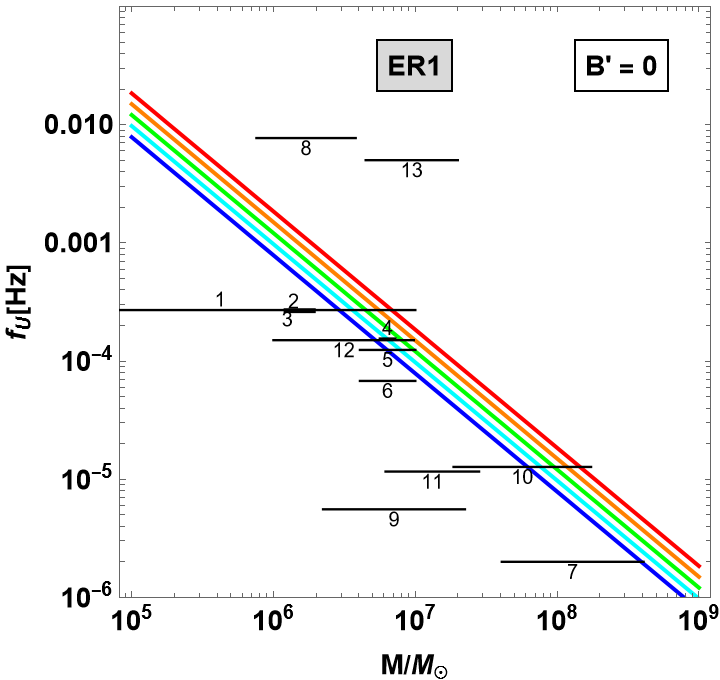}}
    \subfloat[]{\includegraphics[width=0.3\textwidth,height=4.8cm]{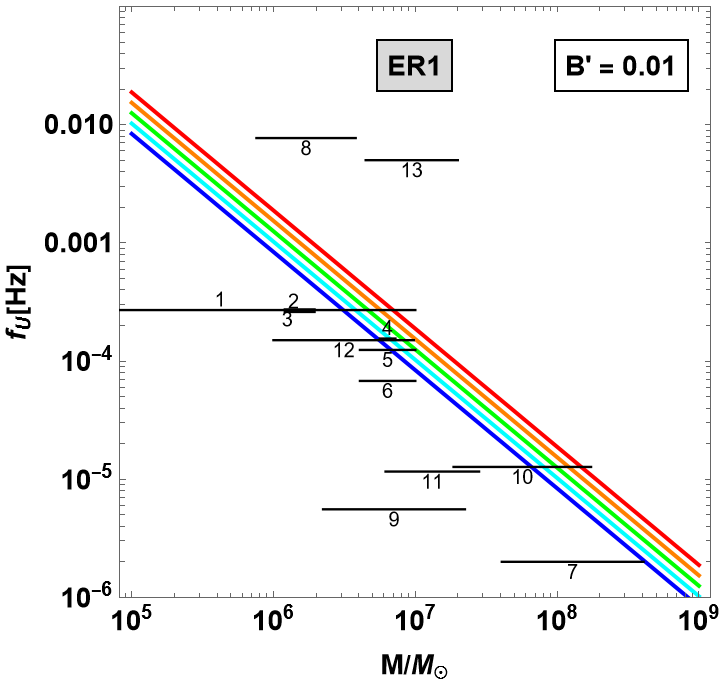}}\\
     \subfloat[]{\includegraphics[width=0.3\textwidth,height=4.8cm]{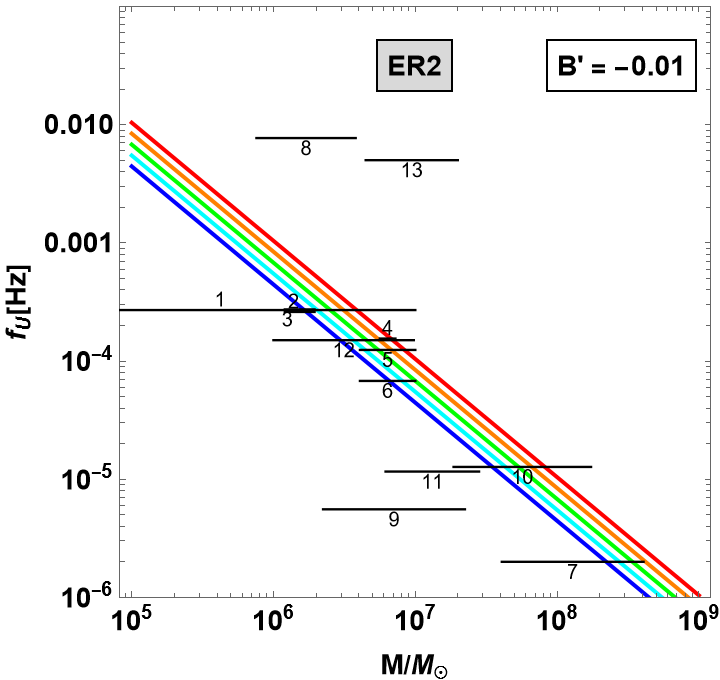}}
    \subfloat[]{\includegraphics[width=0.3\textwidth,height=4.8cm]{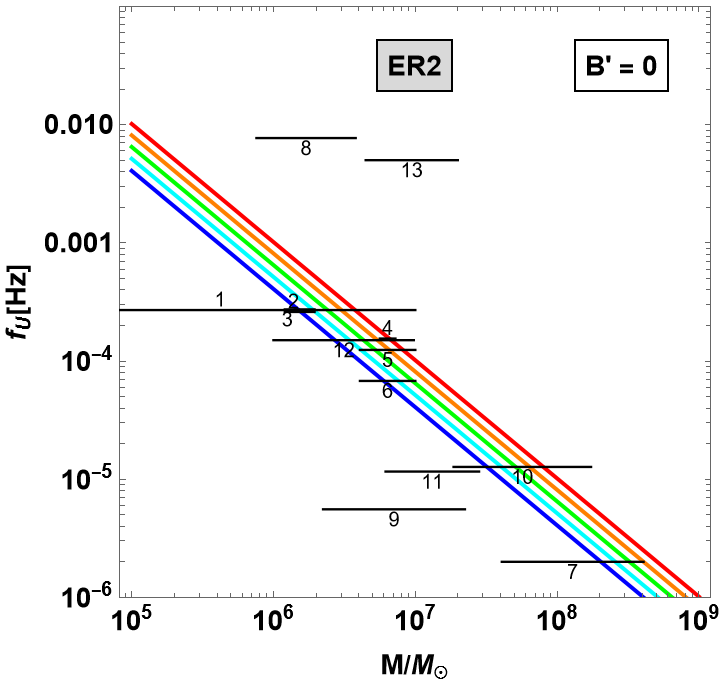}}
    \subfloat[]{\includegraphics[width=0.3\textwidth,height=4.8cm]{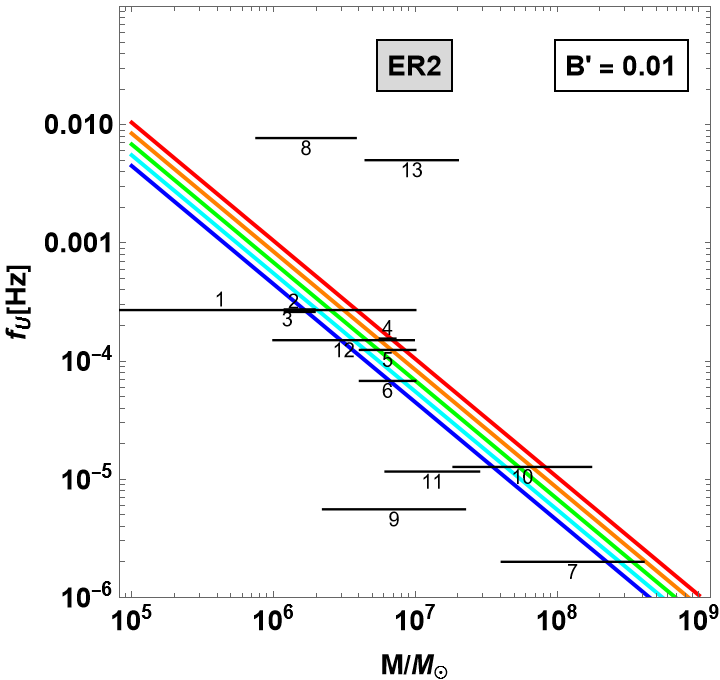}}\\
     \subfloat[]{\includegraphics[width=0.3\textwidth,height=4.8cm]{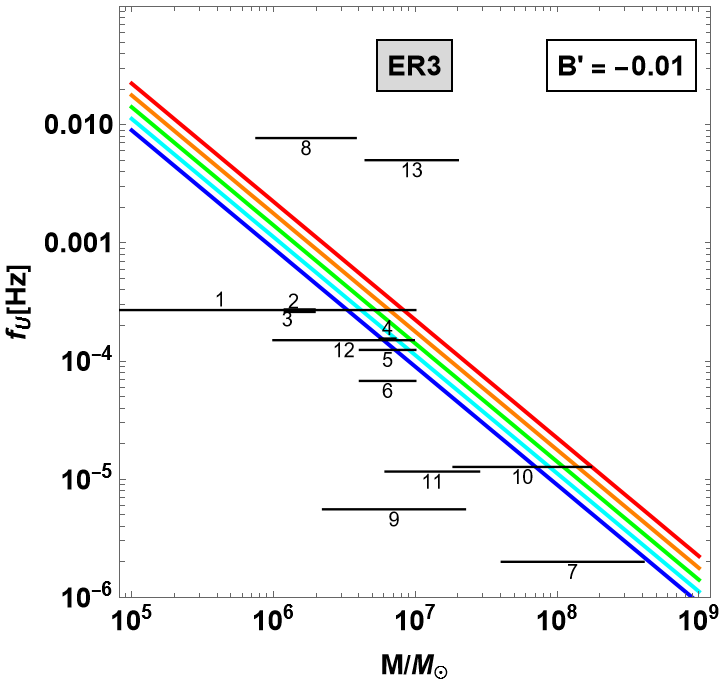}}
    \subfloat[]{\includegraphics[width=0.3\textwidth,height=4.8cm]{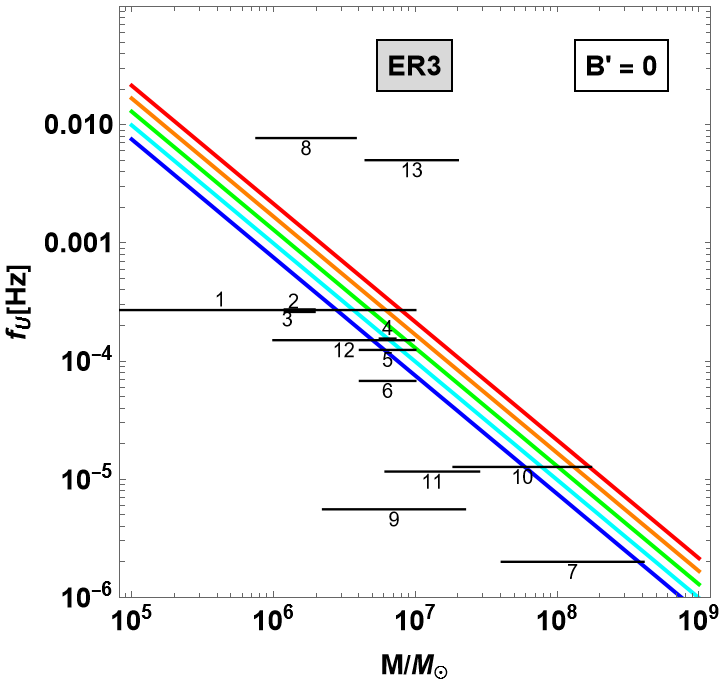}}
    \subfloat[]{\includegraphics[width=0.3\textwidth,height=4.8cm]{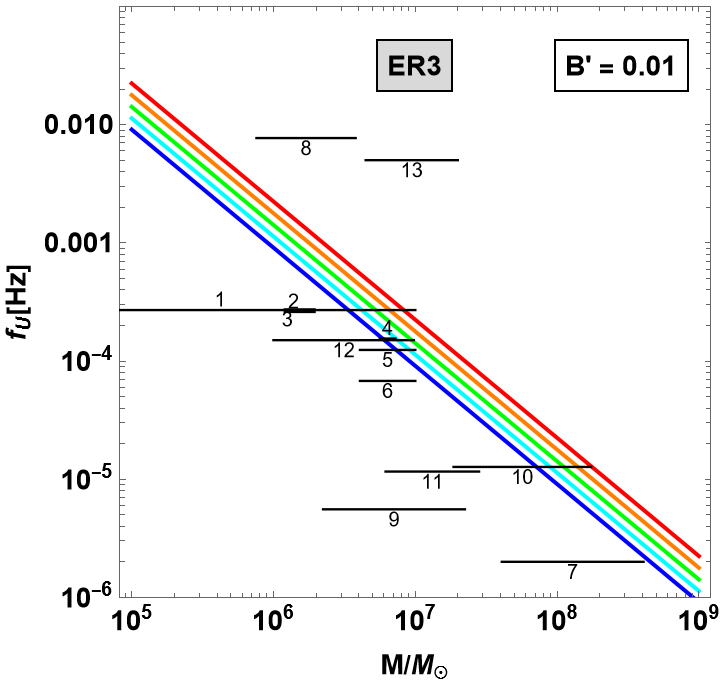}}
    \vspace{0.1em}
   \begin{tikzpicture}
  \matrix[column sep=0.3cm, row sep=0.2cm] {
    \draw[red, thick] (0,0) -- ++(1,0); & \node {$\bm{\kappa}$ = 0,}; &
    \draw[orange, thick] (0,0) -- ++(1,0); & \node {$\bm{\kappa}$ = 5,}; &
    \draw[green, thick] (0,0) -- ++(1,0); & \node {$\bm{\kappa}$ = 10,};& 
    \draw[cyan, thick] (0,0) -- ++(1,0); & \node {$\bm{\kappa}$ = 15,}; &
    \draw[blue, thick] (0,0) -- ++(1,0); & \node {$\bm{\kappa}$ = 20}; & & \\
  };
\end{tikzpicture}
    \caption{\footnotesize{\it Fitting the observed HF QPOs frequencies in different models \textcolor{black}{ER0, ER1, ER2, ER3} of  the upper oscillation frequency $f_U$ at the resonance radii $3:2$ for different values of $\bm{\kappa}$ within $R_s=1000$ for different cases of magnetic field strength $B'$ \textcolor{black}{of supermassive black hole  in AGNs of Tab.\ref{astrodata AGNs}} }}
    \label{AGNs ER0-ER3}
\end{figure}

\begin{figure}[!ht]
    \centering
    \subfloat[]{\includegraphics[width=0.3\textwidth,height=4.8cm]{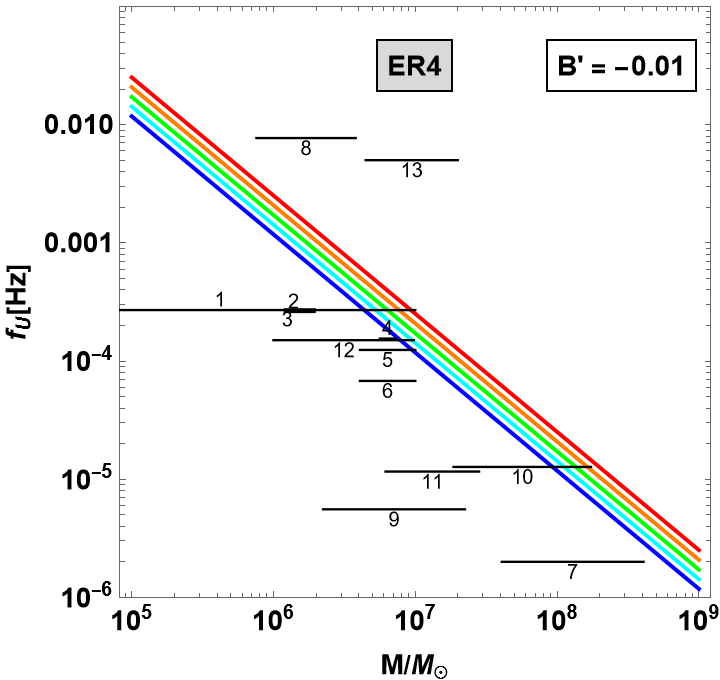}}
    \subfloat[]{\includegraphics[width=0.3\textwidth,height=4.8cm]{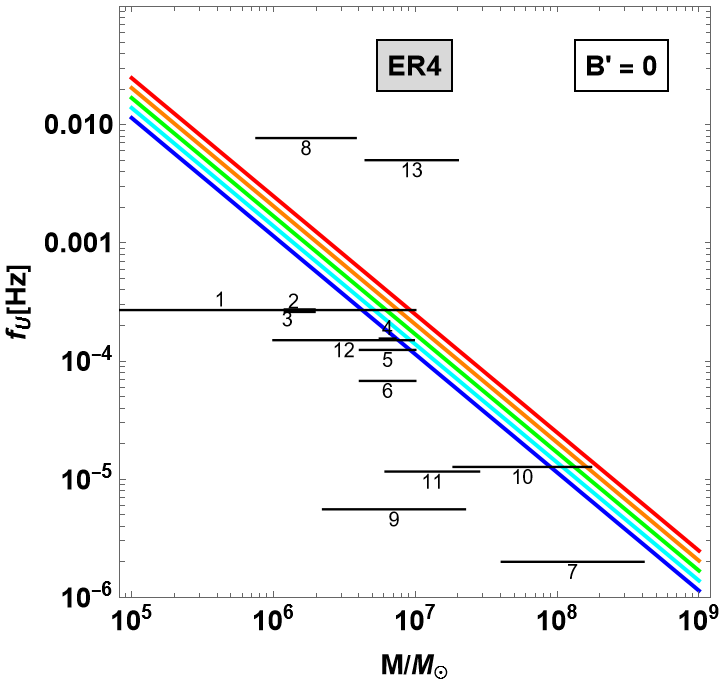}}
    \subfloat[]{\includegraphics[width=0.3\textwidth,height=4.8cm]{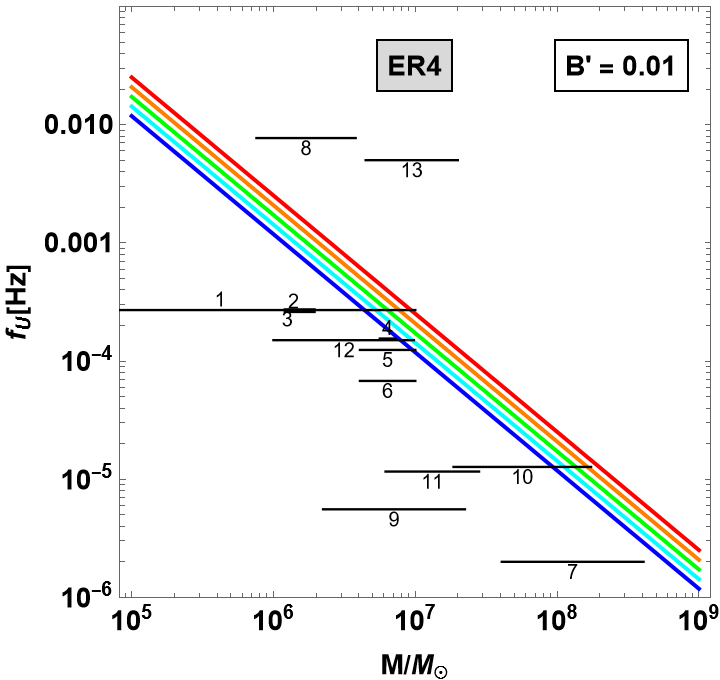}}\\
    \subfloat[]{\includegraphics[width=0.3\textwidth,height=4.8cm]{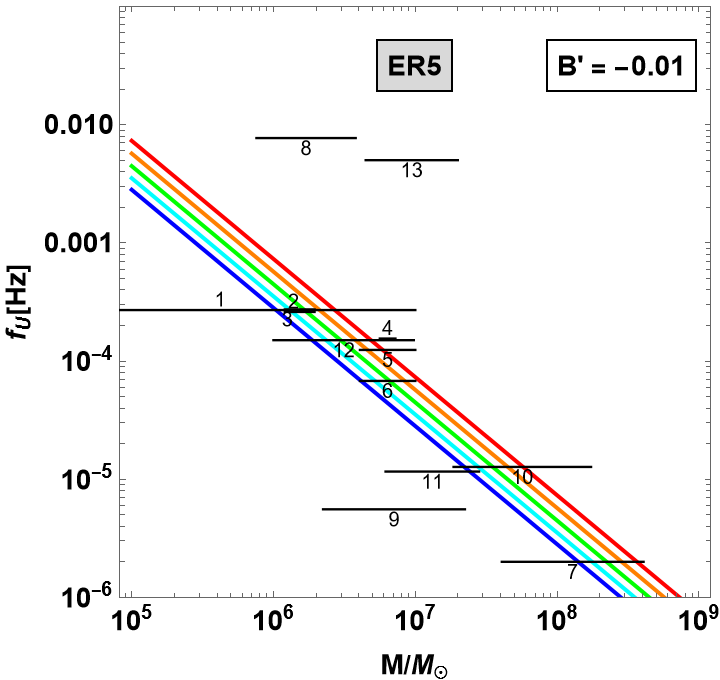}}
    \subfloat[]{\includegraphics[width=0.3\textwidth,height=4.8cm]{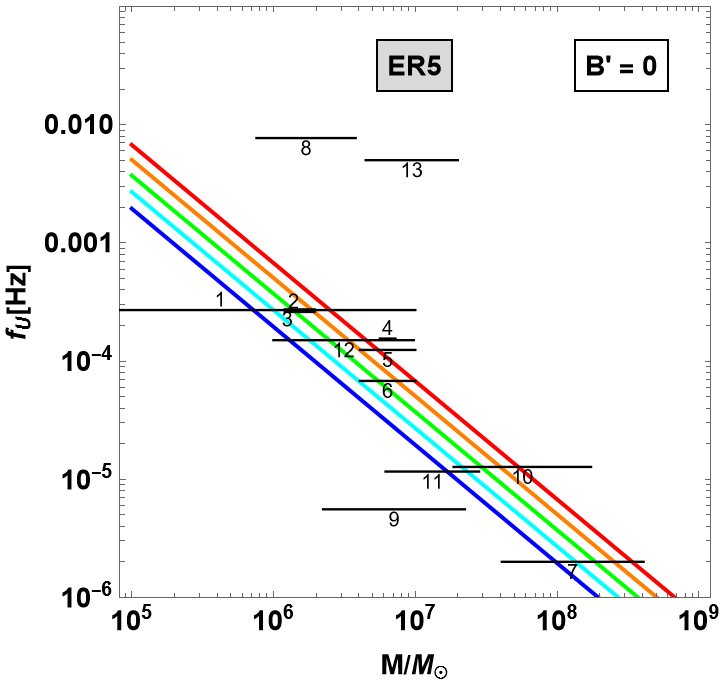}}
    \subfloat[]{\includegraphics[width=0.3\textwidth,height=4.8cm]{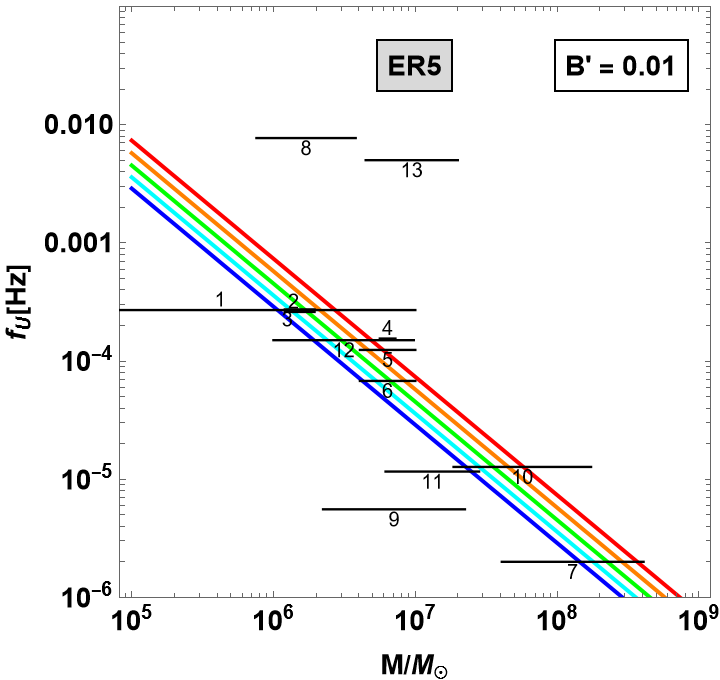}}\\
    \subfloat[]{\includegraphics[width=0.3\textwidth,height=4.8cm]{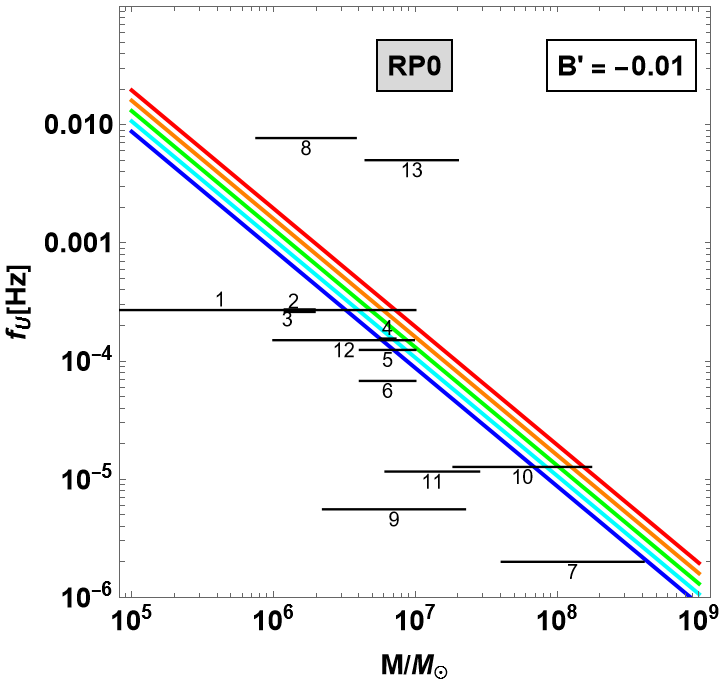}}
    \subfloat[]{\includegraphics[width=0.3\textwidth,height=4.8cm]{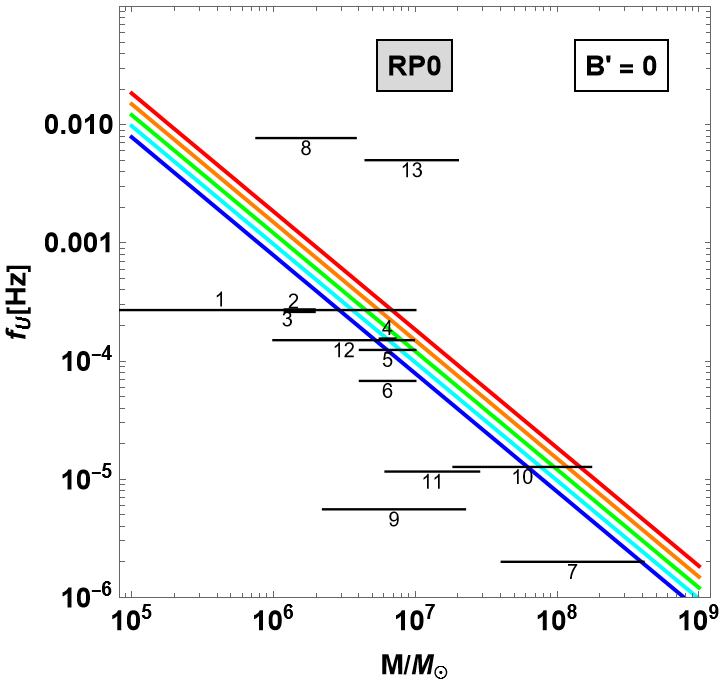}}
    \subfloat[]{\includegraphics[width=0.3\textwidth,height=4.8cm]{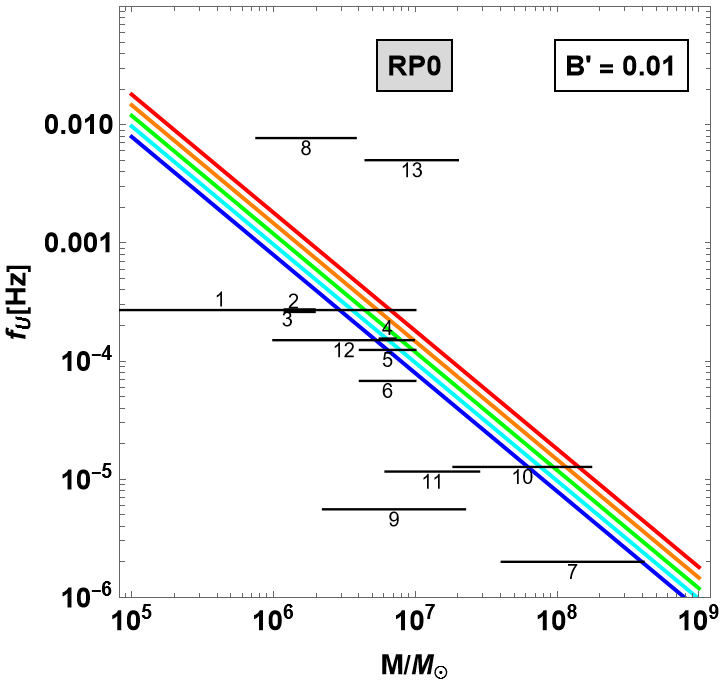}}\\
     \subfloat[]{\includegraphics[width=0.3\textwidth,height=4.8cm]{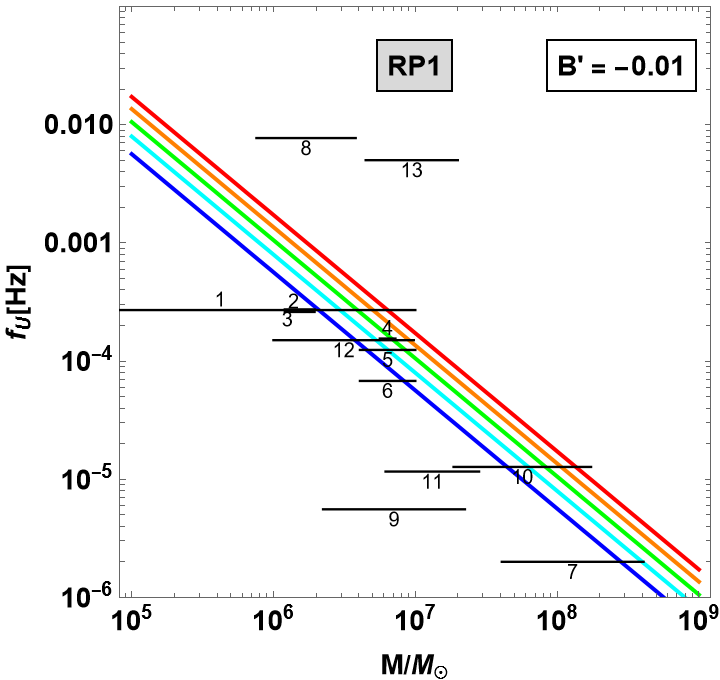}}
    \subfloat[]{\includegraphics[width=0.3\textwidth,height=4.8cm]{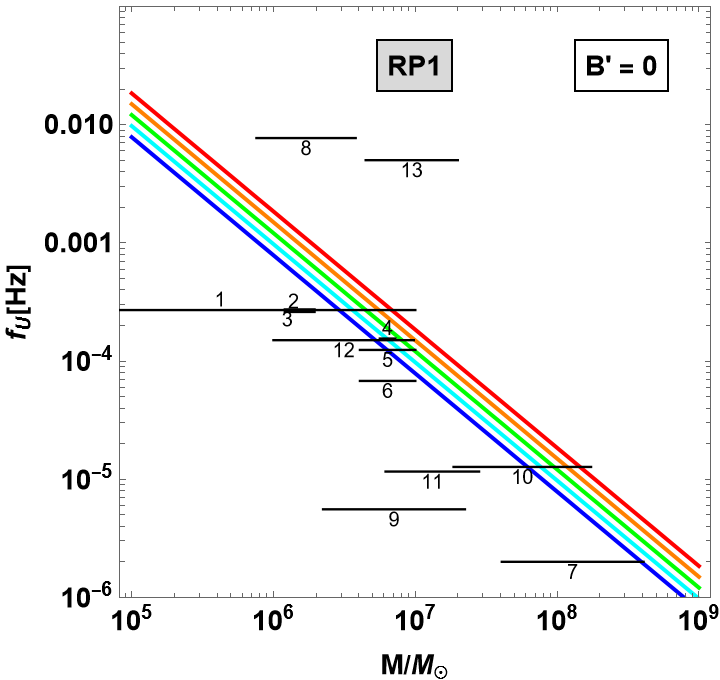}}
    \subfloat[]{\includegraphics[width=0.3\textwidth,height=4.8cm]{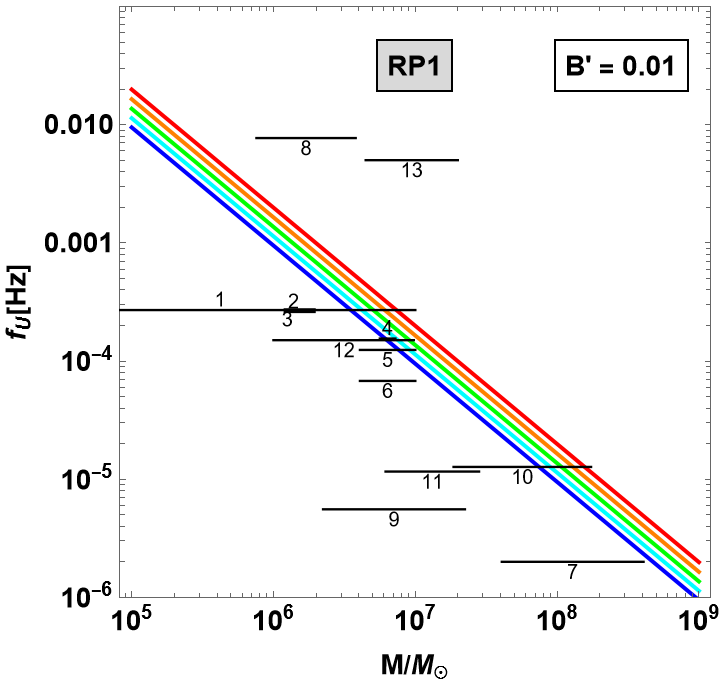}}
    \vspace{0.1em}
   \begin{tikzpicture}
  \matrix[column sep=0.3cm, row sep=0.2cm] {
    \draw[red, thick] (0,0) -- ++(1,0); & \node {$\bm{\kappa}$ = 0,}; &
    \draw[orange, thick] (0,0) -- ++(1,0); & \node {$\bm{\kappa}$ = 5,}; &
    \draw[green, thick] (0,0) -- ++(1,0); & \node {$\bm{\kappa}$ = 10,};& 
    \draw[cyan, thick] (0,0) -- ++(1,0); & \node {$\bm{\kappa}$ = 15,}; &
    \draw[blue, thick] (0,0) -- ++(1,0); & \node {$\bm{\kappa}$ = 20}; & & \\
  };
\end{tikzpicture}
    \caption{\footnotesize{\it  Fitting the observed HF QPOs frequencies in different models \textcolor{black}{ER4, ER5, RP0, RP1} of  the upper oscillation frequency $f_U$ at the resonance radii $3:2$ for different values of $\bm{\kappa}$ within $R_s=1000$ for different cases of magnetic field strength $B'$ \textcolor{black}{of supermassive black hole in AGNs of Tab.\ref{astrodata AGNs}} }}
    \label{AGNs ER4-RP1}
\end{figure}

\begin{figure}[!ht]
    \centering
    \subfloat[]{\includegraphics[width=0.3\textwidth,height=4.8cm]{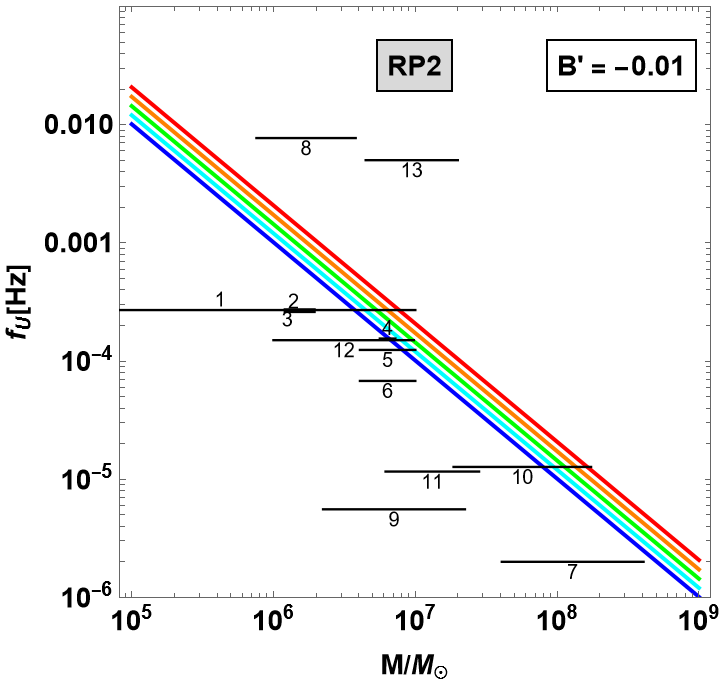}}
    \subfloat[]{\includegraphics[width=0.3\textwidth,height=4.8cm]{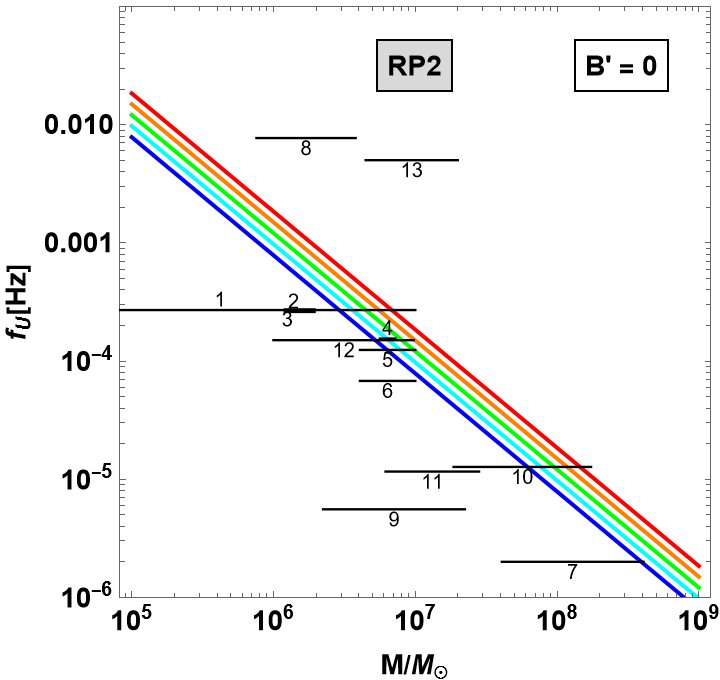}}
    \subfloat[]{\includegraphics[width=0.3\textwidth,height=4.8cm]{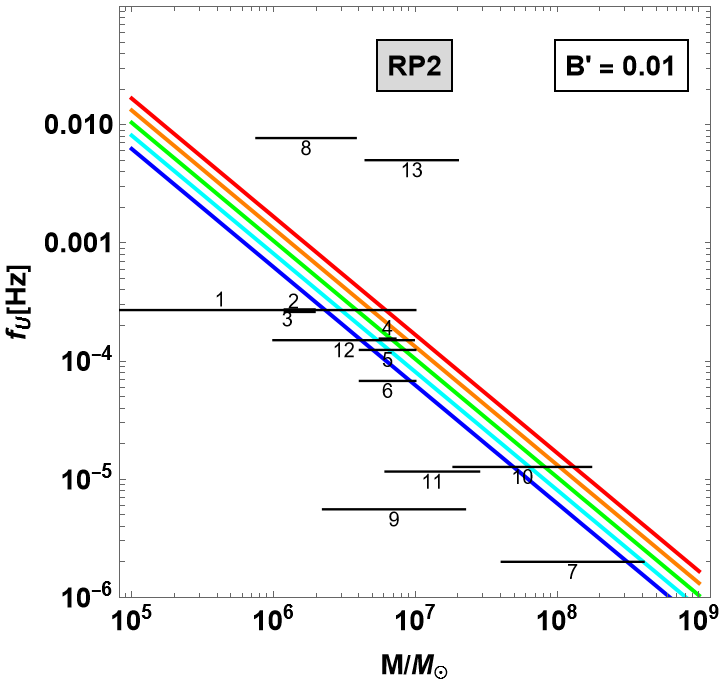}}\\
     \subfloat[]{\includegraphics[width=0.3\textwidth,height=4.8cm]{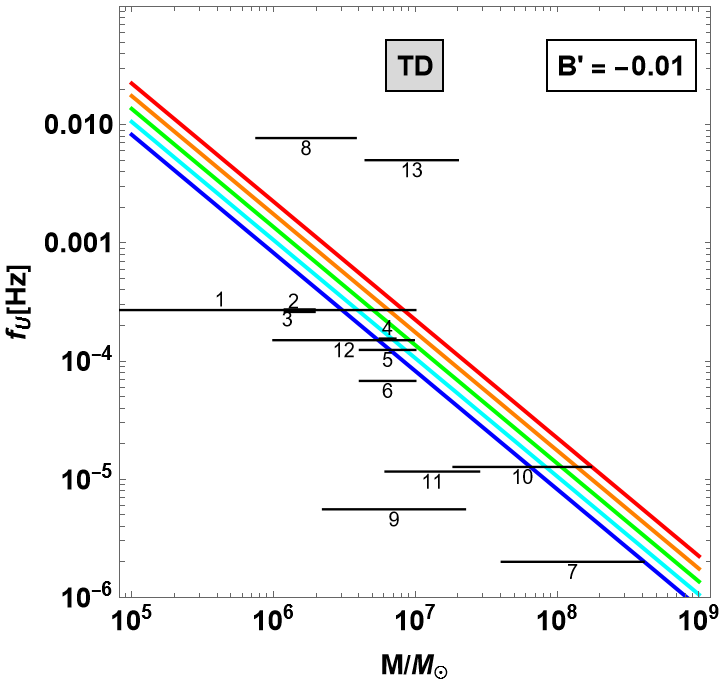}}
    \subfloat[]{\includegraphics[width=0.3\textwidth,height=4.8cm]{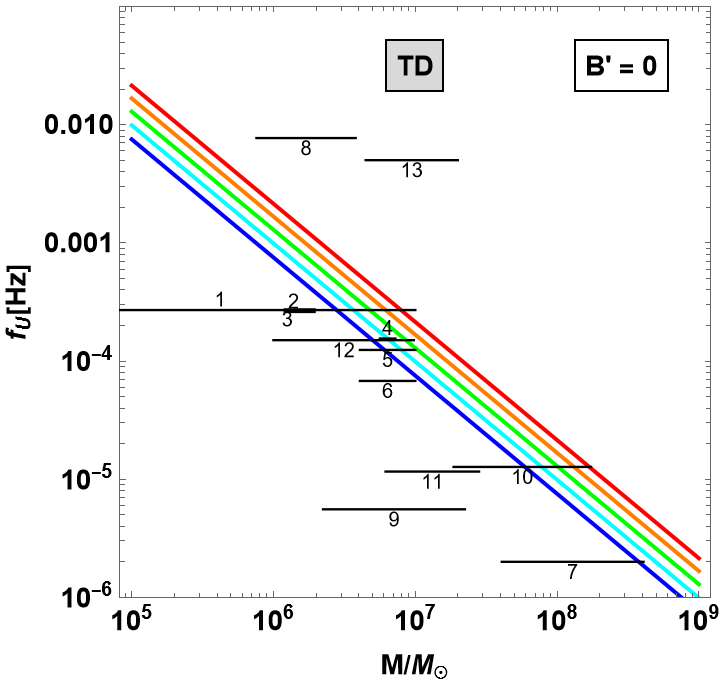}}
    \subfloat[]{\includegraphics[width=0.3\textwidth,height=4.8cm]{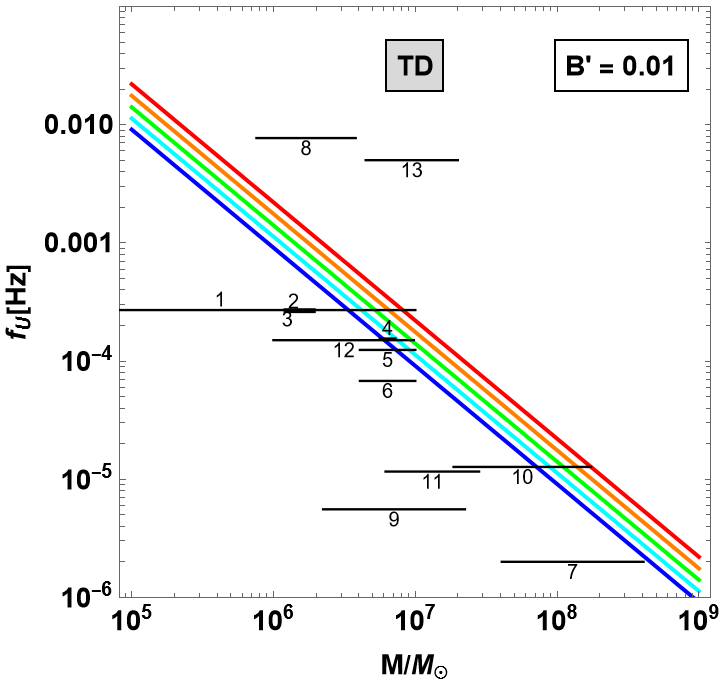}}\\
     \subfloat[]{\includegraphics[width=0.3\textwidth,height=4.8cm]{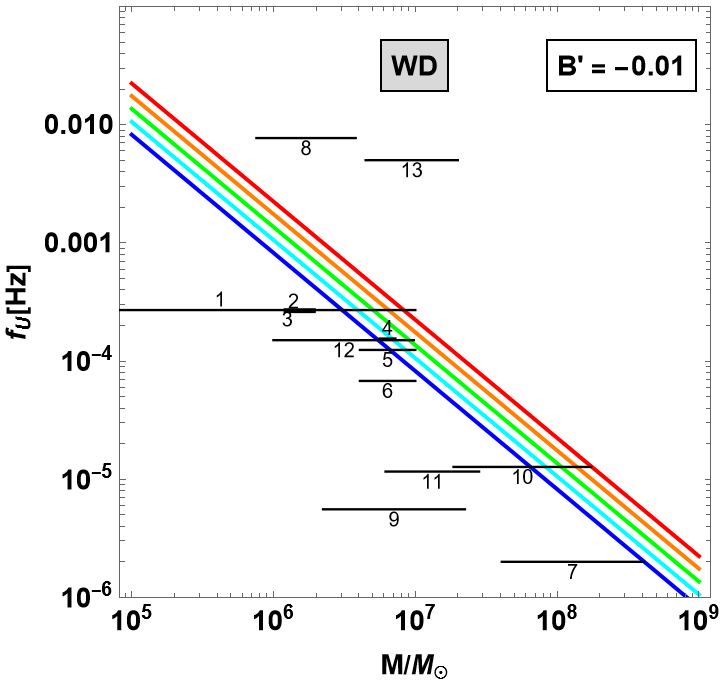}}
    \subfloat[]{\includegraphics[width=0.3\textwidth,height=4.8cm]{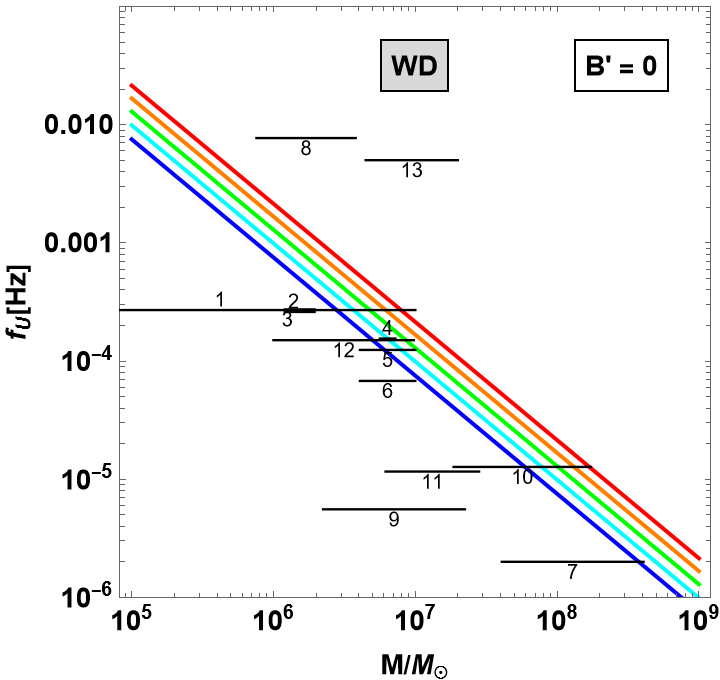}}
    \subfloat[]{\includegraphics[width=0.3\textwidth,height=4.8cm]{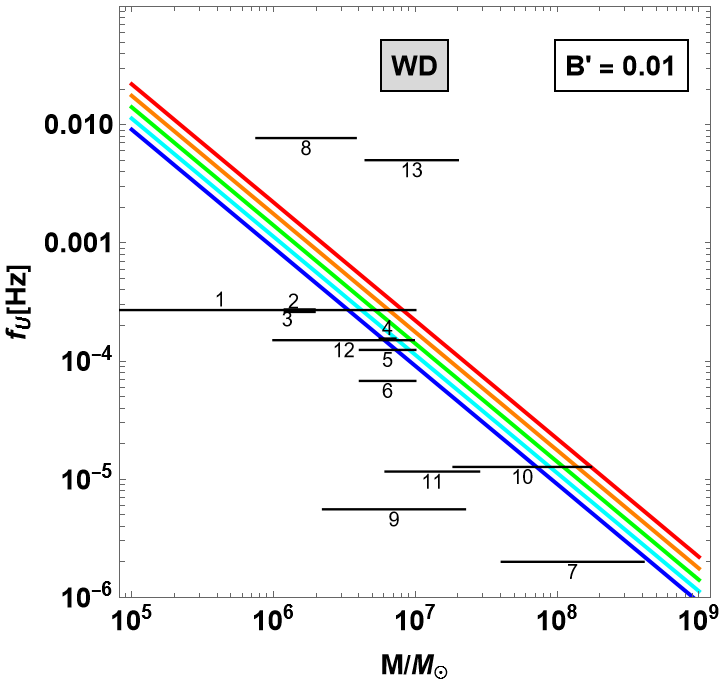}}
    \vspace{0.1em}
   \begin{tikzpicture}
  \matrix[column sep=0.3cm, row sep=0.2cm] {
    \draw[red, thick] (0,0) -- ++(1,0); & \node {$\bm{\kappa}$ = 0,}; &
    \draw[orange, thick] (0,0) -- ++(1,0); & \node {$\bm{\kappa}$ = 5,}; &
    \draw[green, thick] (0,0) -- ++(1,0); & \node {$\bm{\kappa}$ = 10,};& 
    \draw[cyan, thick] (0,0) -- ++(1,0); & \node {$\bm{\kappa}$ = 15,}; &
    \draw[blue, thick] (0,0) -- ++(1,0); & \node {$\bm{\kappa}$ = 20}; & & \\
  };
\end{tikzpicture}
    \caption{\footnotesize{\it  Fitting the observed HF QPOs frequencies in different models \textcolor{black}{RP2, TD, WD} of  the upper oscillation frequency $f_U$ at the resonance radii $3:2$ for different values of $\bm{\kappa}$ within $R_s=1000$ for different cases of magnetic field strength $B'$ \textcolor{black}{of supermassive black hole in AGNs of Tab.~\ref{astrodata AGNs}} }}
    \label{AGNs RP2-WD}
\end{figure}

\textcolor{black}{The fitting of HF QPO upper frequencies with a 3:2 ratio for supermassive black holes in AGNs is depicted in Figs.\ref{AGNs ER0-ER3},\ref{AGNs ER4-RP1}, and \ref{AGNs RP2-WD}, under three cases of $B' $ and various values of ${\bm\kappa}$.} 
\newpage

\textcolor{black}{
Regarding the supermassive black holes in active galactic nuclei, we observe that for weaker attractive or repulsive Lorentz forces, or in the absence of a magnetic field ($B' = \pm 0.01$ and $B' = 0$), only AGNs 8, 9, and 13 are excluded, as listed in Table~\ref{astrodata AGNs}. Most models yield valid fits for the remaining AGNs, although they differ in the CDM density range over which the fit is obtained. In particular, models such as ER1, ER2, ER4, and RP0 reproduce the same AGNs under both $B' = 0$ and $B' = \pm 0.01$, but with different values of $\kappa$. Moreover, the model that fits the majority of AGNs for $B' = \pm 0.01$ is ER5, which covers AGNs 1, 2, 3, 4, 5, 6, 7, 10, 11, and 12. For $B' = 0$, the model ER0 fits exactly the same set of AGNs as ER5 does for $B' = \pm 0.01$. It is worth noting that the majority of these fits occur at higher values of $\kappa$. Consequently, the model that fits the majority of microquasars is ER4 across all considered values of $B'$. In contrast, for AGNs, the best-fitting model depends on the magnetic field: ER5 provides the best fit for $B' = \pm 0.01$, while ER0 gives the best fit for $B' = 0$.}

\textcolor{black}{Our results are in good agreement with those reported in \cite{stuchlik2022geodesic}, which also provide estimates of the amount of dark matter surrounding black holes in both microquasars and supermassive black holes in AGNs. Consistent with our findings, for microquasars the ER4 model yields the best fits to the majority of the data. In the case $B'=0$, low CDM densities are sufficient, and \cite{stuchlik2022geodesic} further demonstrated that small amounts of dark matter can mimic the effects of black hole spin. Within the framework of our analysis, however, the magnetic field strength can act as an additional factor enhancing the role of dark matter around black holes. For AGNs, our results show that ER0 provides the best fits when $B'=0$, while ER5 dominates when $B'=\pm0.01$. This is once again consistent with \cite{stuchlik2022geodesic}, which found agreement across both low and high amounts of dark matter. Importantly, our study extends these results by explicitly demonstrating how magnetic field strength can complement dark matter effects, thereby refining the interpretation of HF QPOs in compact objects.}

\section{Conclusion and Discussion}

Charged particle trajectories close to a BH can generally be classified into three cases: particles can spiral into the BH, tend to infinite space, or orbit stably in circular trajectories around the BH. However, the behavior of these motions differs depending on whether the particles are charged or neutral, as well as on the nature of the spacetime in the region of the BH.

Throughout this study, we consider a Schwarzschild BH embedded in a uniform magnetic field and bordered by a CDM halo. Through our analysis of charged particle dynamics, we identified several arising from both the magnetic field and the CDM. Notably, CDM influences the BH geometry by affecting the site of the event horizon. As the CDM density increases—reflected by higher values of the parameter $\bm{\kappa}$—the event horizon grows. Consequently, the effective mass of the BH increases, resulting in a stronger gravitational pull that drives charged particles inward, effectively reducing the attribution of the magnetic field in that scheme.

Moreover, the location of the ISCO varies depending on the $B’$ strength and CDM density-$\bm{\kappa}$. These two factors act in opposition; a stronger magnetic field transitions the ISCO closer to the BH, while increasing CDM density pushes the ISCO outward. This opposing influence is also observed in the stability region: $B'$ tends to shrink the area of stability, whereas CDM expands it. In both positive and negative magnetic field cases, CDM contributes to enlarging the stability zone of charged particle motion. Indeed, 
at high CDM densities, the effective potential, represented by one or two boundaries depending on the sign of $B'$, tends to merge these boundaries. This causes the trapping of particles near the BH, increasing their likelihood of being captured. 
It's also observed that in the case of a positive uniform magnetic field, particles follow curled trajectories around the BH. The degree of this curvature increases with the magnitude of the magnetic field. However, CDM tends to reduce the curling of these trajectories. At high intensity of $B'$  and off the equatorial plane, ionized particle ejection i.e. jet phenomena can occur. CDM affects this behavior as well: as $\bm{\kappa}$ increases, particles are more strongly trapped near the BH. To overcome this trapping and enable particle escape, a stronger magnetic field is required for relativistic jets.

\textcolor{black}{Furthermore, to highlight the role of Lorentz force polarity in charged particle dynamics, we analyzed the behavior of the ionized Keplerian disk. For the attractive case, the Lorentz force leads to the destruction of the IKD at $B'={-0.01,\,-0.1}$. In particular, for $B'=-0.1$, chaotic motion emerges as particles spiral upward and downward along the uniform magnetic field lines. Both disk destruction and chaotic dynamics become more pronounced with increasing CDM density. At stronger magnetic fields ($B'=-1$), ionized particles undergo oscillations in both directions along the field lines; however, for sufficiently high CDM densities (e.g., $\bm{\kappa}=20$), these oscillations are suppressed near the ISCO, causing particles to plunge into the BH.  
In contrast, for the repulsive case, no disk destruction is observed for any of the considered field strengths or CDM densities. At $B'=0.01$, ionized particles follow curled orbits around the BH, with the curling angle decreasing as $\bm{\kappa}$ increases. For stronger fields ($B'={0.1,\,1}$), the trajectories resemble those in the attractive case, but without leading to IKD destruction.
}

Furthermore, we have examined the harmonic oscillations of charged particles orbiting a magnetized BH surrounded by a CDM halo. Our analysis has shown that the influence of CDM is significant both at small radii, nearby to the BH, and at large distances. In particular, at large distances, the radial frequency is not affected by the magnetic field, whether it is positive or negative, when CDM surrounds the BH. Regarding resonant phenomena, we have studied the variation in the location of the resonance radii corresponding to the 3:2 ratio between higher and lower frequencies across several HF QPO models. We have found that the presence of the CDM halo shifts the resonance radii farther from the BH in all considered models, namely ER, the RP model, the TD, and the WD models.

At the end, we have performed a fitting of the QPO data observed from four microquasars—GRS 1915+105, H1743-322, XTE J1550-564, and GRO J1655-40. \textcolor{black}{In addition, the observational data from supermassive BHs in AGNs (Table~\ref{astrodata AGNs}), was explored and  the effects of varying the CDM density parameter $\bm{\kappa}$ and magnetic field strength across several models: ER (ER0, ER1, ER2, ER3, ER4, ER5), RP (RP0, RP1, RP2), TD, and WD is investegated. For microquasars, the ER4 model consistently provides the best agreement across all magnetic field strengths. In contrast, for supermassive BHs in AGNs, the preferred model depends on the magnetic field polarity: ER0 fits best in the absence of a magnetic field ($B'=0$), whereas ER5 dominates for weak nonzero fields ($B'=\pm 0.01$). These findings suggest that ER4 is the most robust model for microquasars, while ER0 and ER5 are the most relevant for AGNs, thereby offering complementary insights into the origin of HF QPOs in compact objects.}

These findings motivate a deeper investigation into astrophysical processes involving Kerr and accelerating BHs,  which are widely regarded as realistic models of rotating and binary BHs in the universe. In future work, however, we plan to extend our analysis of HF QPOs by incorporating the effects of magnetic fields and rotational velocities directly into the fitting process, for both microquasars and AGNs. This direction is particularly motivated by the results presented in \cite{stuchlik2022large}, which serve as a valuable reference point for initiating such an extended study. A compelling open question arises: to what extent does DM contribute to the luminosity and structure of accretion disks around rotating BHs, in conjunction with the well-established influence of magnetic fields and their varying intensities?

\section*{Acknowledgements}
\paragraph{}H. El M would like to acknowledge networking
support of the COST Action 
 CA 22113 - Fundamental challenges in theoretical physics (Theory and Challenges), 
CA 21136 - Addressing observational tensions in cosmology with systematics and fundamental physics (CosmoVerse), and 
CA 23130 - Bridging high and low energies in search of quantum gravity (BridgeQG). He also thanks IOP for its support.

Z. A expresses gratitude for the financial support he receives from the National Center for Scientific and Technical Research (CNRST) of Morocco.

This work was carried out under the project UIZ 2025 Scientific Research Projects: {\tt PRJ-2025-81}.
\bibliographystyle{unsrt}
\bibliography{main}
\end{document}